%% file: main.tex
\documentclass[a4paper]{aa}
\usepackage{graphics,epsfig,txfonts}
\usepackage[utf8]{inputenc}
\usepackage[section]{placeins}
\usepackage{amsmath}
\usepackage{natbib}
\usepackage[normalem]{ulem} 

\usepackage{color}
\usepackage{xcolor}

\bibpunct{(}{)}{;}{a}{}{,}

\def \MSUN{{\rm M}_{\odot}}
\def\vmeas {\ifmmode{\mathrm{v}} \else {$\mathrm{v}$}\fi} 
\def\vsys {\ifmmode{V_\mathrm{sys}} \else {$V_\mathrm{sys}$}\fi} 
\def\vrot {\ifmmode{V_\mathrm{rot}} \else {$V_\mathrm{rot}$}\fi} 
\def\PAkin {\ifmmode{PA_\mathrm{{kin}}} \else {$PA_\mathrm{{kin}}$}\fi}  
\def\PAphot {\ifmmode{PA_\mathrm{{phot}}} \else {$PA_\mathrm{{phot}}$}\fi}  
\def\re {\ifmmode{R_\mathrm{{e}}} \else {$R_\mathrm{{e}}$}\fi}   
\def \rsoft {\ifmmode{r_\mathrm{{soft}}} \else {$r_\mathrm{{soft}}$}\fi} 
\def\lec{{\leq\,}}

\title{The extended Planetary Nebula Spectrograph (ePN.S) early-type galaxy survey: The specific angular momentum of ETGs}
\author{Claudia Pulsoni\inst{1}
     \and {Ortwin Gerhard\inst{1}}
     \and {S. Michael Fall\inst{2}}
     \and {Magda Arnaboldi\inst{3}}
     \and {Ana I. Ennis\inst{4,5,6,7}}
     \and {Johanna Hartke\inst{8,9,10}}
     \and {Lodovico Coccato\inst{3}}
     \and {Nicola R. Napolitano\inst{11,12,13}}
     }
     
\titlerunning{The specific angular momentum of ETGs}
\authorrunning{C. Pulsoni et al.}

\institute{
Max-Planck-Institut f\"ur extraterrestrische Physik, Giessenbachstra{\ss}e, 85748 Garching, Germany
\and Department of Physics and Astronomy, Johns Hopkins University, 3400 N. Charles Street, Baltimore, MD 21218
\and European Southern Observatory, Karl-Schwarzschild-Str. 2, 85748 Garching, Germany
\and Facultad de Ciencias Astronómicas y Geofísicas de la Universidad Nacional de La Plata, and Instituto de Astrofísica de La Plata
(CCT La Plata – CONICET, UNLP), Paseo del Bosque S/N, B1900FWA La Plata, Argentina
\and Consejo Nacional de Investigaciones Científicas y Técnicas, Godoy Cruz 2290, C1425FQB, Ciudad Autónoma de Buenos Aires, Argentina
\and Waterloo Centre for Astrophysics, University of Waterloo, 200 University Ave W, Waterloo, Ontario N2L 3G1, Canada 
\and Perimeter Institute for Theoretical Physics, Waterloo, Ontario N2L 2Y5, Canada
\and European Southern Observatory, Alonso de Córdova 3107, Santiago de Chile, Chile
\and Sub-Department of Astrophysics, Department of Physics, University of Oxford, Denys Wilkinson Building, Keble Road, Oxford OX1 3RH, UK
\and Finnish Centre for Astronomy with ESO (FINCA), University of Turku, FI-20014 Turku, Finland
\and School of Physics and Astronomy, Sun Yat-sen University, Zhuhai Campus, 2 Daxue Road, Xiangzhou District, Zhuhai, P. R. China
\and CSST Science Center for Guangdong-Hong Kong-Macau Great Bay Area, Zhuhai, China, 519082
\and INAF – Osservatorio Astronomico di Capodimonte, Salita Moiariello 16, 80131 - Napoli, Italy
} 
\date{\today}

\abstract{
    Mass and angular momentum are key parameters of galaxies. Their co-evolution establishes an empirical relation between the specific stellar angular momentum $j_*$ and the stellar mass $M_*$ that depends on morphology. 
}{
    In this work, we measure $j_*$ in a sample of 32 early type galaxies (ETGs) from the ePN.S survey, using the full two-dimensional kinematic information. We present local $\lambda$ profiles and projected $j_*$ profiles in apertures. 
    We derive the distribution of these galaxies on the total $j_*-M_*$ plane and determine the ratio between the stellar $j_*$ and the specific angular momentum of the host dark matter halo. 
}{
    We use integral-field-spectroscopic data in the central regions (1-2 effective radii, $R_e$) and planetary nebula (PN) kinematics in the outskirts (out to a mean $6R_e$). In the $j_*$ determination, we account for misaligned rotation and for the differences between light-weighted $j_*$ and mass-weighted $j_*$, estimating also the effects of gradients in the mass-to-light ratio driven by variations in the initial-mass-function. We use simulated ETGs from the IllustrisTNG simulation TNG100 to correct for the limited radial coverage of the PN data and to account for projection effects on $j_*$.
}{
    The radially extended, two-dimensional kinematic data show that the stellar halos of ETGs do not contain large stellar mass fractions of high $j_*$. The $j_*$-profiles of fast-rotator ETGs are largely converged within the range of the data. For slow rotators, $j_*$ is still rising and is estimated to increase beyond $6R_e$ by up to 40\%, using simulated galaxies from TNG100. More than $60\%$ of their stellar halo angular momentum is in misaligned rotation. We find that the ePN.S ETG sample displays the well-known correlation between $j_*$, $M_*$, and morphology: elliptical galaxies have systematically lower $j_*$ than similar mass S0 galaxies. However, fast and slow rotators lie on the same relation within errors with the slow rotators falling at the high $M_*$ end.
    A power-law fit to the mass-weighted $j_*-M_*$ relation gives a slope of $0.55\pm0.17$ for the S0s and $0.76\pm0.23$ for the ellipticals, with normalisation about 4 and 9 times lower than spirals, respectively. The estimated retained fraction of angular momentum at $10^{10}\leq M_*\leq 10^{10.5}M_\odot$ is $\sim25\%$ for S0s and $>10\%$ for ellipticals, and decreases by $\sim1.5$ orders of magnitude at $M_*\sim10^{12}M_\odot$.
    }{
    Our results show that ETGs have substantially lower $j_*$ than spiral galaxies with similar $M_*$. Their $j_*$ must be lost during their evolution, and/or retained in the hot gas component and the satellite galaxies that have not yet merged with the central galaxy.
    }


\begin{document}

\maketitle

\section{Introduction}

\subsection{Specific angular momentum and galaxy formation}

Galaxies acquire angular momentum primordially by gravitational torques induced by the large scale tidal fields \citep{1969ApJ...155..393P}. Then, by approximate conservation of the angular momentum, the collapsing star-forming gas builds up the galaxy rotation \citep{1979Natur.281..200F, 1980MNRAS.193..189F}. The progenitors of present-day early type galaxies (ETGs) are mostly fast rotating, disk-like objects \citep[e.g.,][]{2017MNRAS.468.3883P, Lagos2017}. 
Subsequently, during their evolution, ETGs can undergo several physical processes that lead to a gain or loss of specific angular momentum (sAM), that is the angular momentum per unit mass of the stellar component $j_*$. 

Since $z\sim1$, massive ETGs, with $M_* \gtrsim 10^{10.5}M_\odot$, tend, on average, to decrease rotational support (\citealt{2017ApJ...837...68C, 2020MNRAS.494.5652W}) due to the effect of mergers \citep[e.g.,][]{2009MNRAS.397.1202J,2014MNRAS.444.3357N, Lagos2018_MergersAM}. Gas poor mergers increase the mass and the velocity dispersion of the stars, while destroying ordered rotation \citep[e.g.,][]{2009MNRAS.397.1202J} and redistributing AM mostly to the dark matter halos by dynamical friction \citep[e.g.,][]{1988ApJ...331..699B}. 
The presence of gas, either in the satellite or in the host galaxy, instead leads to a net spin-up of the merger remnant \citep[e.g.,][]{2014MNRAS.444.3357N,2017MNRAS.468.3883P}. These different formation paths are thought to be at the base of the evolution of the class of slow rotators (SRs) from the fast rotators (FRs) between $z=1$ and 0 \citep[][]{2017MNRAS.468.3883P, Lagos2017, 2018MNRAS.480.4636S}. 

Another parameter impacting $j_*$ is the epoch of latest gas accretion. Tidal torques theory predicts that gas infalling at later times has higher sAM \citep{1996MNRAS.282..436C}. In addition, simulations also predict a change with time in the main accretion mode of galaxies, from filamentary at $z\gtrsim1$ to gas cooling from a hydrostatic halo which is more efficient in spinning up galaxies \citep[e.g.,][]{2018MNRAS.481.4133G}. Therefore, galaxies that form most of their stars early and that do not accrete gas at recent times, for example because prevented by AGN feedback, have systematically lower $j_*$ \citep{Lagos2017, Rodriguez-Gomez2022}. 

The co-evolution of mass and AM establishes an empirical relation between the sAM of the stellar component, $j_{*}$, and the total stellar mass $M_{*}$. \cite{Fall1983} found that galaxies distribute according to a power-law $j_{*}\propto M_{*}^{\alpha}$ with $\alpha\sim0.6$, also called the Fall relation. The proportionality constant tightly correlating with the bulge-to-total mass fraction or the Hubble type 
\citep{2016MNRAS.463..170C,  Fall2018}. This value of $\alpha$ is remarkably close to the expected $\alpha = 2/3$ for dark matter halos from tidal torque theory \citep{1969ApJ...155..393P, 1979MNRAS.186..133E}. Therefore, the observed $j_*-M_*$ relation for disk galaxies has been interpreted as resulting from the approximate conservation of primordial angular momentum of the stellar component, which is similarly torqued as the dark matter halo \citep[][]{Fall1983,RomanowskyFall2012}. 

The $j_*-M_*$ relation for spiral galaxies is now well established for a large range of masses \citep[][and references therein]{2018A&A...612L...6P, 2021A&A...647A..76M, 2023MNRAS.518.6340D}, facilitated by the fact that for exponential disks the $j_{*}(\lec r)$ converges rapidly beyond $2R_e$.
ETGs are found to roughly follow a parallel sequence to the spirals, with approximately five times lower $j_{*}$ in ellipticals \citep{Fall2013} and eight times lower $j_{*}$ in bulge-only galaxies \citep{Fall2018}.
However, in this case, the measurement of the total $j_*$ is challenging and consequently the $j_*-M_*$ relation for these galaxies is far less explored than for the late-types.

\subsection{Measuring $j_*$ in ETGs}

The case of massive ETGs is of particular interest since their evolution is dominated by mergers which have a strong effect on $j_*$. However, as mentioned above, the inclusion of ETGs in the $j_*-M_*$ diagram is challenging. Following the pioneering work of \cite{Fall1983}, the only work to-date that has attempted such a measurement by integrating velocity profiles of ETGs out to large radii is \citet{RomanowskyFall2012}. As discussed by these authors, the issue resides in the larger S{\'e}rsic indices of ETGs compared to disk galaxies, which imply that a larger fraction of their light, and therefore of their total AM, is distributed in the external regions. 
Hence, accurate measurements of $j_*$ in ETGs require extended kinematic measurements, out to radii that are inaccessible to stellar absorption-line spectroscopy.
These are possible only through alternative kinematic tracers such as planetary nebulae \citep[PNe, e.g.,][]{2009MNRAS.394.1249C} or globular clusters  \citep[GCs, e.g.,][]{2010A&A...513A..52S,2011ApJS..197...33S}. 

Extended kinematic studies of ETGs \citep{2016MNRAS.457..147F, Pulsoni2018, 2021MNRAS.504.4923D} revealed that these galaxies can display a large variety of kinematic behaviors, including embedded disks, strongly rotating outskirts, twisting velocity fields and multiple rotating components.
The presence of these features in both FRs and SRs suggests that ETGs stellar halos are often triaxial. 
The kinematic diversity in ETG stellar halos emphasizes the importance of an approach based on a two-dimensional kinematic mapping to estimate their total $j_*$, sufficiently extended to trace the variations of rotation amplitudes and direction with radius. 

Another complication is the estimate of the projection effects on $j_*$, because of the three-dimensional geometry of ETGs compared to disk-dominated systems and their complex kinematics at large radii. \cite{RomanowskyFall2012} tackle this issue using randomly-oriented, simple axisymmetric models with cylindrical velocity fields. These assumptions, however, are not necessarily valid for ETGs and might bias the determination of $j_*$. 
For example, the velocity fields of regularly rotating FRs are often characterised by a "spider" morphology, with rotation amplitude decreasing above and below the projected major axis \citep[e.g.,][]{2011MNRAS.414.2923K}. Then assuming a cylindrical morphology with constant rotation amplitude above and below the major axis systematic overestimates $j_*$.


\subsection{This paper}

This paper is part of the extended Planetary Nebula Spectrograph (ePN.S) survey which uses PNe to sample the kinematics of the stellar halos in ETGs  \citep{2017IAUS..323..279A}. PNe are established probes of the stellar population in ETG halos and are good kinematic traces of the bulk of the host-galaxy stars \citep[e.g.][]{2009MNRAS.394.1249C, 2013MNRAS.432.1010C}.  
To-date, several studies demonstrate that the PNe spatial distribution follows the surface brightness of the host galaxy and that their kinematics is directly comparable to integrated light measurements \citep{1995ApJ...449..592H, 1996ApJ...472..145A, 2001ApJ...563..135M, 2009MNRAS.394.1249C, 2013A&A...549A.115C}.

The goal is to use the ePN.S kinematic data out to large radii to measure $j_*$ in 32 ETGs using the full two-dimensional kinematic information.
This increases by a factor of four the sample of ETGs of \citet{RomanowskyFall2012} for which the sAM has been calculated from similarly extended velocity data.\footnote{For the other 32 ETGs presented in that work, the approximation $j_t\propto R_e V_{rot}(2R_e)$ was used.}. To do this, we complemented the PN kinematics with absorption line kinematics from integral-field-spectroscopy (IFS) in the central regions available in the literature or newly extracted from archive MUSE cubes. 
We correct for projection effects by using simulated galaxies from the IllustrisTNG cosmological simulation as physically motivated models, which have been found to reproduce well the $j_*-M_*$ relation and its dependency on morphology \citep{2023MNRAS.518.6340D, Rodriguez-Gomez2022}.

Previous $j_*$ determinations for ETGs are based on assuming constant mass-to-light ratios with radius.
In this paper, we examine this assumption by considering both blue photometric bands (i.e., $B$, $V$, or $g$) and the infrared emission at 3.6 $\mu$m, which is a good proxy for the stellar mass \citep[e.g.,][]{Forbes2017_Spitzer}. We also explore the effects of IMF gradients on the distribution of ETGs in the $j_*-M_*$ diagram using results from stellar population studies from the literature.

The paper is structured as follows. Section~\ref{sec:thedata} describes the data used in this work (Sects.~\ref{sec:ePNS_survey}, ~\ref{sec:kinematic_data_center}, \ref{sec:photometric_data}, and Sect.~\ref{sec:additional_data}) and the procedures to reconstruct the 2D velocity fields and images (Sects.~\ref{sec:Vfields} and \ref{sec:photometry}). We derive differential $\lambda(R)$ profiles in Sect.~\ref{sec:lambda_profiles} and aperture projected $j_*(R)$ profiles in Sect.~\ref{sec:ePN.S_jp}. Section~\ref{sec:ePN.S_jp_M*} contains the main observational result of this paper, the dependence of the projected $j_*$ on the stellar masses of ETGs.
In Sect.~\ref{sec:IllustrisTNG} we compare the distribution of projected $j_*$ of simulated TNG100 ETGs with the ePN.S galaxies (Sect.~\ref{sec:jp_TNG_VS_ePNS}) and estimate the correction for the limited radial coverage of the PN to estimate the total projected sAM (Sect.~\ref{sec:fract_j_beyond_6Re}). In Sect.~\ref{sec:deprojecting_j}, the TNG100 ETGs are used to evaluate projection effects on $j_*$. In Sect.~\ref{sec:ePNS_jt_M*_relation} we derive the total $j_*-M_*$ relation for the ePN.S ETGs and in Sect.~\ref{sec:retained_retention_fraction_fj} we estimate the retained fraction $f_j$ of halo sAM as a function of $M_*$. Finally in Sect.~\ref{sec:conclusions} we draw our conclusions.

\section{The data}\label{sec:thedata}

\subsection{The ePN.S survey and the ETG sample}\label{sec:ePNS_survey}

The ePN.S survey aims to investigate the kinematics, the dynamics, the angular momentum, and the mass distribution in the halos of ETGs using PNe as kinematic tracers where the surface brightness is too low for absorption-line spectroscopy. 
The advantage of using PNe over other tracers is that they sample the stellar kinematics in ETG halos \citep{1995ApJ...449..592H, 1996ApJ...472..145A, 2001ApJ...563..135M, 2009MNRAS.394.1249C, 2013A&A...549A.115C}, out to very large radii \citep{2015A&A...579A.135L,2018A&A...616A.123H,2022A&A...663A..12H}. 

The ePN.S survey targets a sample of 32 nearby ETGs with absolute magnitudes $-22.38 > M_K > -26.02$, distances $\leq25$ Mpc, and covering a wide range of internal parameters (i.e. luminosity, central velocity dispersion, ellipticity, boxy/diskyness, see Fig.~2 in \citealt{2017IAUS..323..279A}). Thus the sample includes a representative subset of nearby bright ETGs. Compared to a magnitude-limited sample of ETGs such as, for example, Atlas3D \citep{A3D_I_Cappellari2011}, the ePN.S galaxies are on average more massive and have lower ellipticities (see Fig.~11 in \citealt{Pulsoni2018}).

The ePN.S ETGs include 24 fast (FRs) and 9 slow rotators (SRs) according to the classification of \cite{2011MNRAS.414..888E}, such that SRs have $\lambda_e \leq 0.31 \varepsilon$. In this paper, we follow their definition and refer to FRs as the ensemble of fast rotating ellipticals and S0s, but we also refer separately to the fast rotating ellipticals as E-FRs. The ePN.S sample also contains the two major mergers remnants NGC1316 and NGC5128, which are interesting cases for studying angular momentum transport to the galaxy outskirts by dynamical friction \citep[e.g.,][]{1987ApJ...319..575B,1988ApJ...331..699B, 1994MNRAS.267..401N}.

The survey is based on PN observations mostly done with the Planetary Nebula
Spectrograph (PN.S) at the William Herschel Telescope in La Palma \citep{2002PASP..114.1234D}, but also includes two catalogs from Counter Dispersed Imaging with FORS2@VLT, and six further catalogs from the literature (references in Table 1 in \citealt{Pulsoni2018}), for a total of 32 ETGs. 
The catalogs contain a total of 8636 PNe, making the ePN.S the largest kinematic survey to-date of extra-galactic PNe in the outer halos of ETGs. The data cover 4, 6, and 8 effective radii ($R_e$) for, respectively, 85\%, 41\%, and 17\% of the sample, and with median extension of 5.6$\re$ (see \citealt{Pulsoni2018}). 

The procedure of outlier removal and construction of Bona Fide PNe catalogs is described in \cite{Pulsoni2018}. We refer to that paper for a detailed kinematic analysis of the ePN.S sample and to the procedure to derive smoothed velocity and velocity dispersion fields.

\subsection{Kinematic data in the central regions}\label{sec:kinematic_data_center}

In order to achieve a complete two-dimensional map of the ETG kinematics, we combine the PN smoothed mean velocity and velocity dispersion fields with two-dimensional kinematics maps from IFS for the central regions ($R\lesssim1-2 R_e$), where the PN detection is incomplete. A good fraction (24/32) of the ePN.S galaxies is part of the Atlas3D survey \citep{2004MNRAS.352..721E, A3D_I_Cappellari2011}, which made available the full two-dimensional velocity fields\footnote{http://www-astro.physics.ox.ac.uk/atlas3d/}. In addition to these, NGC3115 has available MUSE IFS data from \cite{2016A&A...591A.143G}.

For NGC0584, NGC1316, NGC1399, and NGC4594 we used reduced data cubes from the ESO science archive\footnote{Based on observations made with ESO Telescopes at the La Silla Paranal Observatory under program IDs 097.A-0366, 094.B-0298, 60.A-9303}. The analysis was carried out using the GIST pipeline \citep{GISTpipeline}.
As a first step, the pipeline shifts the spectra to rest-frame, and applies any necessary spatial masks to the data. In most cases we used a target signal-to-noise ratio of $200-250$ to increase the signal-to-noise ratio of the data using Voronoi binning \citep{VoronoiBinning}, and then derived the stellar kinematics using PPxF \citep{ppxf} and MILES templates \citep{sanchezblazquez06, falconbarroso11}. 
We restrict our analysis to $4800-6000$ \r{A}  to avoid emission lines that affect the calculation of the velocity dispersion \citep[e.g.,][]{2018A&A...609A..78B} and we masked any strong emission lines and residual sky lines. 
The mean velocity and velocity dispersion profiles for these galaxies are shown in Appendix~\ref{sec:MUSE_kin_profiles}. The full 2D velocity fields will be made available in a future publication (Ennis et al. in prep.).

For 2/32 ePN.S ETGs, that is NGC3923 and NGC4742, IFS kinematic data are not available. Major and minor axis slit data from \cite{1998MNRAS.294..182C} show that NGC3923 has negligible rotation in the center, therefore most of the contribution to its sAM comes from the outskirts. Therefore for NGC3923 we only use the PN kinematics. 
For NGC4742 we use the major axis slit data from \cite{1983ApJ...266...41D}.

Finally, the central regions of NGC5128 have been observed with several MUSE programs. However, due to the presence of the extended dust lane in the centre of the galaxy, it was not possible to derive kinematics from the archival MUSE data covering the central $\sim 6.5$ arcminutes. On the other hand, this galaxy has the richest PN catalog of the sample, with 1222 PNe distributed from 2 to more than 50 kpc, i.e. from $\sim 0.5$ out to almost 20 $R_e$. Because of the excellent coverage of the PN data and because most the AM of this galaxy is distributed at large radii (see Fig.~\ref{fig:jp_profiles}), we do not consider additional kinematic data for this galaxy and derive its sAM from the PN kinematics only.

We also use SLUGGS kinemetry results from \citet{2016MNRAS.457..147F}, that is rotation velocity, kinematic position angle, and velocity dispersion profiles available for 18/32 ePN.S galaxies. These data are based on kinematic maps from observations using slitlets and extend out to typically $\sim3R_e$. The kinemetric profiles are used to bridge the radial gap between IFS velocity fields and PN data when necessary, as described in Sect.~\ref{sec:Vfields}.

Table~\ref{tab:kinematics_AND_photometry_data} summarises the kinematic data used for each galaxy.

\subsection{Photometric data}\label{sec:photometric_data}

We use the most radially extended photometric data available in the literature, which are typically in the optical $B$, $V$, or $g$ bands \citep[e.g.,][see Table~\ref{tab:kinematics_AND_photometry_data}]{1990A&AS...86..429C,2009ApJS..182..216K, 2017ApJ...839...21I,2017A&A...603A..38S}. For most of the ePN.S galaxies, extended ellipticity and photometric position angle profiles are also available. 
For NGC3489, NGC4339, NGC4742, and NGC5128, whose ellipticity and position angle profiles are not available in the literature, we assume constant ellipticity and position angle with radius, equal to the average $\langle \varepsilon\rangle$ and $\langle \PAphot{} \rangle $ values listed in Table~1 in \cite{Pulsoni2018}. In particular, for NGC5128 we use the ellipticity estimated by \cite{2022A&A...657A..41R}.

We also consider light profiles extracted from the infrared (IR) $3.6\mu$m imaging with the Spitzer Space Telescope and published by \cite{Forbes2017_Spitzer}. The Spitzer data are available for 20/32 ePN.S ETGs and typically cover radii out to 100-200 arcsec, depending on the galaxy. Hence, we use the S{\'e}rsic fits from \cite{Forbes2017_Spitzer} to extrapolate the light profiles to larger radii, out to the radial extent of the ePN.S data. Ellipticity and position angle profiles at $3.6\mu$m are unfortunately not provided. Therefore, we assume that the shape of the isophotes at these longer wavelengths are the same as in the bluer bands.
Table~\ref{tab:kinematics_AND_photometry_data} summarizes and collects the references of the photometric data used in this paper.

\subsection{Distances, effective radii, and stellar masses}\label{sec:additional_data}

The distances of the ePN.S galaxies are derived from the surface brightness fluctuation method and are listed and referenced in Table 1 of \cite{Pulsoni2018}, together with the adopted effective radii $R_e$. These are circularised $R_e$, i.e., the semi-major axis of the ellipse enclosing half of the galaxy light multiplied by the squared-root of the axis-ratio.
The stellar masses $M_*$ are estimated from their total absolute K-band magnitudes $M_K$ obtained from the 2MASS extended source catalog \citep{2003AJ....125..525J}, assuming the distances above. We corrected for the over-subtraction of the sky background by the 2MASS data reduction pipeline \citep{2012PASA...29..174S} using the formula $M_{K_{corr}}=1.07M_K+1.53$ provided by \cite{2013ApJ...768...76S}, and converted to stellar masses following the relation $\log_{10}M_{*}=10.39-0.46(M_{K_{corr}}+23)$ from \citet{2019MNRAS.484..869V}. This is based on the mass-to-light ratio from the stellar population modelling of \cite{2013MNRAS.432.1862C}, converted to a \citet{ChabrierIMF} initial mass function (IMF). These $M_*$ values are in good agreement with the stellar masses derived using a mass-to-light ratio dependent on the $(B-V)_0$ color as in \citet[][where $(B-V)_0$ is the extinction corrected color from Hyperleda\footnote{http://leda.univ-lyon1.fr/}]{Fall2013}, and with the stellar masses obtained by \citet{Forbes2017_Spitzer} from the Spitzer 3.6$\mu$m luminosity, but converted from a \cite{KroupaIMF} to a Chabrier IMF by a factor 0.92 \citep{2014ARA&A..52..415M} and adjusted to the adopted distances (see Fig.~\ref{fig:ComparisonStellarMasses} in Appendix). The mean variation between $M_*$ values obtained from the three methods for the same galaxies is $\sim0.05$ dex, which we consider as uncertainty on $M_*$. This does not include the uncertainty on the distances, which is typically of the order of $0.1$ mag \citep[e.g.,][]{2009ApJ...694..556B}. We will consider the effect of IMF variations among and within galaxies on $M_*$ in Sect.~\ref{sec:ePN.S_jp_profiles_mass_variableIMF}.
Galaxy types are taken from the Hyperleda catalog. 

\subsection{Reconstructing 2D velocity fields}\label{sec:Vfields}

\begin{figure}
    \centering
    \includegraphics[width=0.8\linewidth]{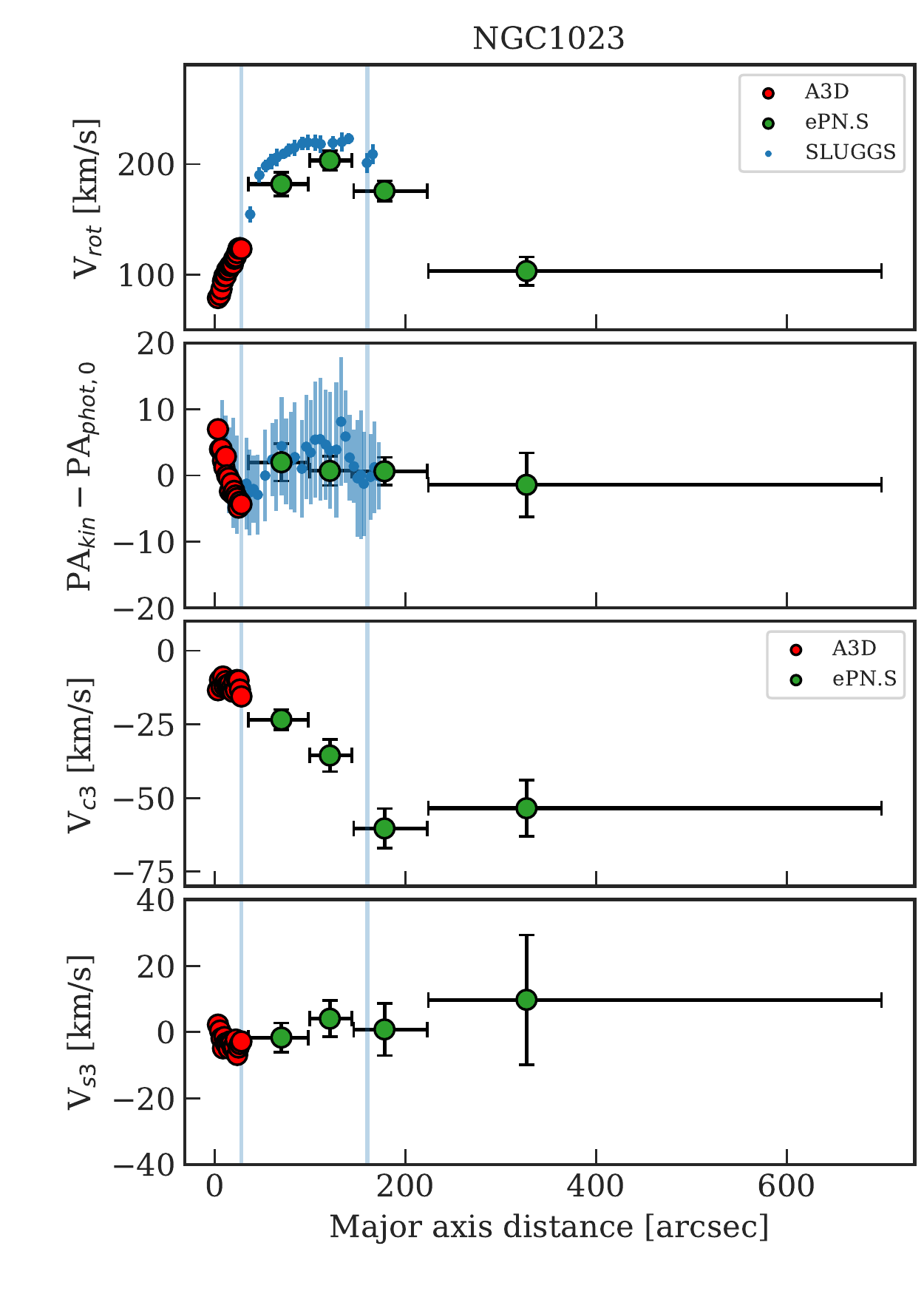}
    \includegraphics[width=\linewidth]{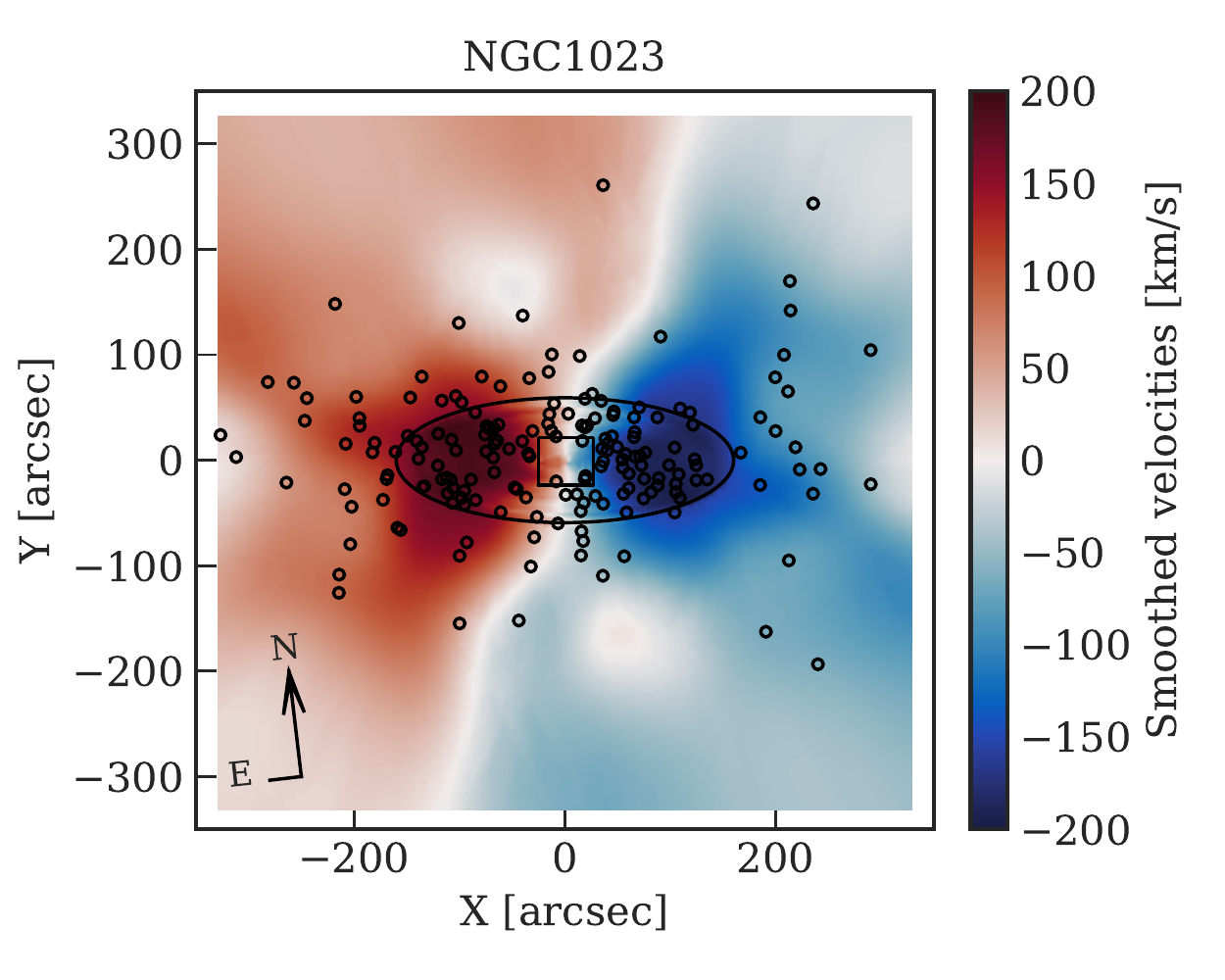}

    \caption{\textbf{Top:} Fitted kinematic parameters in elliptical bins as a function of major axis distance $a$ for NGC1023 (see eq.\ref{eq:harmonic_expansion}). Results on the Atlas3D velocity field are shown in red, those on the ePN.S smoothed velocity field in green. Blue symbols show rotational velocities and kinematic position angles from kinemetry on SLUGGS data \citep{2016MNRAS.457..147F}. These agree within errors with the $V_\mathrm{rot}$ and $\PAkin$ values measured from Atlas3D. ePN.S values also agree within errors except for the first PN velocity point at 70 arcsec, where the PN systematically underestimate the mean velocity (\citealt{Pulsoni2018}, Sect.~7.2). In the radial range marked with vertical lines, we therefore use the $V_\mathrm{rot}(a)$ and $\PAkin(a)$ profiles from SLUGSS, and $V_{\rm s3}(a)$ and $V_{\rm c3}(a)$ from the linear interpolation of the values fitted on the Atlas3D and ePN.S velocity fields.
    \textbf{Bottom}: The 2D rotation velocity field of NGC1023 reconstructed from Atlas3D, SLUGGS, and ePN.S., using Eq.~\ref{eq:harmonic_expansion}. The velocity field is rotated so that the photometric major axis ($PA_\mathrm{phot} = 83.3$ deg) is horizontal. The final mean velocity field is given by the Atlas3D data in the center (within the black square), the reconstructed 2D field from SLUGGS at intermediate radii (within the black ellipse), and the ePN.S data at large radii. The PN positions are highlighted by open black circles.}
    \label{fig:KinParams_VelField}
\end{figure}

The IFS and ePN.S mean LOS velocity fields are divided in elliptical annuli with constant ellipticity $\langle \varepsilon\rangle$ and position angle $\langle \PAphot{} \rangle $. The velocities in each bin are fitted with the model
\begin{equation}
\begin{split}
 V(a,\phi) & = V_{0}(a) + V_\mathrm{rot}(a) \cos[\phi-\PAkin(a)]+ \\
 & + V_{\rm s3}(a) \sin[3\phi-3\PAkin(a)] + \\
                    & + V_{\rm c3}(a) \cos[3\phi-3\PAkin(a)],
\end{split}
\label{eq:harmonic_expansion} 
\end{equation}
where $a$ is the semi-major axis of the bin and $\phi$ the eccentric anomaly
\begin{equation}
    \phi = \arctan[(y_n/(1 - \langle\varepsilon\rangle)x_n)],
\end{equation} 
see also \citealt{Pulsoni2018}. The coordinates $(x_n, y_n)$ of the velocity fields are rotated such that $x_n$ is aligned with the photometric major axis given by $\langle PA_\mathrm{phot}\rangle$.
The rotation velocity $V_\mathrm{rot}$, the kinematic position angle $\PAkin$\footnote{The kinematic quantities $\PAkin$ and $V_\mathrm{rot}$ are comparable to the results from a kinemetry fit to IFS data \citep{2006MNRAS.366..787K}. 
However, we refrain from applying the kinemetry analysis, designed to fit IFS kinematic maps, to the PN velocity fields which have a much lower spatial resolution and signal-to-noise ratio, and prefer the model in Eq.~\eqref{eq:harmonic_expansion} with fewer free parameters.}, the amplitudes of the third order harmonics $V_{\rm s3}$ and $V_{\rm c3}$, and the constant $V_0$ are free parameters. From $V_0(a)$, we estimate the systemic velocity of each field as the weighted sum
\begin{equation}
 V_{\rm sys} = \frac{\sum_{\rm bins} V_0/ (\Delta V_0)^2}{\sum_{\rm bins} 1/(\Delta V_0)^2},
\label{eq:V_sys} 
\end{equation}
where $\Delta V_0$ are the errors on $V_0(a)$. For ePN.S fields, $\Delta V_0$ are derived from Monte Carlo simulations as described in \citealt{Pulsoni2018}; for the IFS fields, $\Delta V_0$  are the errors on the fitted parameters.

The 2D velocity fields are reconstructed by combining together the IFS data in the center and the ePN.S smoothed velocity fields at large radii. This is done by:
\begin{enumerate}
    \item subtracting each velocity field by its systemic velocity $V_\mathrm{sys}$;
    \item interpolating them onto a regular grid of pixels of coordinates $(x_n,y_n)$ to create 2D mean velocity $V_n(x_n, y_n)$ and velocity dispersion maps $\sigma_n(x_n, y_n)$;
    \item "bridging" the radial gap between IFS and PN kinematics using SLUGGS data when available, or a smooth interpolation between the IFS and PN fields.
\end{enumerate}

For the mean velocity fields, $V_n(x_n, y_n)$ are estimated in the radial gap between the IFS and the ePN.S maps using Eq.~\eqref{eq:harmonic_expansion}, with the parameters $V_{\rm rot}$,  $\PAkin$, $V_{\rm s3}$, and $V_{\rm c3}$ given by a linear interpolation between the fitted values on the IFS data in the center and the ePN.S data at large radii. If available, we use the $V_{\rm rot}$ and $\PAkin$ profiles from SLUGGS instead of the interpolated values. 
For the velocity dispersion fields, $\sigma_n(x_n,y_n)$ are given by the velocity dispersion values from the IFS maps in the center and from ePN.S in the outskirts. In the radial gap, we assume $\sigma$ constant in elliptical annuli and use a liner interpolation between the $\sigma(a)$ profiles from the IFS and ePN.S data, or SLUGGS velocity dispersion profiles. 
We checked that our linear interpolation between kinematic parameters is consistent with long-slit data whenever available at these intermediate radii. 
Table~\ref{tab:kinematics_AND_photometry_data} summarises the kinematic data and details the procedure used to reconstruct the velocity fields of each galaxy.

Figure~\ref{fig:KinParams_VelField} illustrates the reconstruction of the 2D velocity field for an example galaxy. In the top panel, the fitted kinematic parameters on IFS and ePN.S data are plotted as a function of $a$
together with kinemetry of SLUGGS data from \citet{2016MNRAS.457..147F}.
The final reconstructed velocity field $V_n(x_n,y_n)$ is shown in the bottom panel: this is given by the IFS map from Atlas3D at the center, by the ePN.S smoothed velocity field at large radii, and by the reconstructed velocity field using SLUGGS data at intermediate radii. 

\begin{figure}
    \centering
    \includegraphics[width=0.8\linewidth]{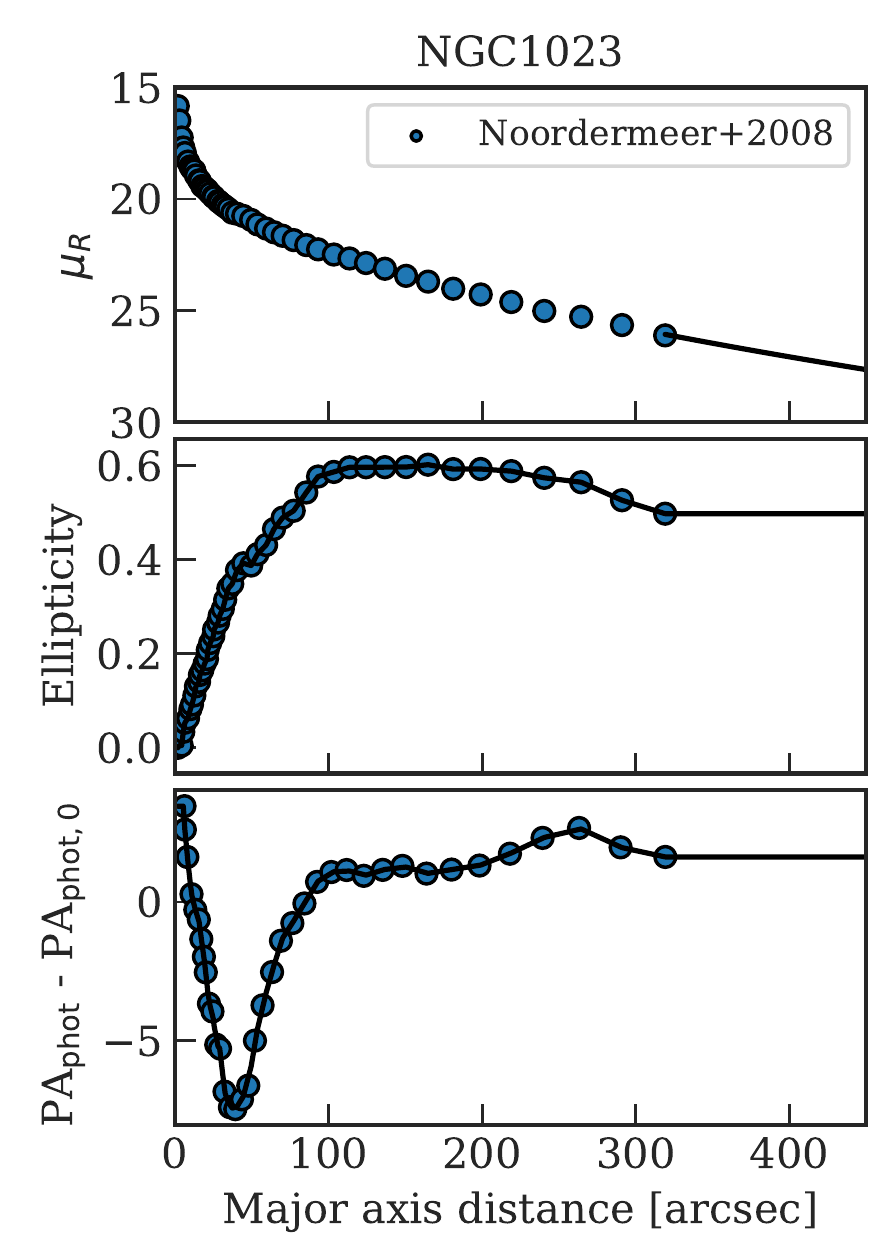}
    \includegraphics[width=\linewidth]{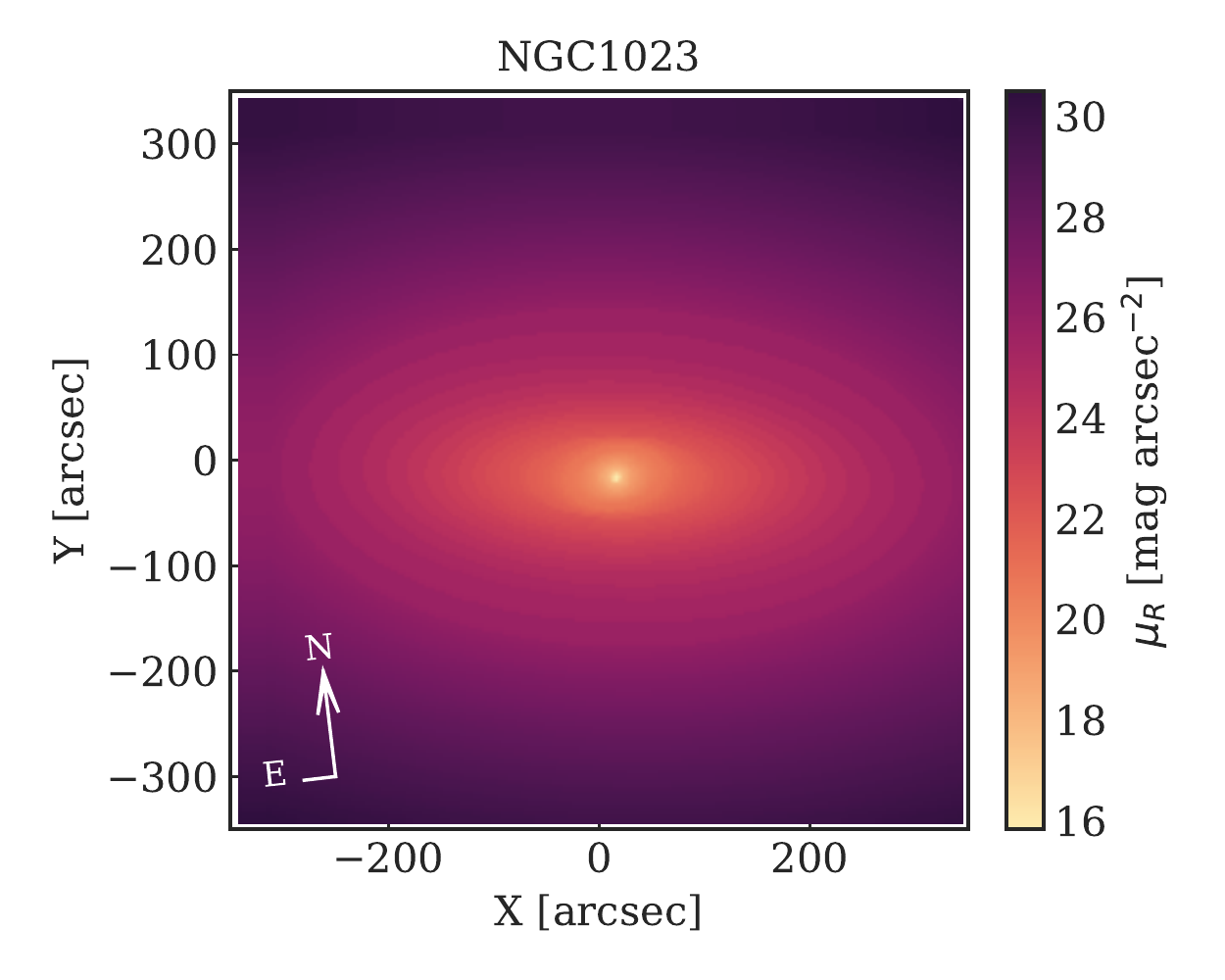}

    \caption{\textbf{Top:} Photometric profiles for NGC1023 extrapolated to large radii. \textbf{Bottom:} The galaxy image of NGC1023 reconstructed on a grid of coordinates from the photometric profiles above.
    }
    \label{fig:example_imageSB}
\end{figure}

\subsection{Reconstructing galaxy images}\label{sec:photometry}

The spatial distribution of light $f(x_n,y_n)$ in the galaxies is reconstructed from the photometric profiles $\mu (a)$, $\varepsilon (a)$, and $\PAphot{}(a)$ available in the literature (references in \citealt{Pulsoni2018}), where $\mu(a)$ is the surface brightness profile along the galaxy semi-major axis $a$, $\varepsilon (a)$ the ellipticity profile, and $\PAphot{}(a)$ is the photometric position angle profile. This is done by assuming that the galaxy isophotes can be approximated by perfect ellipses with ellipticity and position angle given by $\varepsilon (a)$ and $\PAphot{}(a)$, and to which we assign surface brightness $\mu(a)$. 
Thus, for each galaxy, we create an image $f_n(x_n,y_n) = 10^{-\mu_n/2.5}$ that represents a map of weights for the pixels of coordinates $(x_n,y_n)$. 

In case the photometric profiles are not extended enough in radius to cover the extent of the PN velocity field, we extrapolate them as follows. The surface brightness profile $\mu (a)$ is fitted with a S\'ersic profile \citep{1993MNRAS.265.1013C} which is extrapolated to large radii. In these outer regions, the ellipticity $\varepsilon (a)$ and position angle $\PAphot{}(a)$ are assumed to be constant and equal to the outermost measurement available.
Figure~\ref{fig:example_imageSB} shows an example of a reconstructed galaxy image from photometric profiles. The extrapolation of the profiles to large radii is shown with solid lines.

\section{Local $\lambda(R)$ profiles}\label{sec:lambda_profiles}

\begin{figure}[h]
    \centering
    \includegraphics[width=\linewidth]{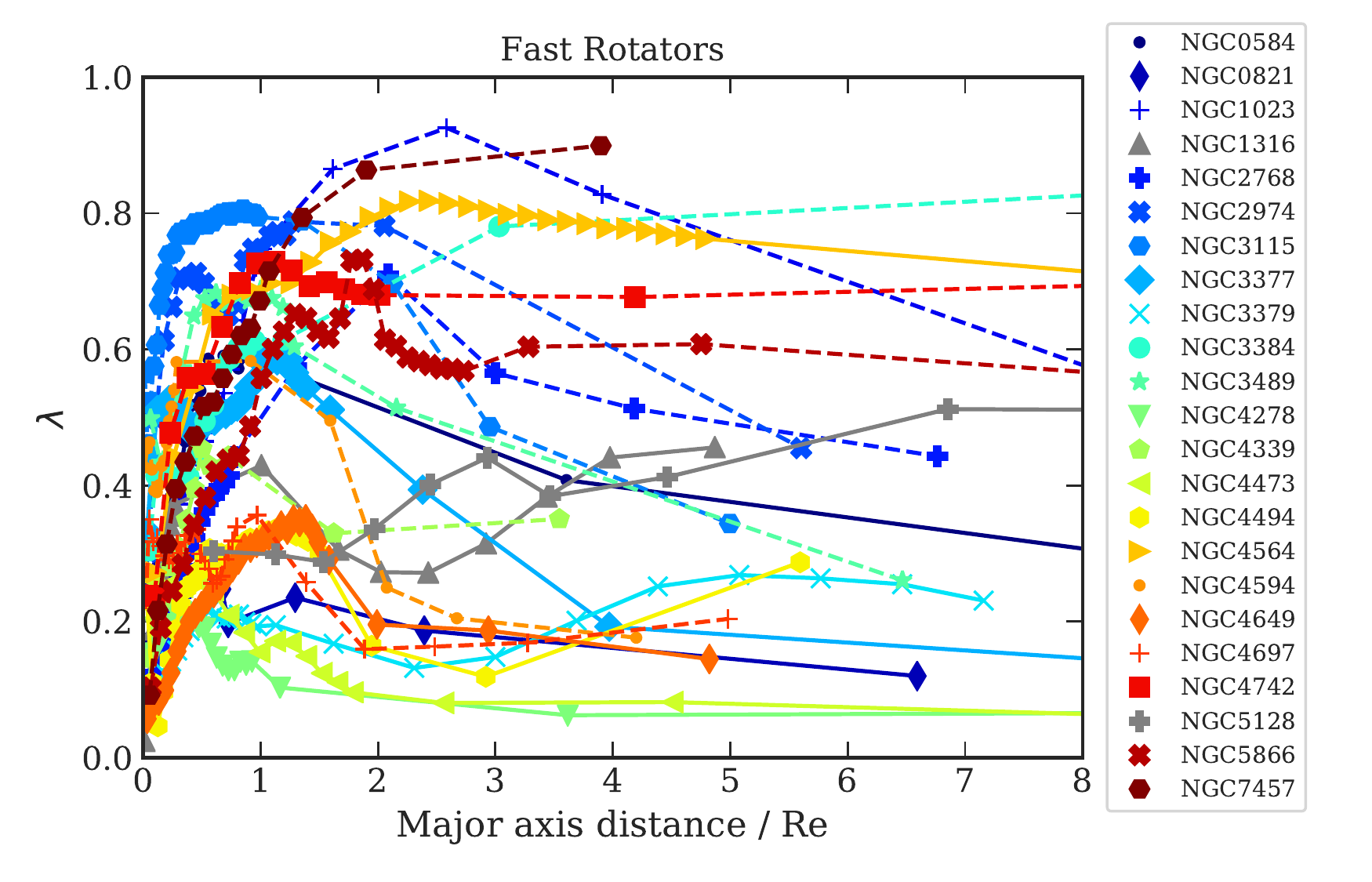}
    \includegraphics[width=\linewidth]{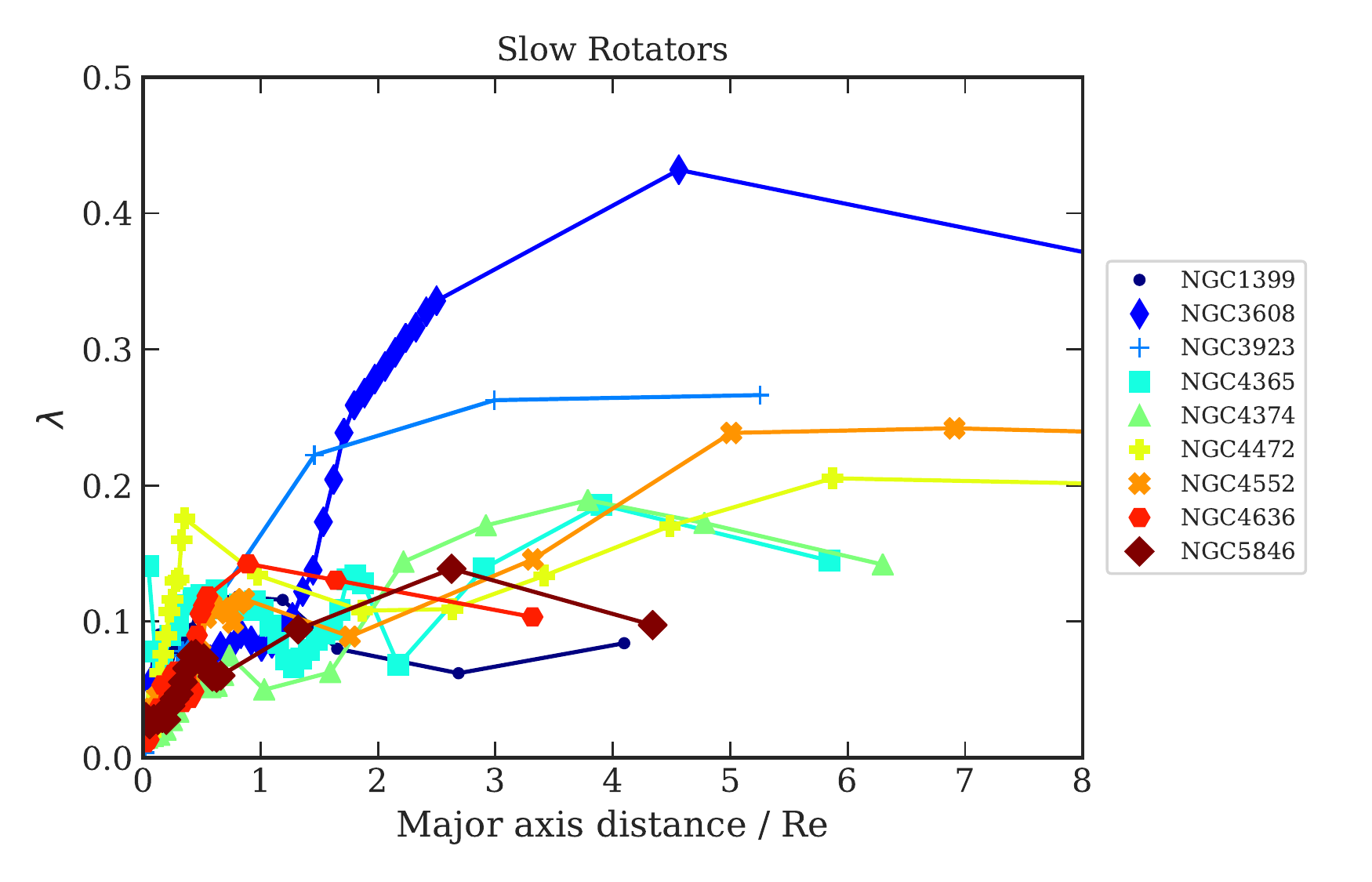}
    \caption{Local $\lambda$ profiles of FRs (\textbf{top panel}) and SRs (\textbf{bottom panels}). S0 galaxies are shown with dashed lines, elliptical FRs with solid lines. The two mergers NGC1316 and NGC5128 are highlighted in gray.}
    \label{fig:lambda_profiles}
\end{figure}
The angular momentum-like parameter $\lambda$, first introduced by \citet{2007MNRAS.379..401E}, is a commonly used proxy for quantifying the projected rotation field. 
It is defined as 
\begin{equation}
    \lambda(a) = \frac{\sum_{n} f_n R_{n} |V_{n}|}{\sum_{n} f_n R_{n} \sqrt{\sigma_{n}^2 + V_{n}^2}}, 
\label{eq:lambda_def}
\end{equation}
where $R_{n}=\sqrt{x_n^2+y_n^2}$ is the circular radius of the n-th pixel of coordinates $(x_n, y_n)$; $V_n$ and $\sigma_n$ are the mean line-of-sight velocity and velocity dispersion; $f_n$ is the flux. 

Galaxies are divided into elliptical radial annuli with constant flattening $\langle \varepsilon \rangle$ and major axis position angle $\langle PA_\mathrm{phot} \rangle$ (reported in Table 1 of \citealt{Pulsoni2018}).
The aperture value of $\lambda_e$ within $1R_e$ is used to divide ETGs into FRs and SRs (see Sect.~\ref{sec:ePNS_survey}). The local $\lambda(a)$ integrated within elliptical annuli of mean semi-major axis $a$ instead quantifies the local rotational support. Here, $V_n$ and $\sigma_n$ are given by the reconstructed velocity fields (Sect.~\ref{sec:Vfields}), and $f_n$ are the fluxes from the reconstructed images (Sect.~\ref{sec:photometry}).
Prior to this work, \cite{2009MNRAS.394.1249C} used PNe to derive the $\lambda(a)$ profiles of a sub-sample of the ePN.S ETGs. Their approach, which substitutes the weighting by $f_n$ with the weighting by the completeness-corrected number density of PNe, gives very similar results.

Figure \ref{fig:lambda_profiles} shows the local $\lambda(a)$ profiles of the FRs (top panel) and the SRs (bottom panel). The $\lambda(a)$ profiles of the FRs show considerable diversity: after peaking at about $1-2R_e$, they either stay constant or decline more or less steeply with $a$. This diversity of $\lambda(a)$ profiles results in a large range of stellar halo rotational support for this class of galaxies. We also note that, in the ePN.S sample, galaxies with the lowest $\lambda(\sim6R_e)$ are mostly E-FRs, while S0s often contain extended stellar disks or rapidly rotating halos with $\lambda(\sim6R_e)$ reaching values of 0.6-0.8 (solid versus dashed lines in Fig.~\ref{fig:lambda_profiles}). 
SRs modestly increase their $\lambda$ with $a$ and often can exceed $\lambda=0.2$. Hence some of the ePN.S SRs host stellar halos with moderate rotation.
The two mergers NGC1316 and NGC5128 are highlighted in Fig.~\ref{fig:lambda_profiles} with gray lines. Both galaxies display moderate rotational support with $\lambda\sim 0.3-0.4$, increasing mildly with radius.

A similar variety of $\lambda(a)$ profiles has been found in simulated galaxies \citep{Pulsoni2020, Schulze2020_Magneticum}. Also for the TNG100 ETGs \cite{Pulsoni2020} did not find a dependence of stellar halo $\lambda$ on $M_*$, except at $M_* > 10^{11.5}M_\odot$ where strongly rotating outskirts with $\lambda>0.6$ are not present. Similarly, we do not detect any dependence of the stellar halo rotational support on stellar mass but the small size of the ePN.S sample at $M_* > 10^{11.5}M_\odot$ does not allow to draw conclusions on trends at high masses. 
Figure~\ref{fig:lambda_profiles} quantifies the variety of kinematic behaviors of stellar halos explored in \citet{Pulsoni2018} and emphasizes the importance of extended kinematic measurements to quantify the angular momentum content in these systems, as the kinematic properties measured in the central $1-2R_e$ do no necessarily extrapolate to large radii.

\section{The projected specific angular momentum of the ePN.S galaxies}\label{sec:ePN.S_jp}

\begin{figure*}[h]
    \centering
    \resizebox{\hsize}{!}{\includegraphics{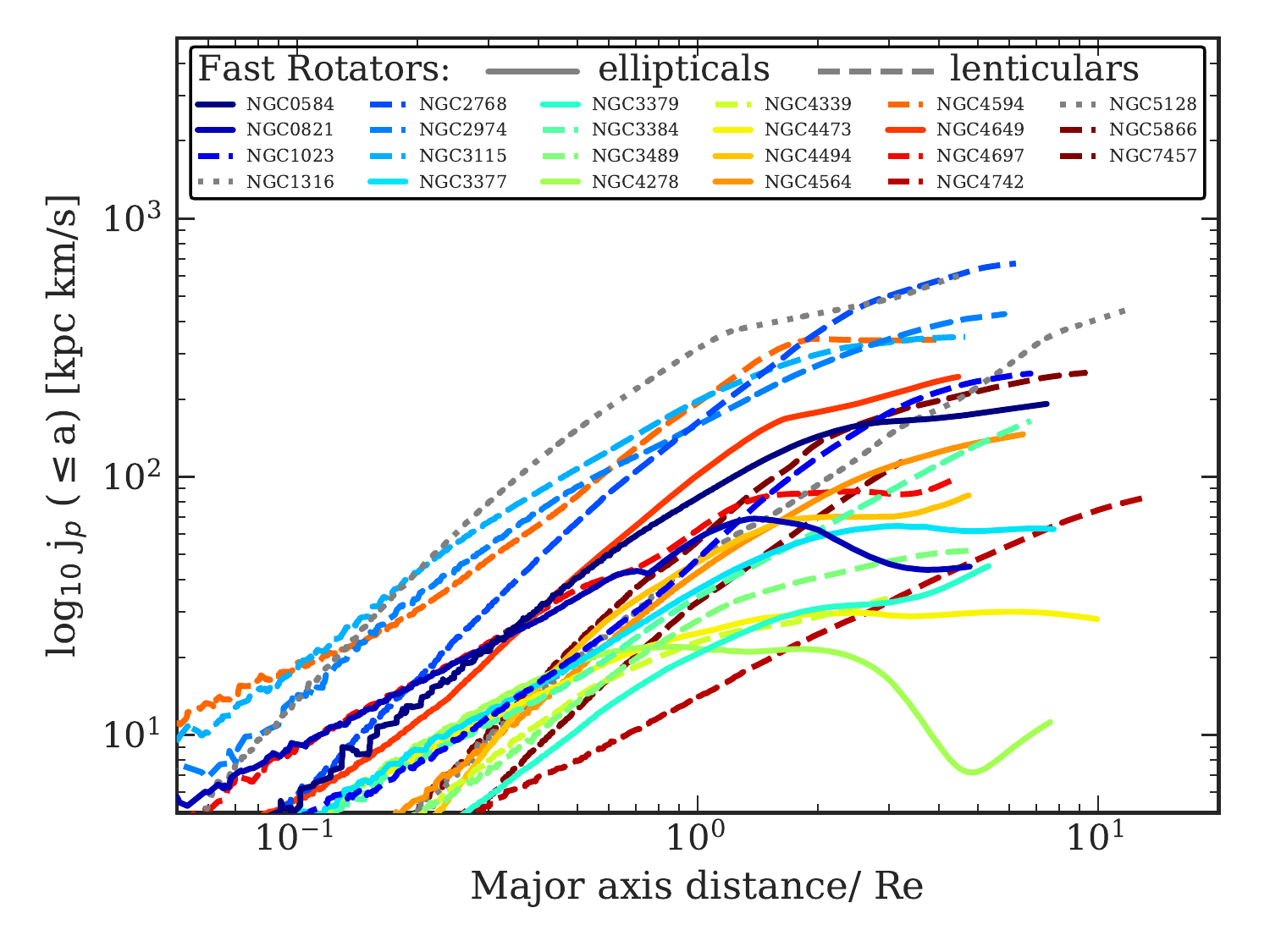}
    \includegraphics{images/MedianCumulativejpProfiles_normalised6Re.pdf}}
    \resizebox{\hsize}{!}{\includegraphics{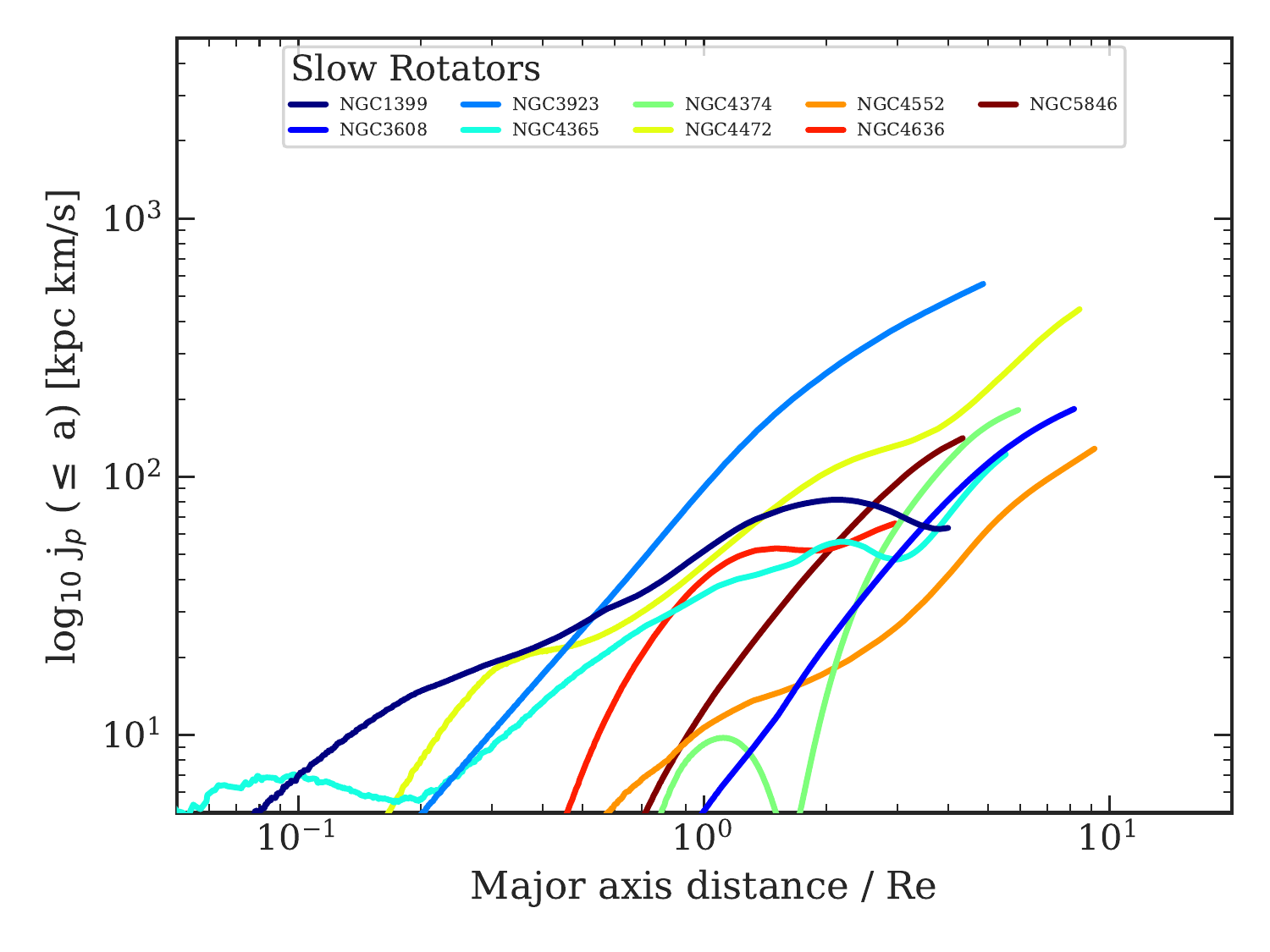}
    \includegraphics{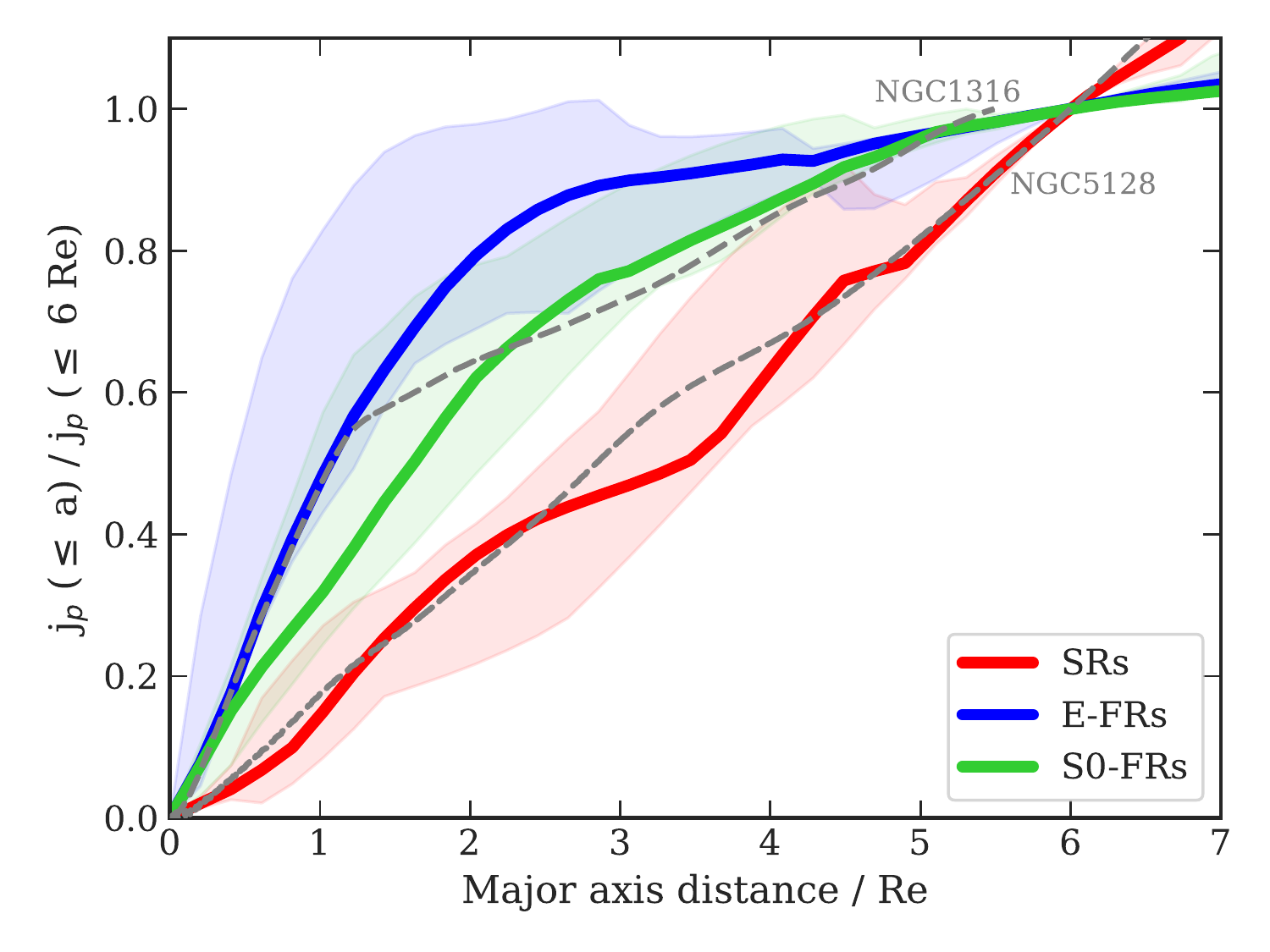}}
    \caption{Light-weighted aperture sAM profiles for fast and slow rotators (\textbf{left panels}): ellipticals and S0s are distinguished by solid and dashed lines, respectively. The two mergers, NGC1316 and NGC5128, are shown with dotted gray lines.
    The \textbf{right} panels shows the median $j_p(\lec R) / j_p(\lec6R_e)$ \textbf{(top)} and the median cumulative angular momentum $J_p(\lec R) / J_p(\lec6R_e)$ profiles \textbf{(bottom)} for galaxies divided into elliptical FRs, S0s, and SRs. The two recent major mergers NGC1316 and NGC5128 are highlighted with gray lines. }
    \label{fig:jp_profiles}
\end{figure*}

In this work, we capitalise on the full two-dimensional kinematic data to determine the total projected sAM $j_p$. This allows us to account for deviations from cylindrical geometry of the velocity fields, non-axisymmetric features, and misaligned rotation in the total $j_p$ budget. 
In a coordinate system centered on the galaxy, with the $x$-axis aligned with the projected photometric major axis given by $\langle PA_\mathrm{phot} \rangle$ and the $z$-axis aligned with the line-of-sight, we define a projected sAM vector of a galaxy $\overrightarrow{j_p}$ as\footnote{Note the different definition of $j_p$ compared to \cite{RomanowskyFall2012}, see also Sect.~\ref{sec:ePN.S_jp_M*} and App.~\ref{sec:appc}.}:

\begin{equation}
    \overrightarrow{j_p} \equiv \frac{\overrightarrow{J_p}}{M_{*}} = \frac{\int \overrightarrow{R}\times\hat{z} V(x,y) \; \Sigma(x,y) dx\;dy}{\int \Sigma(x,y)dx\;dy},
    \label{eq:jp_def1}
\end{equation}
where $\hat{z}$ is the unit vector aligned with the line-of-sight, $\overrightarrow{R}$ is the position vector of coordinates $\mathrm{(x,y)}$ in the plane of the sky, $V(x,y)$ is the line-of-sight velocity, and $\Sigma(x,y)$ the surface stellar mass density of the galaxy. 
The integral in Eq. \eqref{eq:jp_def1} can be approximated with a sum over surface elements with coordinates $(x_n,y_n)$, that is the pixels in our kinematic maps, with mean line-of-sight velocity, $V_{n}$, and stellar mass, $\Sigma_n \Delta x_n\Delta y_n$ : 

\begin{equation}
    \overrightarrow{j_p} \simeq \frac{\sum_n \overrightarrow{R}_{n}\times\hat{z} V_{n} \; (M_*/L_b)_n \;f_{b,n}\; \Delta x_n\;\Delta y_n}{\sum_n (M_*/L_b)_n \; f_{b,n}\; \Delta x_n\;\Delta y_n},
    \label{eq:jp_def2}
\end{equation}
where we substituted the stellar mass density per unit area with the flux in a photometric band indexed with $b$, multiplied by the corresponding mass-to-light ratio: $\Sigma_n =  (M_*/L_b)_n \; f_{b,n}$. The modulus of $\overrightarrow{j_p}$ is 
\begin{equation}
    j_p = \sqrt{j_{p,x}^2 + j_{p,y}^2}.
    \label{eq:jp_def3}
\end{equation}

The aperture profile $j_p(\lec a)$ is derived using Eq.~\eqref{eq:jp_def2} and summing over elliptical apertures of increasing $a$. 
The local $j_p(a)$ is instead derived by summing over pixels within elliptical annuli. 
Unless otherwise stated, we indicate with $j_p$ the total (aperture) projected sAM, integrated out to the outermost available measurement, at a mean $6 R_e$ (median $5.6 R_e$) with a range $[3, 13] R_e$. 

If the mass-to-light ratio is a constant quantity within galaxies, the $(M_*/L_b)$ term cancels out in Eq.~\eqref{eq:jp_def2} and the pixels $(x_n,y_n)$ are weighted only by their fluxes. 
In this case, we define a "light-weighted" $j_{p,\mathrm{light}}$, calculated by weighting the local $j_p(x,y)$ with light profiles in blue optical bands, similar to previous work on sAM in ETGs (Sect.~\ref{sec:ePN.S_jp_profiles_light}).

Estimating the gradients of $(M_*/L_b)$ is difficult because it requires constraining how the stellar population properties change out to large radii. To construct a "stellar-mass-weighted" integrated $j_{p,\mathrm{mass}}$, we use IR fluxes in the Spitzer $3.6\mu$m band as proxy for the stellar mass distribution (Sect.~\ref{sec:ePN.S_jp_profiles_mass_constantIMF}). 

Finally, a variation of the IMF with radius as indicated by several studies of massive ETGs \citep[see][for a review]{2020ARA&A..58..577S} would have significant impact on the $(M_*/L_b)$ gradients, although with still large uncertainties. In Sec.~\ref{sec:ePN.S_jp_profiles_mass_variableIMF}, we estimate these effects on $j_p$ using the results of \citet{2023MNRAS.518.3494B}. We denote with $j_{p,\mathrm{mass+IMF}}$ the mass-weighted total $j_p$ obtained with this model for the IMF-gradients.

\subsection{Aperture $j_p(\lec a)$ profiles: weighting by light}\label{sec:ePN.S_jp_profiles_light}

We start our analysis with the light-weighted case, following a similar approach as \citep[][]{RomanowskyFall2012}. We derive light-weighted $j_p(\lec a)$ profiles in apertures using Eq.~\eqref{eq:jp_def2} and a constant $(M_*/L_b)$ throughout the galaxies. We choose light profiles in blue optical bands, which are typically the most radially extended as the contribution from the sky background is lower at these wavelengths: for example, the sky in the $I$ band is 3 mag more luminous than in $B$ \citep{2003A&A...400.1183P}. The blue fluxes are used to determine the $f_n(x_n, y_n)$ maps.

The left panels of Fig.~\ref{fig:jp_profiles} show the resulting $j_p(\lec a)$ profiles for the ePN.S FRs and SRs, plotted out to the mean major-axis distance of the outermost kinematic bin (see \citealt{Pulsoni2018}). The top right panel shows the median $j_p(\lec a)$ profiles normalised at $j_p(\lec 6R_e)$. Galaxies are divided among lenticulars, E-FRs, and SRs. 
The bottom right panel displays instead the median projected angular momentum profiles $J_p(\lec a)$ normalised at $J_p(\lec 6R_e)$. $J_p(\lec a)/J_p(\lec 6R_e)$ is a cumulative quantity that increases monotonically with radius; therefore the $J_p(\lec a)/J_p(\lec 6R_e)$ profiles are of more immediate interpretation than the
\begin{equation}
    \frac{j_p(\lec a)}{j_p(\lec 6R_e)} = \frac{J_p(\lec a)}{J_p(\lec 6R_e)} \frac{M_*(\lec 6R_e)}{M_*(\lec a)}
\end{equation} 
profiles, which depend on the relative rate at which both $J(\lec a)$ and $M_*(\lec a)$ increase with radius. 

The aperture $j(\lec a)$ of most FRs are monotonically increasing functions that tend to plateau beyond $2R_e$.
By assuming that the aperture values of $j_p$ and $J_p$ measured at $6R_e$ are good approximations for the galaxy-integrated quantities, we estimate that the E-FRs reach a median 48\% of total $j_p$ within $1R_e$ and a median 90\% within $3R_e$. 
The median $j_p$ profile of the S0s increases more slowly with major axis distance $a$, reaching $30\%$ of $j_p(\lec 6R_e)$ at $1R_e$ and 77\% at $3R_e$, eventually plateauing beyond $4R_e$. Hence for both FR classes the $j(\lec a)$ profiles are nearly converged within the radial range of the PN data.

The difference between E-FRs and S0s is also visible, although less marked, in the median total $J_p(\lec a) / J_p(\lec 6R_e)$ profiles and can be explained by the different distributions of rotational support for the two classes in Fig.~\ref{fig:lambda_profiles}: most elliptical FRs have less rotation in their stellar halos compared to S0s, which often rotate rapidly to large radii.
The median $J_p(\lec a) / J_p(\lec 6R_e)$ profiles are shallower compared to the sAM profiles: E-FRs and S0s contain 50\% of their AM within 1.7 and 2.2 $R_e$, respectively, and both reach 80\% within 4$R_e$. The fact that the $J_p(\lec a) / J_p(\lec 6R_e)$ seem to gently flatten at large radii suggests that only a small fraction of the total AM is left in the outskirts.
Within our sample, we do not observe significant differences between low and high mass FRs in both their median $j_p(\lec a) / j_p(\lec 6R_e)$ or $J_p(\lec a) / J_p(\lec 6R_e)$. 

SRs exhibit markedly different $j_p$ profiles and radial AM distributions. Their more extended mass distributions and the fact that these galaxies rotate faster at larger radii (see Fig.~\ref{fig:lambda_profiles}) determine much steeper outer profiles than for FRs. The median $j_p(\lec a) / j_p(\lec 6R_e)$ and $J_p(\lec a) / J_p(\lec 6R_e)$ do not converge within the radial coverage of the ePN.S data, as Figure~\ref{fig:jp_profiles} clearly shows.
Thus a non-negligible fraction of the total angular momentum of these galaxies is distributed at larger radii. The correction from $j_p(\lec 6R_e)$ to the total, galaxy-integrated $j_p$ is estimated in Sect.~\ref{sec:fract_j_beyond_6Re} using cosmological simulations.
 
\begin{figure}
    \centering
    \includegraphics[width=\linewidth]{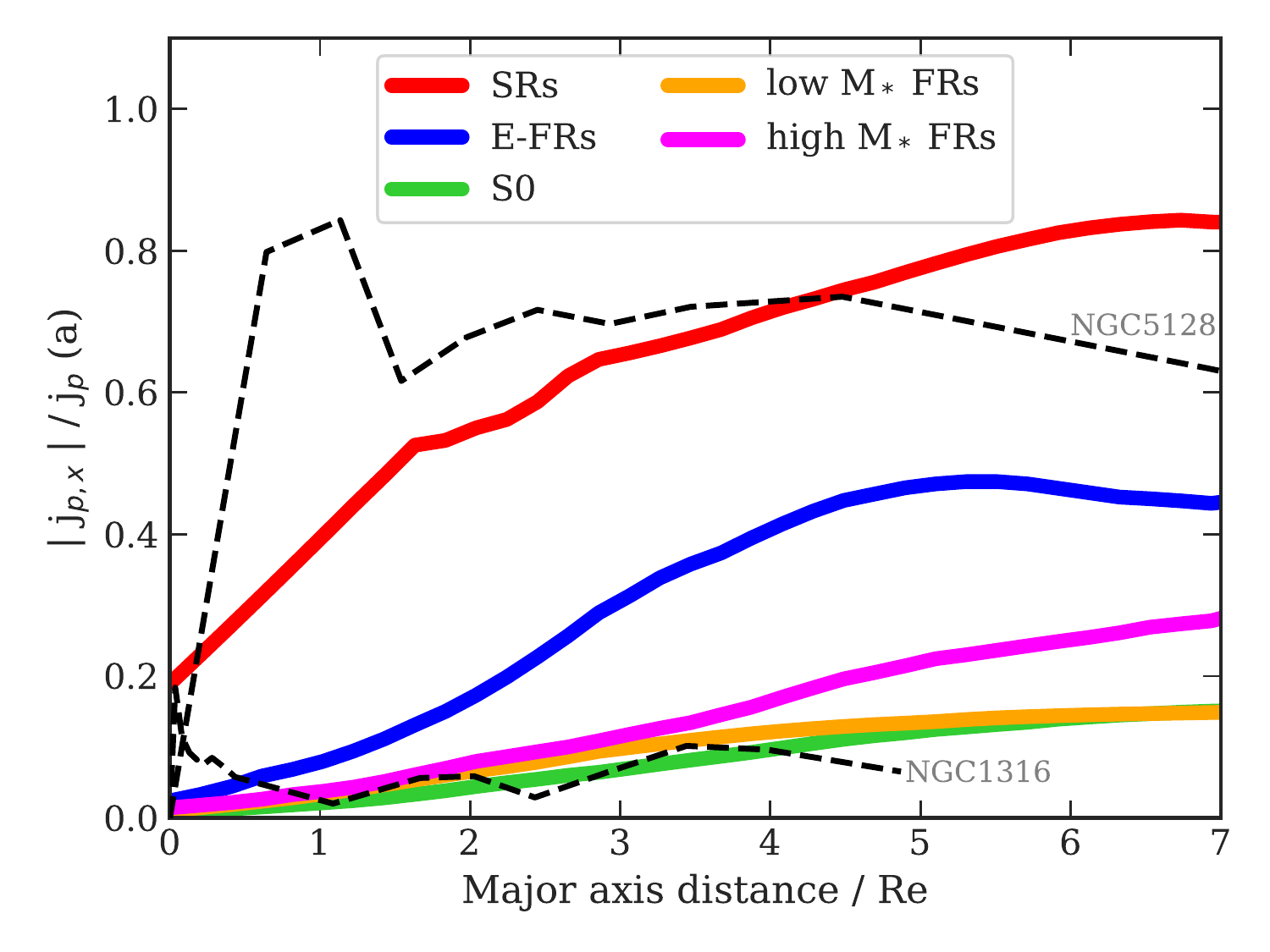}
    \caption{Median contribution to the local j from misaligned rotation as a function of major axis distance for different groups of galaxies as in the legend. The median profiles have been smoothed to highlight the radial trends. The profiles for the two mergers NGC1316 and NGC5128 are shown separately.}
    \label{fig:Contribution_jxjy}
\end{figure}

\subsubsection{Contribution by minor axis and misaligned rotation}

Figure~\ref{fig:Contribution_jxjy} illustrates the different two-dimensional distributions of the local angular momentum in different classes of galaxies and highlights their different dynamical structure at large radii. It shows the median ratio of the $j_{p,x}$ component (see Eq.~\ref{eq:jp_def3}) to the local $j_p(a)$ as a function of $a$, where the $x$-axis is aligned with the projected major axis of the galaxy. While the component $j_{p,y}$ is determined by rotation along the projected major axis, non-zero $j_{p,x}$ signals the presence of kinematic twists or misalignments contributed by minor axis rotation. 
The median profiles are derived for the ePN.S SRs, FRs divided between low mass and high mass at $M_* = 10^{10.6}M_\odot$, S0s, and E-FRs. 

The contribution of $j_{p,x}$ to the local $j_p(a)$ through Eq.~\eqref{eq:jp_def3} depends on the rotator class as well as on $M_*$ in the ePN.S sample. 
In low mass FRs and in S0s, the contribution from "off axis" rotation to $j_p(a)$ is negligible: $j_p(a)$ comes mostly from major-axis rotation. In high mass FRs, the contribution of $j_{p,x}$ increases with $a$, from 10\% at $2R_e$ to $30\%$ at $6R_e$. 
The ratio $\| j_{p,x}\| / j_p(a)$ is even larger for the E-FRs, which increases from $20\%$ at 2$R_e$ to 50$\%$ at 6$R_e$.  However, even in this case, the contribution from off-axis rotation to the total AM is small, as most of the total AM is dominated by the central 3$R_e$ (see Fig.~\ref{fig:jp_profiles}).
On the other hand in SRs, both components $j_{p,x}$ and $j_{p,y}$ are equally important to the total AM budget. For these systems extended 2-D kinematic information is essential for measuring their total angular momentum.

The two major mergers show a different distribution of $j_p$ with $a$ in Fig.~\ref{fig:jp_profiles}, although in both cases the profiles do not seem to plateau within the radial coverage of the PN data. NGC1316 $j_p$ increases steeply within the central 1$R_e$ and most of its $j_p$ is contributed by the major-axis rotation. The $j_p$ and $J_p$ profiles of NGC5128 rise more slowly with radius, meaning that a larger fraction of its AM is distributed at large radii more similarly to SRs. In this galaxy, a large contribution to $j_p$ comes from minor-axis rotation, as shown in Fig.~\ref{fig:Contribution_jxjy}.

\subsection{Aperture $j_p(\lec a)$ profiles: weighting by stellar mass}\label{sec:ePN.S_jp_profiles_mass}

Evaluating the stellar mass associated to the light emitted by the galaxies at each radius would require modelling of the star formation history through spectral analysis. The stellar population mix at each radius determines the age and metallicity distribution which fix the $(M_*/L_b)$ ratio, modulo an assessment of IMF which establishes the overall normalization of the $(M_*/L_b)$ \citep[e.g.,][]{2019MNRAS.487.3776P}.
For a constant IMF, stellar population gradients in ETGs imply $(M_*/L_b)$ ratios about $20-30\%$ larger in the center than at $0.5 R_e$ \citep[e.g.,][]{2019MNRAS.489.5612D, 2021MNRAS.507.2488G}. However, recent studies find that ETGs have significant IMF gradients with radius, with enhanced fractions of low-mass stars in the central regions and standard (Kroupa- or Chabrier-like) IMF beyond $\sim0.5 R_e$ \citep{2015MNRAS.447.1033M, 2018MNRAS.477.3954P,2019MNRAS.489.4090L}. The presence of these low-mass stars, which contribute only a few percent to the bolometric light of an old stellar population \citep[see, e.g., figure 4 from][]{2013ARA&A..51..393C}, can increase the $(M_*/L_b)$ by a factor of more than 2 in the center \citep{2019MNRAS.489.5612D}. Although a direct determination of the stellar mass distribution is beyond the scope of this paper, in this section we aim at evaluating the overall effect of stellar population and IMF gradients on $j_p$.

\subsubsection{Constant IMF}\label{sec:ePN.S_jp_profiles_mass_constantIMF}
We start by assuming a constant IMF.
A good proxy for the stellar mass distribution is the IR-light emission, as the fluxes at these wavelengths are dominated by the emission from the old stars. Hence, compared to bluer wavelengths, they are less sensitive to the emission from younger stars with lower $(M_*/L_b)$ ratios.
For this investigation, we considered light profiles from Spitzer $3.6\mu$m imaging published by \cite{Forbes2017_Spitzer}. 
These data are available for 20/32 ePN.S galaxies and cover their central $\sim 3 R_e$. At larger radii, the flux is assumed to continue following the extrapolation of the S{\'e}rsic profile fitted to the inner data.

\begin{figure}
    \centering
    \includegraphics[width=\linewidth]{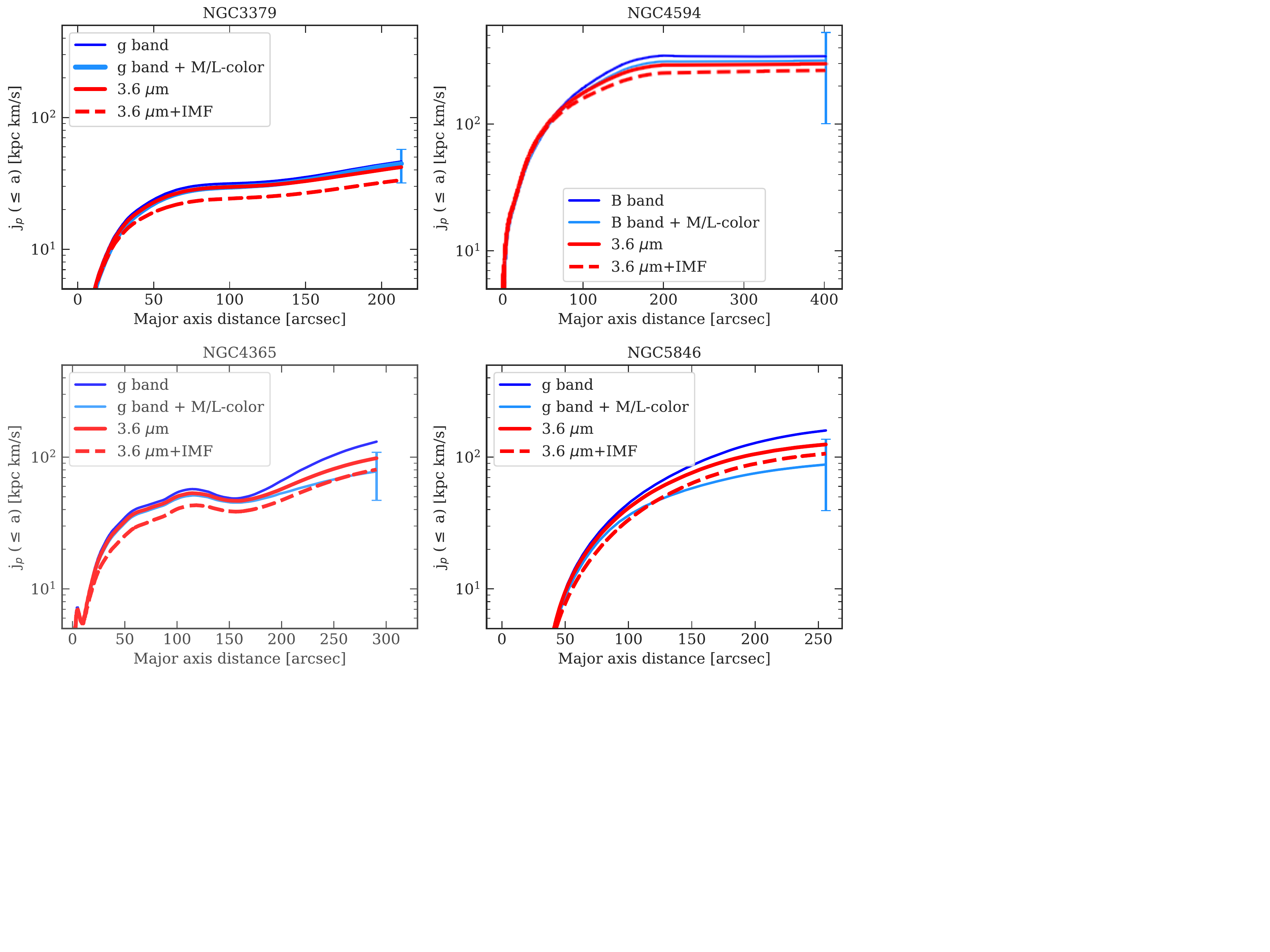}
    \caption{Light-weighted and mass-weighted $j_p(\lec a)$ profiles in four example galaxies, two FRs (\textbf{top}) and two SRs (\textbf{bottom}). For each galaxy, we show the profiles weighted by the blue-band fluxes, by the IR fluxes, by the mass-to-light ratio profile given by the colors, and by the IR fluxes corrected for IMF-driven gradients in the mass-to-light ratio. The error bar shows the error on $j_p$ derived from the errors on the colors and the dispersion in the  mass-to-light ratio versus color relations. }
    \label{fig:jp_profiles_massVSlight}
\end{figure}

Where they overlap, the $3.6\mu$m profiles are typically steeper than those in the blue bands. Therefore, in the calculation of $j_p$ weighted with IR fluxes, the central regions become more strongly weighted. This results in aperture $j_p$ profiles with similar shapes but lower amplitudes compared to those weighted with blue-band fluxes. Figure~\ref{fig:jp_profiles_massVSlight} shows the comparison for four example galaxies. The total $j_p$ values
(measured integrating within the radial coverage of the PN data) are lower by a mean 13\%, independent of stellar mass. Therefore, we use this factor to correct the total $j_p$ from light-weighted to mass-weighted in galaxies that lack Spitzer $3.6\mu$m profiles.

Another approach is to consider spatially resolved mass-to-light-ratio versus color relations, typically calibrated in the central $1-2 R_e$ for large samples of galaxies \citep{2019A&A...621A.120G, 2021MNRAS.507.2488G}, and apply them to extended color profiles for the ePN.S galaxies. Unfortunately, color profiles that cover the radial extent of the kinematic data are available only for 12 ePN.S galaxies \citep{2011ApJS..197...21H, 2014ApJ...791...38W, 2016ApJ...820...42I,2017ApJ...839...21I,2017A&A...603A..38S,2022FrASS...952810R} to which we applied the relations from \citet[][which assume a Chabrier IMF]{2019A&A...621A.120G} to derive the mass-to-light ratio profiles for the corresponding blue-band fluxes. The resulting $j_p$ profiles also in this case typically show similar shapes and lower amplitudes compared with the blue-light-weighted profiles. The total $j_p$ is consistent with the determinations from the IR-light-weighting within the errors on the colors and on the mass-to-light-ratio versus color relations, as shown in Fig.~\ref{fig:jp_profiles_massVSlight}.

\subsubsection{With IMF gradients}\label{sec:ePN.S_jp_profiles_mass_variableIMF}

We estimate the effects of IMF-driven gradients in the $(M_*/L_b)$ ratio using the results of \cite{2023MNRAS.518.3494B}, who measure M/L and IMF gradients in spatially resolved spectra of ETGs. In agreement with previous studies, they find that IMF-driven M/L gradients are substantial in the central $1 R_e$, where the IMF typically goes from standard at $R>0.5 - 1R_e$ to bottom-heavier in the center, and provide mean mass excess profiles $\alpha (a) = M_\mathrm{*, variable \; IMF}(a)\;/\;M_\mathrm{*, fixed \; IMF}(a)$ separately for E-FRs, S0s, and SRs in bins of luminosity in the r-band and central velocity dispersion $\sigma_0$.
We convert their values based on Kroupa IMF to a Chabrier IMF by dividing them by a fraction 0.92
(any constant factor is unimportant in the computation of $j_p$ but relevant for correcting the stellar masses, see below).

To associate the mean $\alpha(a)$ profiles from \cite{2023MNRAS.518.3494B} to the ePN.S ETGs, we sort our galaxies in similar bins of luminosity and velocity dispersion. We use $M_r$ magnitudes in the AB system from \cite{2013MNRAS.432.1709C} for the Atlas3D galaxies, from \cite{2014ApJS..212...18B} for NGC4594, from \cite{2021MNRAS.504.2146B} for NGC3115, from  \cite{2016ApJ...820...42I,2017ApJ...839...21I} for NGC1316 and NGC1399, and from \cite{1978ApJ...223..707S} for NGC1344, NGC3923, NGC4742, and NGC5128. 
The central $\sigma_0$ values quoted in \cite{2023MNRAS.518.3494B} are not corrected for the seeing effects and the velocity dispersion profiles are shown only for $R>1$kpc. Hence, for a fairer comparison with these data, we use as $\sigma_0$ the velocity dispersion at $a = 1$ kpc.
The five galaxies NGC3377, NGC3489, NGC4339, NGC4742, and NGC7457, that is the five least massive ETGs in the sample, have too low $\sigma_0$ to fall in any of Bernardi et al.'s bins. Therefore, for these objects we do not use an IMF correction.
For the other systems, we calculate $j_p$ using Eq.~\eqref{eq:jp_def2} and weighting with the IR fluxes multiplied by the IMF-driven mass excess $\alpha(a)$. The increased mass in the central regions causes an overall decrease in amplitude of the $j_p$ profiles which is mildly mass dependent (see Sect.~\ref{sec:ePN.S_jp_M*}). The IMF-corrected $j_p$ profiles for four example galaxies are also shown in Fig.~\ref{fig:jp_profiles_massVSlight}.

We also apply IMF corrections to the stellar masses $M_*$ using the mass excess profiles $\alpha (a)$ from \cite{2023MNRAS.518.3494B}. From these we derive

\begin{equation}
    M_\mathrm{*, var\;IMF} / M_\mathrm{*, Chabrier\;IMF} = \frac{2\pi\int  f(a)\;\alpha(a)\; a\; da}{2\pi\int f(a) \; a\;  da},
\label{eq:Correct_Masses_IMF}
\end{equation}
where $f(a)$ are the fluxes in the blue-bands. Using the IR fluxes in Eq.~\eqref{eq:Correct_Masses_IMF} gives slightly larger $M_\mathrm{*, var\;IMF} / M_\mathrm{*, Chabrier\;IMF}$ ratios by $\sim2$ percent.
Finally we correct stellar masses derived in Sect.~\ref{sec:additional_data} assuming a fixed Chabrier IMF, by multiplying the values $M_*$ by the ratio $M_\mathrm{*, var\;IMF} / M_\mathrm{*, Chabrier\;IMF}$.

\subsection{Errors on the total projected sAM $j_p$} \label{sec:errors_jp}

The uncertainties on the measured total $j_p$ come from the uncertainties on the stellar mass distribution and on the mean velocities. The first are accounted for in the different determinations of $j_p$ using light-weighting or mass-weighting, which give mean differences of the order of 15\%.

We estimate the errors on the measured $j_p$ from the uncertainties on the mean velocities, which are largely dominated by the errors on the PN velocity fields.
We use Monte Carlo simulations to evaluate the effect of these errors on $j_p$. 
We build simulated PN catalogs by adding to their mean velocity a random value from a Gaussian distribution centered at 0 and with dispersion given by the measurement error and the velocity dispersion added in quadrature (see the discussion in Sect.~3 of \cite{Pulsoni2018}). The simulated catalogs are used to produce the simulated mean PN velocity fields. Each simulated PN velocity field is then complemented with a Monte Carlo simulation of the IFS kinematics and of the interpolated velocity field in the center, using their corresponding errors on the velocities. The uncertainties on $j_p$ is then the standard deviation of the distribution of $j_p$ values from the Monte Carlo simulations. These are a median $\sim 10\%$ for the S0s and a median $20\%$ for the E-FRs and SRs, and similar for light-weighted and mass-weighted values. For the galaxies without IR fluxes available, for which we estimated the mass weighted $j_p$ by reducing the light-weighted $j_p$ using a mean shift (see Sect.~\ref{sec:ePN.S_jp_profiles_mass_constantIMF}), the error on the mass-weighted $j_p$ is taken to be the sum in quadrature of the error on the light-weighted $j_p$ and the standard deviation around the applied mean shift.

\subsection{The $ j_{p} - M_{*} $ diagram}\label{sec:ePN.S_jp_M*}

\begin{figure}
    \centering
    \includegraphics[width=\linewidth]{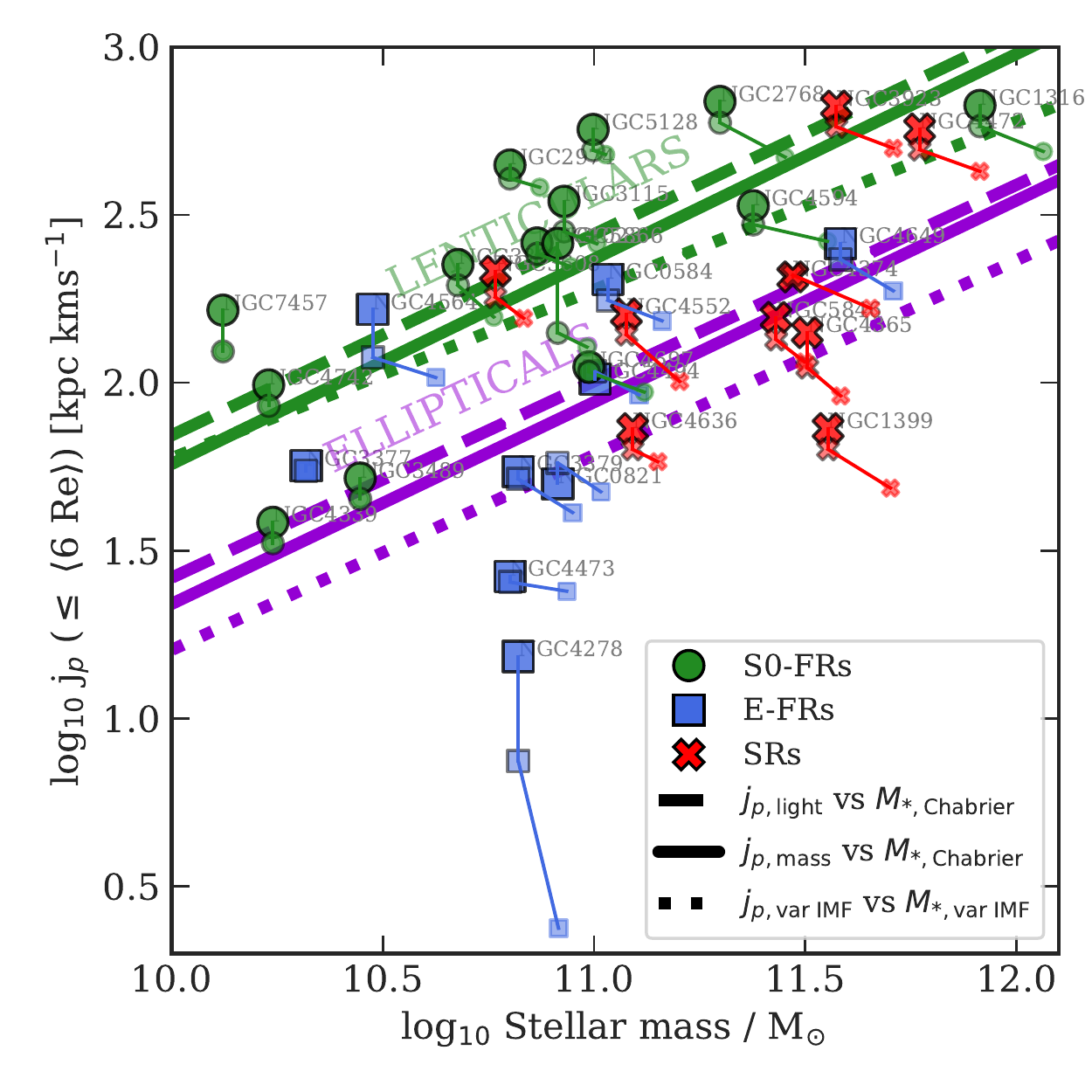}
    \caption{Projected sAM as a function of stellar mass for the ePN.S galaxies. Lenticulars, fast, and slow rotators ellipticals are shown with different colors and symbols as in the legend. We show with progressively decreasing sizes $j_{p,\mathrm{light}}$ vs $M_*$, $j_{p,\mathrm{mass}}$ vs $M_*$, and $j_{p,\mathrm{mass+IMF}}$ vs $M_\mathrm{*, var\;IMF}$. Arrows connect different measures for the same galaxies.
    Dashed lines show the power-law fit to $j_{p,\mathrm{light}}$ vs $M_*$; solid lines show the fit to $j_{p,\mathrm{mass}}$ vs $M_*$; dotted lines show the fit to $j_{p,\mathrm{mass+IMF}}$ vs $M_\mathrm{*, var\;IMF}$. 
    Purple lines show the fits to the elliptical galaxies (E-FRs+SRs), while green lines show the fits to the lenticulars.
    }
    \label{fig:jpM_diagram}
\end{figure}

Figure~\ref{fig:jpM_diagram} shows the relation between the total projected sAM and the stellar mass of the ePN.S galaxies. 
We show the blue luminosity-weighted $j_{p,\mathrm{light}}$, the mass-weighted $j_{p,\mathrm{mass}}$ using IR fluxes, and the mass-weighted $j_{p,\mathrm{mass+IMF}}$ corrected for IMF-gradients, as calculated in the previous Sects.
In the first two cases, $j_p$ is plotted against $M_*$ derived as described in Sec.~\ref{sec:additional_data} assuming a constant Chabrier IMF. The $j_{p,\mathrm{mass+IMF}}$ values are instead shown against $M_\mathrm{*, var\;IMF}$, that is $M_*$ corrected for IMF gradients (see Sect.~\ref{sec:ePN.S_jp_profiles_mass_variableIMF}). 

In all three cases, galaxies are found to have $j_p$ values increasing with stellar mass. The lenticulars show systematically higher $j_p$ compared to the ellipticals (E-FRs and SRs) of similar $M_*$. The dependence on morphology is in agreement with previous work \citep[e.g.,][]{RomanowskyFall2012, Fall2013, Fall2018}.
The nine SRs in Fig.~\ref{fig:jpM_diagram} appear to follow the relation traced by the fast rotating ellipticals for large $M_*$. Even though for the SRs the measured $j_p$ likely underestimates the total projected $j_p$ as their $j_p(\lec a)$ profiles are not converged (see Fig.~\ref{fig:jp_profiles}), we estimate in Sect.~\ref{sec:fract_j_beyond_6Re} that the integration of $j_p$ out to $15R_e$ increases the values for the SRs by only 0.18 dex with respect to the FRs.

Assuming a power-law relation of the form 
\begin{equation}
    \log_{10} j_p/ j_0 = A \; \left[\log_{10} M_*/M_\odot - 11\right]
    \label{eq:j-M_powerlaw}
\end{equation}
and performing a separate fit to the lenticular and elliptical galaxies, we find a slope close to $0.6$ for most cases, as shown in Table~\ref{tab:fit_jp}. 
Weighting $j_p$ by the stellar mass does not strongly change the slope $A$ compared to the light-weighting case, but systematically decreases the normalisation $j_0$. Only for the S0 galaxies does the correction for IMF-driven gradients in the mass-to-light ratio introduce a tilt in the slope, mostly driven by the four low mass S0s for which we did not perform a correction for IMF gradients (see Sect.~\ref{sec:ePN.S_jp_profiles_mass_constantIMF}).
The fitted parameters $A$ and $\log j_0$ and their errors are collected in Table~\ref{tab:fit_jp}.
The errors are given by the sum in quadrature of the errors on the fit and the standard deviation of the distribution of parameters given the errors on the $j_p$ values. These are obtained by fitting Eq.~\eqref{eq:j-M_powerlaw} on Monte Carlo simulations of the $j_p$ values extracted from Gaussian distribution centered on $j_p$ and sigma equal to their errors.

\begin{table}[]
    \centering
    \begin{tabular}{lcc}
\hline\hline\noalign{\smallskip}
  \multicolumn{1}{l}{group of data} &
  \multicolumn{1}{c}{A} &
  \multicolumn{1}{c}{$\log_{10}j_0$}  \\
\noalign{\smallskip}\hline\noalign{\smallskip}
    \textbf{Ellipticals}  &&\\ 
    $j_{p,\mathrm{light}} - M_*$ & 0.59$\pm$0.25  &  2.0$\pm$0.1\\
    $j_{p,\mathrm{mass}} - M_*$ & 0.60$\pm$0.25 &  1.94$\pm$0.11 \\
    $j_{p,\mathrm{mass+IMF}} - M_\mathrm{*,var\;IMF}$ &0.58$\pm$0.29  &  1.78$\pm$0.14\\
\noalign{\smallskip}\hline\noalign{\smallskip}
    \textbf{Lenticulars}  &&\\
    $j_{p,\mathrm{light}} - M_*$ &  0.60$\pm$0.15  &  2.44$\pm$0.08 \\
    $j_{p,\mathrm{mass}} - M_*$ &0.61$\pm$0.16  &  2.37$\pm$0.08\\
    $j_{p,\mathrm{mass+IMF}} - M_\mathrm{*,var\;IMF}$ & 0.49$\pm$0.15  &  2.27$\pm$0.07 \\
    \end{tabular}
    \caption{Results of the fit of $j_p$ versus $M_*$ with the power-law in   Eq.~\eqref{eq:j-M_powerlaw}.}
    \label{tab:fit_jp}
\end{table}

We conclude this section noting that the definition of $j_p$ used in this work is different from the similarly called quantity $j_p$ defined in \citet{RomanowskyFall2012}. As commented by these authors in their appendix, their $j_p$ does not directly quantify the projection of the total angular momentum $j_t$ on the plane of the sky, but represents an intermediate step in the derivation of $j_t$ from long slit observations along the projected semi-major axis, assuming cylindrical rotation (see their Eq.~3). Therefore, a direct comparison with these previous (systematically higher) estimates of $j_p$ is not straightforward.

\section{The projected sAM of the IllustrisTNG ETGs}\label{sec:IllustrisTNG}

The new generation of cosmological hydrodynamical simulations are able to produce a rich variety of galaxy morphologies and to resolve the dynamical and stellar population properties of galaxies. 
The increased resolution and the inclusion of efficient star formation feedback has proved to be fundamental to overcome the "angular momentum catastrophe" \citep[e.g.,][]{1995MNRAS.275...56N, 1999ApJ...519..501S, 2007MNRAS.374.1479G} and reproduce realistic galaxies with angular momentum content that matches the observations \citep{ Genel2015, Teklu2015ApJ...812...29T, Zavala2016, Lagos2017, Lagos2018_MergersAM}. 

IllustrisTNG is a suite of cosmological magneto-hydrodynamical simulations that form and evolve galaxies in a $\Lambda$CDM universe. They include prescriptions for star formation and evolution, chemical enrichment of the inter-stellar medium, gas cooling and heating, black hole and supernova feedback  \citep{2018MNRAS.475..676S, 2018MNRAS.475..624N, 2018MNRAS.475..648P,2018MNRAS.477.1206N, 2018MNRAS.480.5113M}.
IllustrisTNG generates a population of galaxies with good mixture of morphological galaxy types \citep{2019MNRAS.483.4140R} and whose $j_*-M_*$ relation, and its dependence on morphology, is in good agreement with observations \citep{2023MNRAS.518.6340D, Rodriguez-Gomez2022}. 
Furthermore, the TNG ETGs were demonstrated to reproduce many of the kinematical and morphological properties of observed ETGs out into the stellar halos. They show similar changes in rotational support and flattening with radius and similar fractions of ETGs displaying kinematic twists \citep{Pulsoni2020}.
In this work, we use the simulated TNG ETGs as models of the ePN.S galaxies to estimate the fraction of the total angular momentum distributed outside the radial coverage of the ePN.S data and to account for projection effects.

\subsection{Sample selection and derivation of physical quantities}\label{sec:TNG100_sample_selection_and_derivation_of_physical_quantities}

The IllustrisTNG simulations adopt a universal Chabrier IMF, consistent with our choice for the ePN.S $M_*$.
We select the sample of simulated $z=0$ ETGs from the entire TNG100 volume as in \cite{Pulsoni2020}, with stellar masses in the enlarged range
$10^{10.1} \leq M_{*}/\MSUN{} \leq 10^{12}$, red colors $(g-r)\geq 0.05\log_{10}(M_{*}/\MSUN{}) + 0.1 {\; \rm mag}$, and with effective radii $R_e{}\geq2\rsoft$. 
Stellar masses and effective radii are derived considering all the stellar particles bound to the galaxies.
As in \cite{Pulsoni2020}, we perform a further selection in the $\lambda_e-\varepsilon_e$ diagram (where $\lambda_e$ is integrated within an elliptical aperture of semi-major axis $\re{}$ and $\varepsilon_e$ is the ellipticity at $1\re{}$) excluding a fraction of elongated galaxies, whose properties are inconsistent with observations. These criteria selected a sample of 1327 galaxies, of which 1047 are FRs and 280 are SRs. The stellar mass function and ellipticity distribution of the TNG100 ETGs are in reasonable agreement with Atlas3D but, compared to the ePN.S sample, they contain a larger fraction of galaxies high ellipticity and lower masses \citep{Pulsoni2020}.

Each simulated ETG is observed along 100 random line-of-sight directions. For each of these projections, we derive projected $j_p(\lec a)$ profiles, ellipticity $\varepsilon(a)$ profiles, and rotational support $\Lambda(\lec a)$ profiles. 
The ellipticity profiles are derived using the 2-D inertia tensor as in \cite{Pulsoni2020}.
The projected $j_p(\lec a)$ is determined by applying Eq.~\eqref{eq:jp_def2} to the discrete velocities of the particles, summed within elliptical apertures and weighted by the particle stellar masses. We find that using the stellar particles instead of the smoothed velocities (as derived in \cite{Pulsoni2020}) gives very similar results (Fig.~\ref{fig:TNG_jvfields_jparticles} in Appendix), but we adopt the first approach as it is computationally faster. For the same reason, we define a $\Lambda$ parameter quantifying the galaxy rotational support similar to the $\lambda$ parameter in Eq.~\eqref{eq:lambda_def}. This is given by the ratio of the projected angular momentum per unit mass and radius and the square-root of the line-of-sight kinetic energy per unit mass:

\begin{equation}
    \begin{split}
        \Lambda & = \frac{\sqrt{l_{p,x}^2 + l_{p,y}^2}}{\sqrt{K}}\\
\mathrm{where} &\\
        \overrightarrow{l}_{p} & = \frac{\sum_n \overrightarrow{R}_{n} \times \hat{z} v_{z,n} m_n}{\sum_n |\overrightarrow{R}_{n}| m_n } \qquad \mathrm{and} \qquad
        K  = \frac{\sum_n m_n v_{z,n}^2}{\sum_n m_n}.
    \end{split}
\label{eq:Lambda}
\end{equation}
Here $\hat{z}$ is the (random) line-of-sight direction, $x$ and $y$ are the particle coordinates on the projection plane. $\overrightarrow{R}_{n}$ is the 2D position vector of each stellar particle with mass $m_n$ on the $(x,y)$ plane, and $v_{z,n}$ is the particle line-of-sight velocity with respect to the center of mass of the galaxy.
Here we consider $\Lambda(\lec a)$ calculated within elliptical apertures of semi-major axis $a$.
The advantage of using $\Lambda$ is that the former can be easily derived from the discrete velocities of the particles and does not necessarily require the intermediate step of producing mean velocity and velocity dispersion fields.
And, indeed, applying Eq.~\eqref{eq:Lambda} to the mean velocity $V$ and velocity dispersion fields $\sigma$ (where in $K$ the $v_{z,n}^2$ is substituted with the $V_{rms} = V^2 + \sigma^2$) delivers nearly identical results (see Fig.~\ref{fig:TNG_Lambdavfields_Lambdaparticles} in Appendix). This allows us to also consistently derive $\Lambda$ for the ePN.S galaxies from their velocity fields (see Fig.~\ref{fig:lambda_profiles_example_galaxy} in Appendix).

\subsection{Comparison with the ePN.S ETGs}\label{sec:jp_TNG_VS_ePNS}

\begin{figure}
    \centering
    \includegraphics[width=\linewidth]{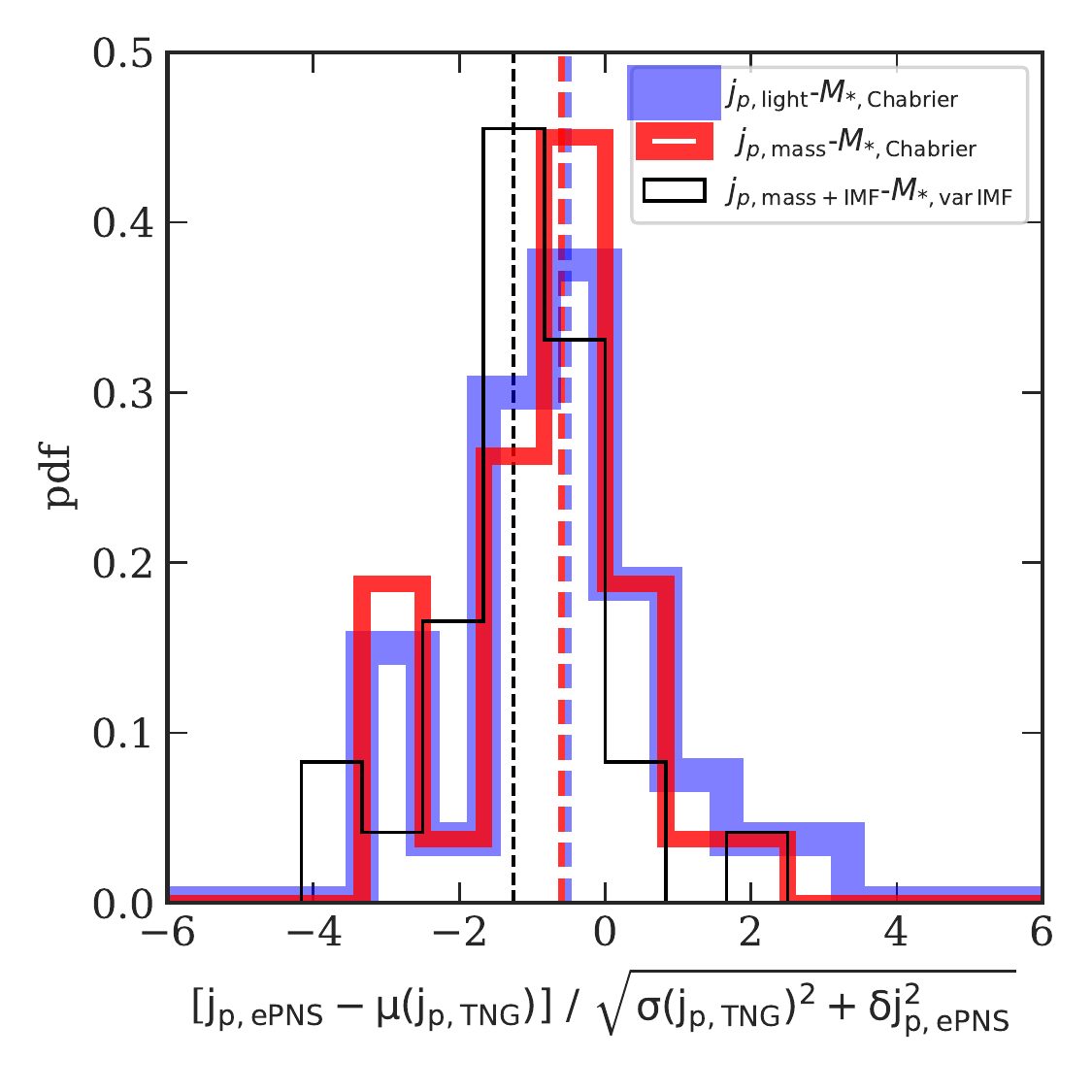}
    \caption{Comparison between the projected sAM of each ePN.S galaxy $j_{p,ePN.S}$ and the median of the distribution of $j_{p}(\lec 6R_e)$ of the ePN.S analogs among the TNG100 ETGs, divided by the sum in quadrature of $1\sigma$-scatter of distribution of the TNG analogs and the error on the ePN.S $j_{p,ePN.S}$. The three histograms correspond to the three $j_p$ determinations for the ePN.S galaxies: the light-weighted $j_{p,\mathrm{light}}$, the mass-weighted $j_{p,\mathrm{mass}}$, and the $j_{p,\mathrm{mass+IMF}}$ corrected for IMF gradients. The $j_p$ values for the TNG100 galaxies are mass-weighted. Vertical dashed lines show the median of the three distributions.}
    \label{fig:jp_ePNSvsTNG}
\end{figure}

To compare the projected sAM of simulated ETGs with that of the ePN.S galaxies, we first need to match the selection function of the two samples (see Sects.~\ref{sec:ePNS_survey} and \ref{sec:TNG100_sample_selection_and_derivation_of_physical_quantities}). Since $j_p$ changes systematically with morphology as well as stellar mass, we match the ePN.S galaxies with simulated galaxies of similar (projected) types and $M_*$. That is, for each ePN.S galaxy we select an ensemble of analogs among the $1327\times100$ random projections of TNG100 ETGs such that they belong to the same rotator class (FRs or SRs) and have similar stellar mass $M_*$, projected ellipticity $\varepsilon$, and rotational support $\Lambda$: $|M_* - M_{*,ePN.S}|\leq \Delta_{M_*}, | \varepsilon - \varepsilon_{ePN.S}| \leq \Delta_\varepsilon, |\Lambda-\Lambda_{ePN.S}| \leq \Delta_\Lambda$. 
The quantities $\varepsilon$ and $\Lambda$ are measured where $\Lambda$ is maximum. This choice is justified by the fact that, although the TNG100 ETGs qualitatively reproduce the $\lambda(a)$ and $\varepsilon (a)$ profiles of observed ETGs \citep{Pulsoni2020}, the simulated galaxies have a less steep distribution of the angular momentum with radius in the central $1-3 R_e$ (see Sect.~\ref{sec:TNG_AMradial_distribution} in the Appendix). 
This is also reflected in the systematically larger radii at which the simulated galaxies reach the peak in rotation compared to observed ePN.S ETGs \citep{Pulsoni2020}. 
Therefore, the most consistent way to compare observations with simulations is to consider $\varepsilon$ and $\Lambda$ where rotation is maximum, instead of considering values measured at fixed multiples of $R_e$.

\begin{figure}
    \centering
    \includegraphics[width=\linewidth]{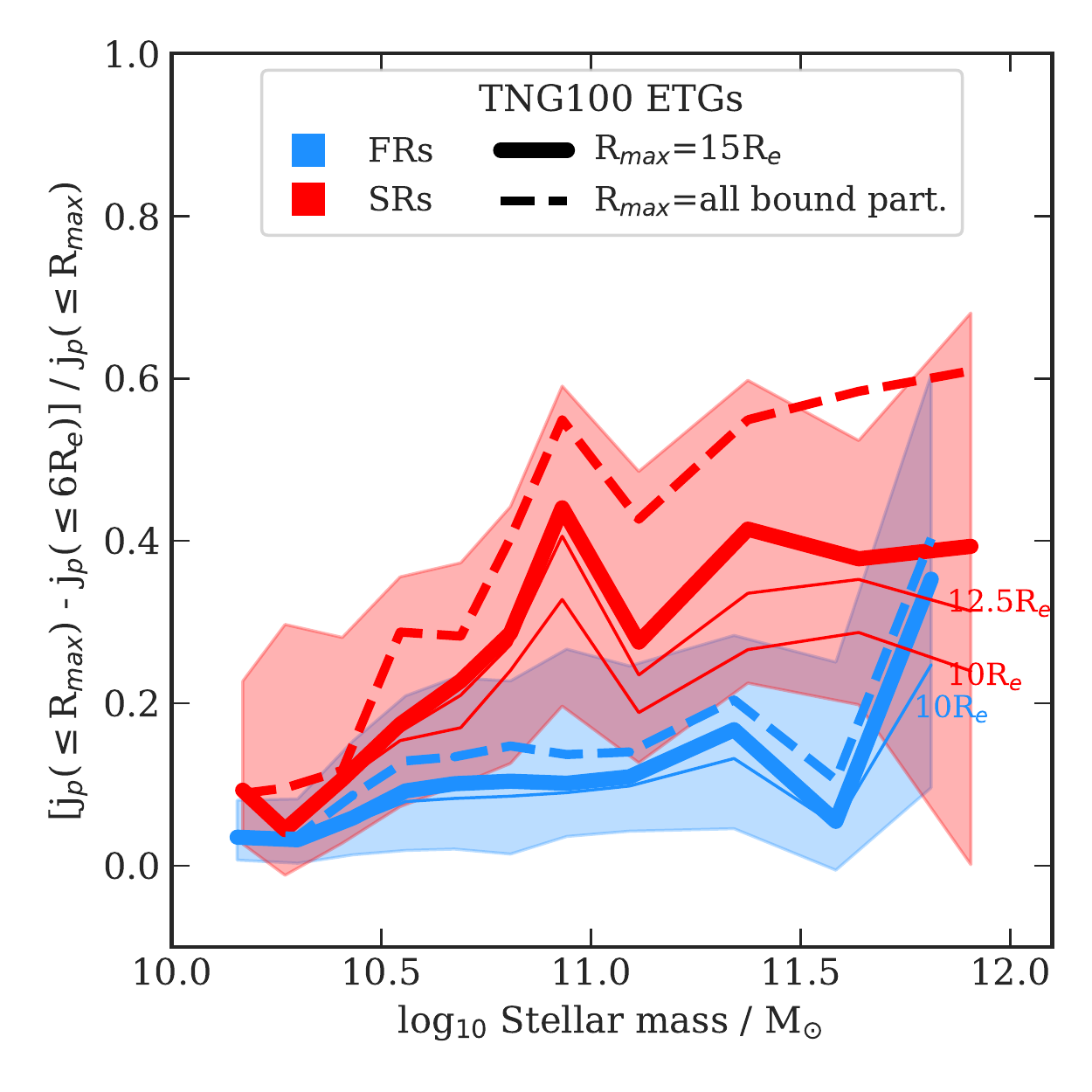}
    \caption{Median difference between the total, galaxy-integrated $j_p$ and $j_p\leq6R_e$, divided by $j_p$, as a function of stellar mass in the TNG100 ETGs. Galaxies are divided into FRs (blue lines) and SRs (red lines). Solid thick lines are for $j_p$ integrated out to $15 R_e$; the shaded areas report the quartiles of the distribution. The dashed lines are for $j_p$ integrated out to the outermost bound particle, while thin solid lines are for apertures of $10$ and $12.5 R_e$.}
    \label{fig:jp_outside6Re}
\end{figure}

Figure~\ref{fig:jp_ePNSvsTNG} shows the distribution of the differences between the $j_p$ of the ePN.S galaxies and the median of the $j_p(\lec 6R_e)$ of the simulated ePN.S analogs, divided by the sum in quadrature of the scatter of the distribution and the error on the ePN.S $j_p$. In this figure $ \Delta_{M_{*}}=0.1\mathrm{dex}\; \MSUN{}, \Delta_\varepsilon=0.05, \Delta_\Lambda=0.05 $, but the results do not depend on the bin size.
The three distributions correspond to the light-weighted ePN.S $j_{p,\mathrm{light}}$, the mass-weighted ePN.S $j_{p,\mathrm{mass}}$, and the $j_{p,\mathrm{mass+IMF}}$ values corrected for IMF-driven gradients in the mass-to-light ratios. In the case of $j_{p,\mathrm{light}}$, we use the blue light-weighted $\Lambda_\mathrm{light}$ values (Eq.~\eqref{eq:Lambda}); in the case of $j_{p,\mathrm{mass}}$ and $j_{p,\mathrm{mass+IMF}}$, we use the mass-weighted $\Lambda_\mathrm{mass}$ through the IR fluxes, which are slightly smaller by a median $2\%$ than the $\Lambda_\mathrm{light}$ (see also Fig.~\ref{fig:lambda_profiles_example_galaxy} in Appendix). We do not correct $\Lambda$ for IMF gradients as $\Lambda$ is only used to match the ePN.S galaxies to the TNG ETGs, and the TNG simulations adopt a constant IMF.
The $j_p$ and $\Lambda$ values of the TNG100 ETGs are instead always mass weighted.

The distribution of the differences is roughly centered at $-0.5\sigma$-scatter around the median for $j_{p,\mathrm{light}}$, indicating that the ePN.S galaxies have slightly lower $j_p$ within $\sim6R_e$ than the TNG100 analogs. Conversely, TNG100 ETGs selected to have $M_*$ and $j_p$ close to the ePN.S galaxies are of "earlier type", with slightly lower rotational support $\Lambda$ and ellipticity $\varepsilon$. For $j_{p,\mathrm{mass+IMF}}$, the median of the distribution shifts to lower values ($-1.25\sigma$). 

Overall, the TNG100 simulation gives a reasonably good description of the angular momentum content of ETGs within $6R_e$. Even though the simulated galaxies have a different distribution of  $j_{p}$ in the central regions compared to observations (see Appendix~\ref{sec:TNG_AMradial_distribution}), they are converged to the ePN.S values at $6 R_e$.

\section{The contribution of the outskirts ($a>6R_e$) to the total $j_p$} \label{sec:fract_j_beyond_6Re}

In Sect.~\ref{sec:jp_TNG_VS_ePNS}, we showed that the TNG100 galaxies have similar values of $j_p$ to the ePN.S galaxies. Assuming that the simulated galaxies have a similar distribution of $j_p$ at large radii as observed ETGs, we can use the TNG100 ETGs to estimate the contribution to the total, galaxy-integrated, $j_p$ from the regions outside the radial coverage of the ePN.S data, typically beyond $a>6 R_e$.

Figure~\ref{fig:jp_outside6Re} shows the median difference between the total, galaxy-integrated $j_p$ and $j_p(\lec6R_e)$ as a function of stellar mass in the simulated FRs and SRs. Low mass galaxies have essentially converged to their total $j_p$ already at $6R_e$, especially the FRs. For FRs with  $M_*>10^{10.5}M_\odot$, $j_p$ increases by a median 10\% beyond $6R_e$, with only the most massive systems with $M_* \gtrsim 10^{11.7}M_\odot$ increasing $j_p$ by $30-40\%$. SRs instead increase $j_p$ considerably in their outskirts, by $\sim40\%$ within $6$ and $15R_e$, and by $\sim 60\%$ if we consider all the bound particles. 

However, including all the bound particles in simulated massive ETGs might lead to an over-estimate of the total $j_p$, as many of these galaxies are centrals in their group halos and the bound particles as identified by the \textsc{subfind} algorithm also include the intra-group/intra-cluster light (ICL). Separating among galaxies and ICL is a non-trivial task and beyond the scope of this work (see \citealt{2022FrASS...972283A} for a review). 
For example, \cite{2018MNRAS.475..648P} use an operative definition of ICL as all the stellar particles beyond a fixed aperture of 100 kpc, which corresponds to $\sim 5R_e$ or less in massive ETGs like NGC4472 or NGC4365 \citep[e.g.,][]{2009ApJS..182..216K}. At these radii, the $j_p(\lec a)$ profiles of the ePN.S SRs still increase steeply with $a$. Therefore as a compromise, we consider all the particles within a radius of $15 R_e$, corresponding to $ \sim300$ kpc for these large galaxies to estimate the total $j_p$. This is approximately the radius at which $j_p$ profiles of the two SRs of \cite{RomanowskyFall2012} converge to their total value, given by a power-law extrapolation of their velocity profiles to infinity. 
The choice of $15R_e$ as radial limit of integration of the sAM in the simulated ETGs does not affect the determination of  $j_p$  in FRs but it is critical for massive SRs, with differences of the order $30\%$ on $j_p$ if we vary the integration limit to the whole extent of the simulated stellar halo or, say, to $10R_e$ (see Fig.~\ref{fig:jp_outside6Re}).

\section{Deprojecting the galaxy angular momentum}\label{sec:deprojecting_j}

Building on the result that the TNG100 ETGs are reasonable models for the observed ePN.S galaxies, we use the simulated ETGs to estimate the projection effects and determine the total true (three-dimensional) sAM $j_t$ from the measured projected sAM $j_p$. To correct for projection effects is a non-trivial task, as it requires a full knowledge of the three dimensional rotational velocity field at all radii as well as the three dimensional distribution of stellar mass. A simple way to structure the problem is to define a "deprojection factor" $C_i$ such that
\begin{equation}
    j_t = C_i \; j_{p},
\label{eq:Cj}    
\end{equation}
where $j_{p}$ is the galaxy integrated projected sAM.
The factor $C_i$ therefore includes all dependencies on inclination, density, and rotation-velocity profiles. 

\cite{RomanowskyFall2012} used a similar parametrization as Eq.~\ref{eq:Cj}. However, in that case, the quantity indicated with $j_p$ does not quantify the projected angular momentum on the plane of the sky and is by definition different from the $j_p$ derived in this work \textbf{in Sect.~\ref{sec:ePN.S_jp}} and in Eq.~\ref{eq:Cj}. Therefore we can not directly apply the $C_i$ factor derived in \cite{RomanowskyFall2012} to our data.

In previous work, $C_i$ was calibrated using theoretical models. By assuming that galaxies are transparent spheroids, with axisymmetric density distributions and cylindrical velocity fields, \cite{RomanowskyFall2012} find that $C_i$ depends primarily on the inclination relative to the line-of-sight and little ($\sim10\%$) on the detailed shape of the rotation profile, while it shows no dependence on the S{\'e}rsic index. 
With these assumptions on the velocity fields, they estimated an inclination-averaged deprojection factor separately for the elliptical and the lenticular galaxies to take into account the inclination bias in the galaxy classification. 

However, the assumptions of axisymmetry and cylindrical velocity fields do not hold for all the ePN.S ETGs. In this sample of galaxies we observe a large variety of velocity fields \citep{Pulsoni2018}. Most FRs display velocity fields with rotation strongly concentrated along the major axis, indicating deviations from the cylindrical geometry. In addition all the ePN.S SRs and 40\% of the ePN.S FRs show kinematic signatures of triaxial stellar halos. This diversity of kinematics and intrinsic structure complicates the derivation of $C_i$ from models.

A way forward is offered by the simulated TNG100 ETGs, which are found to qualitatively reproduce the observed ePN.S velocity fields and ellipticity profiles, suggesting a similar intrinsic structure \citep{Pulsoni2020}, and $j_p$ values (Sect.~\ref{sec:jp_TNG_VS_ePNS}). 
Therefore, by assuming that the TNG100 ETGs are reasonable models of the real ETGs, we can use them to directly measure the projection factor $C_i$ connecting $j_p$ to $j_t$. 

\subsection{The projection factor as a function of inclination}\label{sec:Ci_inclination}

\begin{figure}[h]
    \centering
    \includegraphics[width=1.\linewidth]{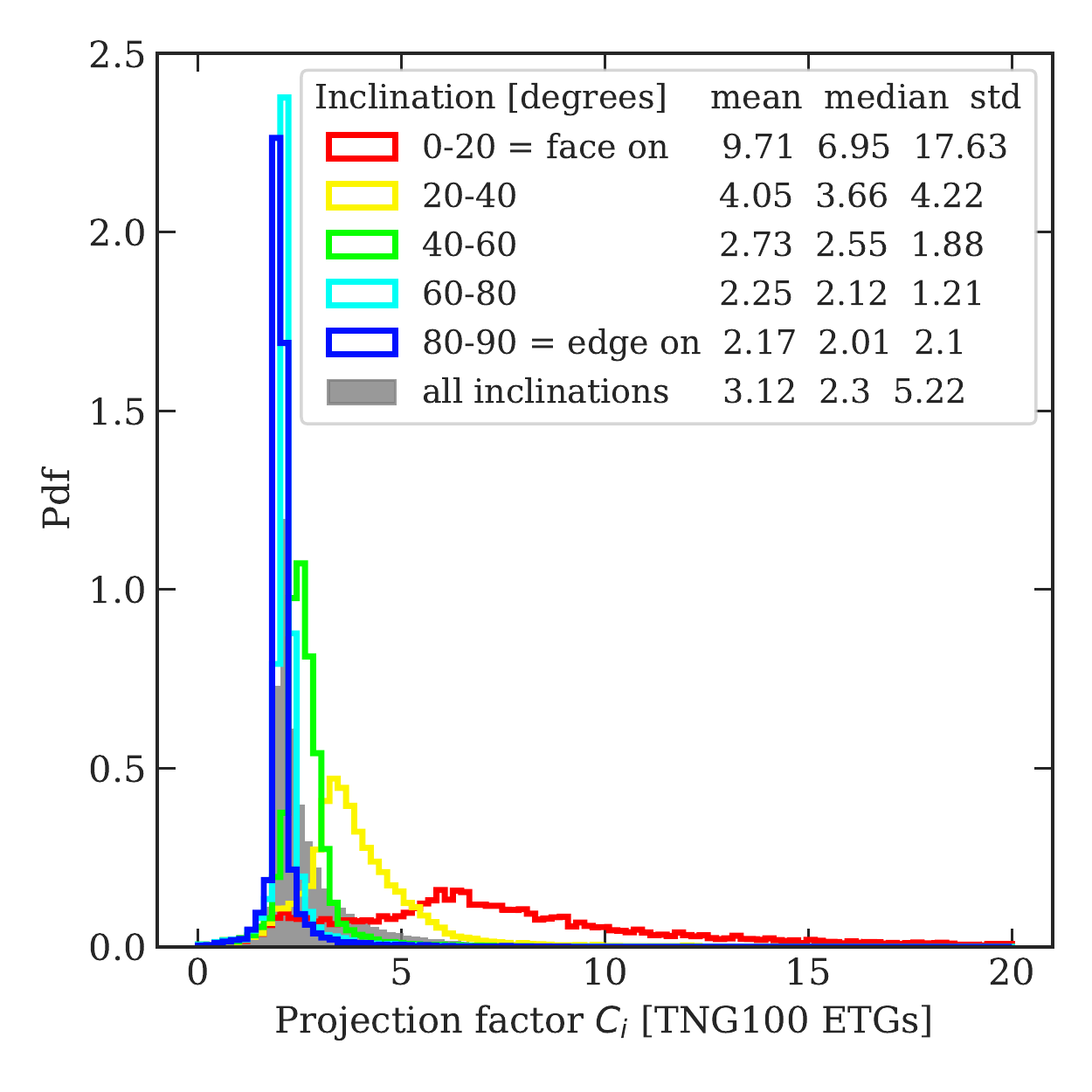}
    \caption{Distribution of $C_i$ for different inclinations measured on 100 random line-of-sight projections of the TNG100 ETGs, as labelled in the legend. The legend also lists the mean, the median, and the standard deviation of each distribution. The gray histogram show the distribution of $C_i$ for all inclinations. }
    \label{fig:Cj_inclinations}
\end{figure}

We derive the deprojection factor $C_i$ using 100 random line-of-sight projections of our sample of 1327 simulated ETGs. For each projection, we derive $C_i = j_p / j_t$ using all the particles within $15R_e$ (see Sect.~\ref{sec:fract_j_beyond_6Re} for the justification). Figure~\ref{fig:Cj_inclinations} shows the distribution of $C_i$ values for different inclinations. For inclinations close to edge-on, $j_p$ differs from $j_t$ by a factor of $\sim2$ with small scatter\footnote{Even in the case of an edge-on disk galaxy $j_p<j_t$, as only the $v_z$ component of the velocity can be measured, not the entire $v_\phi$.}. At lower inclinations, the mean factor $C_i$ increases as a larger fraction of the angular momentum becomes hidden by projection effects. The width of the $C_i$ distribution becomes correspondingly larger. For near-face-on projections, the projected sAM is (on average) only a small fraction of the total. The $C_i$ distribution in fact stretches from values close to $1$, which are for minor-axis rotators, to values above 10. In these cases, $j_t$ can only be recovered with a large uncertainty.

\begin{figure*}[h]
    \centering
    \includegraphics[width=\linewidth]{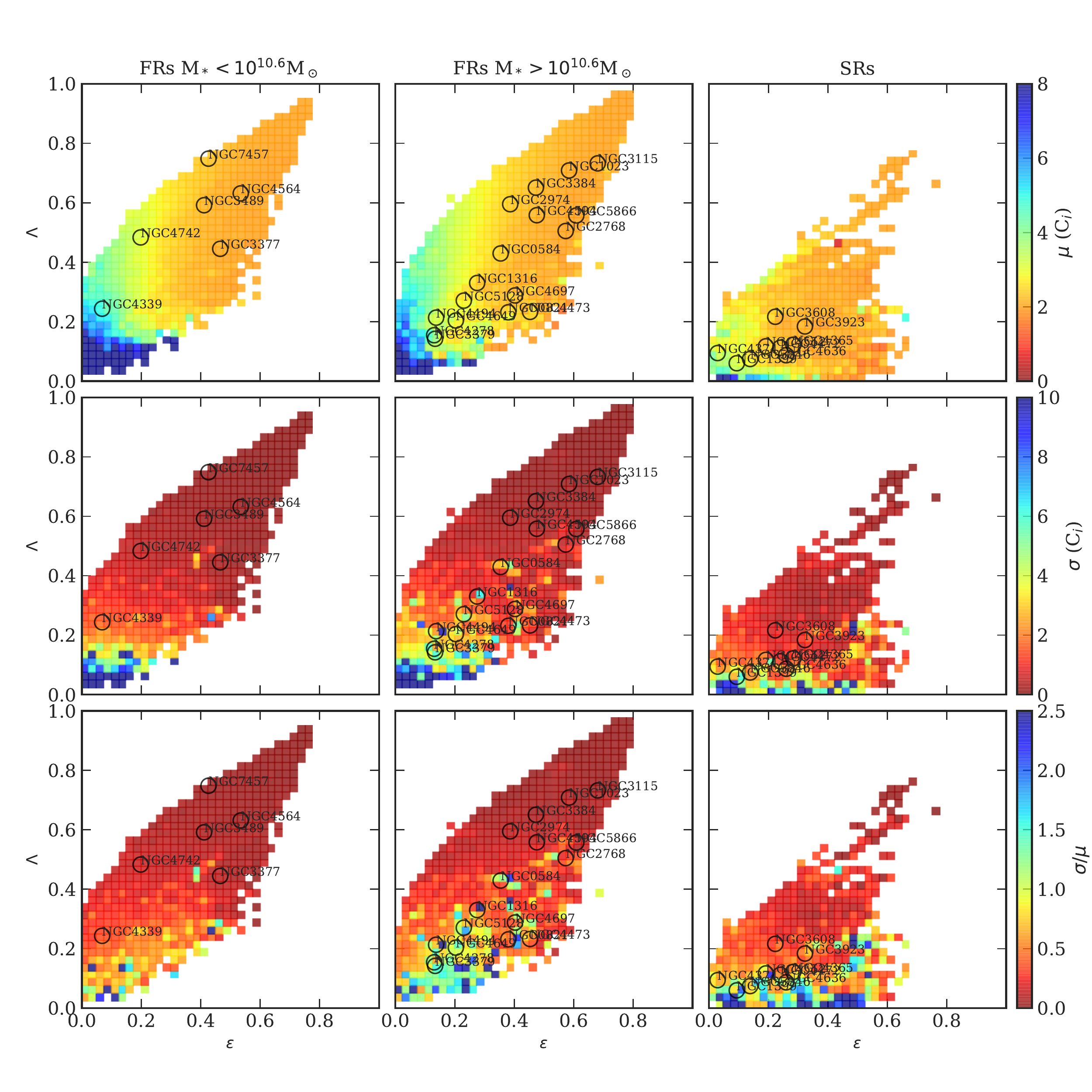}
    \caption{Distribution of $C_i$ in the  $\Lambda - \varepsilon$ plane, where $\Lambda$ and $\varepsilon$ are measured at the location of maximum rotational support for the entire sample of simulated ETGs. Galaxies are divided in low mass FRs (\textbf{left}), high mass FRs (\textbf{center}), and SRs (\textbf{right}). The first row of panels shows the median $\mu(C_i)$ in each bin. The second row shows the root-mean-square scatter around the mean, $\sigma(C_i)$. The third row shows the ratio between scatter and median. The location of the ePN.S galaxies in the $\Lambda_\mathrm{light} - \varepsilon$ plane is also shown with open circles. For each ePN.S ETG, the median $C_i$ value of its "analogs", selected to have similar $\Lambda$, $\varepsilon$, and $M_*$, is used to "de-project" $j_p$ (see Sect.~\ref{sec:ePNS_jt_M*_relation}).}
    \label{fig:Cj_simulations}
\end{figure*}

Figure~\ref{fig:Cj_inclinations} demonstrates the strong dependency of $C_i$ on inclination and therefore the need of an assessment of the galaxy inclination before any attempt of reconstructing $j_t$ from $j_p$. In \cite{RomanowskyFall2012}, the problem is bypassed by considering inclination averaged values for $C_i$ and neglecting possible inclination biases. For the TNG100 ETGs, the median $C_i$ over all the inclinations is $2.3$ (mean 3.12, see the gray histogram in Fig.~\ref{fig:Cj_inclinations}).

\subsection{The projection factor from observed $\Lambda-\varepsilon$ plane}\label{sec:Ci_lambda_epsilon}

Galaxy inclinations for ETGs are difficult to estimate. However, the projected shape and rotational support of galaxies are also a function of inclination. Therefore, one can expect a dependence of $C_i$ on $\Lambda$ and $\varepsilon$, which are directly measurable quantities. Fig.~\ref{fig:Cj_simulations} shows the variation of $C_i$ with these observables for the entire sample of simulated ETGs, not just the ePN.S analogs, with $\Lambda$ and $\varepsilon$ calculated at the location of maximum $\Lambda$
(see Sect.~\ref{sec:jp_TNG_VS_ePNS}). We divide galaxies into SRs, low, and high mass FRs ($M_*>10^{10.6}M_\odot$).

Indeed $C_i$ varies smoothly with the observables $\Lambda$ and $\varepsilon$. At high $\Lambda$ and $\varepsilon$ the median value of $C_i$ is close to 2 and the scatter in the distribution of $C_i$ is very small, of the order 0.2 or less. This is much smaller than the scatter for edge-on ETGs (Fig.~\ref{fig:Cj_inclinations}), because high $\Lambda$ and $\varepsilon$ single out edge-on galaxies with strong intrinsic rotation and flatter shapes. 
At decreasing $\Lambda$ and $\varepsilon$, the values of $C_i$ increase in parallel with decreasing inclinations. There is a difference in the trend of $C_i$ at decreasing $\Lambda$ and $\varepsilon$ for low mass FRs, high mass FRs, and SRs. The distribution of $C_i$ reaches high median $\mu(C_i)$ and scatter values $\sigma(C_i)$ for low mass FRs, up to $\mu(C_i)$ of the order 10 when they are observed close to face-on. For the high mass FRs and SRs the increase in both median and scatter is progressively reduced, indicating a systematic change in the dynamical structure of these galaxies with an increasing contribution to $j_t$ from minor axis rotation (see Fig.~\ref{fig:Contribution_jxjy}). For the SRs, $\mu(C_i)$ does not exceed 4-5.

To conclude, the TNG100 ETGs reveal that 3D sAM can be well predicted given the projected ellipticity ($\varepsilon$) and velocity field ($\Lambda$) of an ETG.

\section{The total sAM of the ePN.S galaxies}\label{sec:ePNS_jt_M*_relation}

\begin{figure*}[h]
    \centering
    \includegraphics[width=0.49\linewidth]{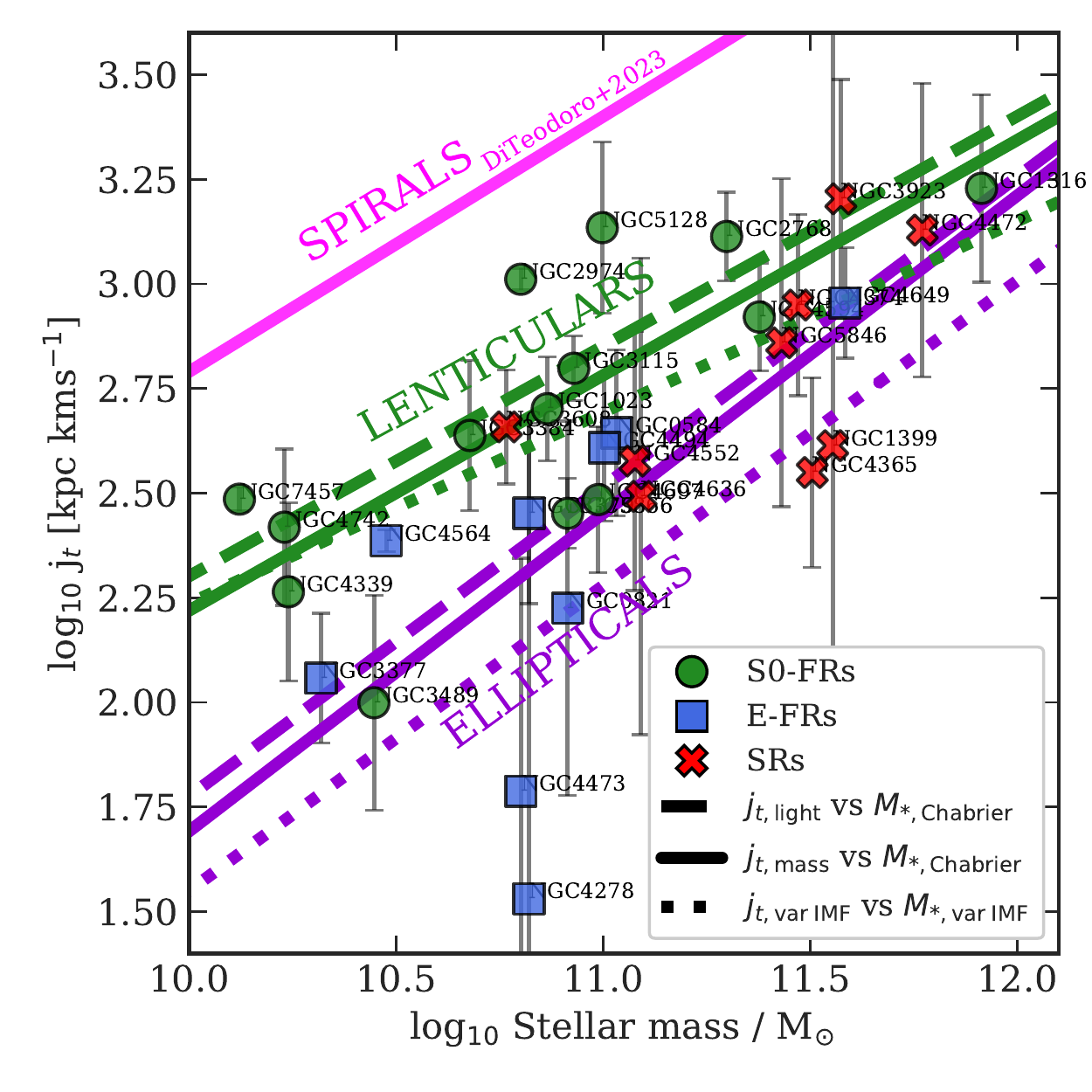}
    \includegraphics[width=0.49\linewidth]{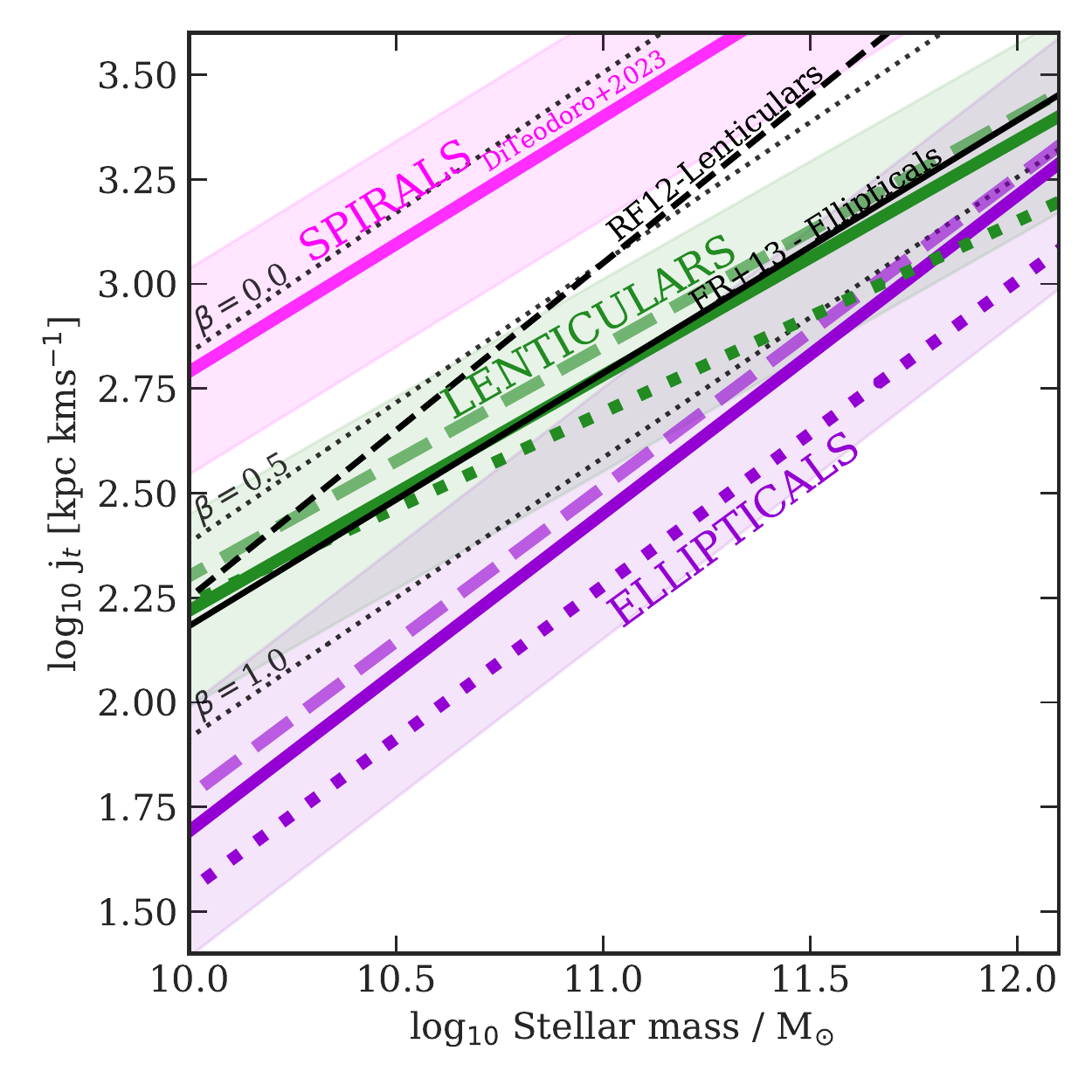}
    \caption{The $j_t - M_*$ plane for the ePN.S galaxies (\textbf{left}) and comparison with previous works (\textbf{right}). 
    \textbf{Left panel:} The total sAM of the ePN.S galaxies as a function of their stellar mass. For clarity we only show the mass-weighted $j_{t,\mathrm{mass}}$ values data-points. Solid lines show the power-law fit to $j_{t,\mathrm{mass}}$ vs $M_*$; dashed lines show the fit to the mass-weighted $j_{t,\mathrm{light}}$ vs $M_*$; dotted lines show the fit to the values corrected for IMF gradients $j_{t,\mathrm{mass+IMF}} - M_\mathrm{*,var\;IMF}$ using the mean mass excess profiles from \cite{2023MNRAS.518.3494B} as discussed in Sect.~\ref{sec:ePN.S_jp_profiles_mass_variableIMF}. Green lines refer to S0 galaxies, purple lines refer to the ellipticals. 
    \textbf{Right panel:} Dashed green and purple lines show the power-law fits to the ePN.S lenticulars and elliptical galaxies, respectively, as in the left panel.
    Dotted black lines show the $j_t-M_*$ relations for different bulge fractions $\beta$ as derived by \cite{Fall2018}. Dashed and solid black lines show the fits of \cite{RomanowskyFall2012} and \cite{Fall2013} to their sample of S0s and ellipticals. In both panels, the magenta line shows the results of 
    \cite{2023MNRAS.518.6340D} on their samples of spiral galaxies.}
    \label{fig:jt_M*_epns}
\end{figure*}

In this section we derive the total sAM $j_t$ for the ePN.S galaxies from the measured projected $j_p$. To do this, we need to estimate the increase of sAM at large radii that is "missed" by the spatial coverage of the ePN.S survey and determine the projection factor as defined in Eq.~\eqref{eq:Cj} to correct for projection effects. 
As discussed above, we can consider the TNG100 ETGs as good, physically motivated, models of the ePN.S ETGs and use them to evaluate the corrections on the observed $j_p$ values.  

For each ePN.S galaxy, whose PN data extend out to $a_\mathrm{max}$, we derive the total sAM using a median correction factor from the TNG ETGs divided in low mass FRs, high mass FRs (threshold mass $10^{10.6}M_\odot$), and SRs, and with similar $\Lambda_{}$ and $\varepsilon$:

\begin{equation}
\begin{split}
j_{t,ePN.S} & \equiv \mathrm{median} \left[ C_i \;\frac{j_p(\lec 15 R_e)}{j_p (\lec a_\mathrm{max})} \right]_{TNG100} j_{p,ePN.S} \\
& = \mathrm{median}\left [\frac{j_t(\lec 15R_e)}{j_p (\lec a_\mathrm{max})} \right]_{TNG100} j_{p,ePN.S}
\end{split}
\label{eq:jt_deproject_jp_ePNS} 
\end{equation}
since $C_i \equiv j_t(\lec 15R_e) / j_p (\lec 15 R_e)$. 
The median $C_i$ adopted for each ePN.S galaxy are shown in Fig.~\ref{fig:Cj_simulations}. The correction from $j_p(\lec a_\mathrm{max})$ to the total $j_p$ is quantified in Fig.~\ref{fig:jp_outside6Re}.
When applying Eq.~\eqref{eq:jt_deproject_jp_ePNS} to the blue light-weighted $j_{p}$ we use correction factors based on the blue light-weighted $\Lambda_\mathrm{light}$ values, while for $j_{p,\mathrm{mass}}$ and $j_{p,\mathrm{mass+IMF}}$ we use the mass-weighted $\Lambda_\mathrm{mass}$.

\subsection{The $ j_{t} - M_{*} $ diagram}\label{jt-M*_morphology}

The left panel of Fig.~\ref{fig:jt_M*_epns} shows the $j_t - M_*$ relation for the ePN.S galaxies. 
The ePN.S sample displays the well-known increase of sAM with stellar mass, and the correlation with morphology. Elliptical galaxies have significantly lower average $j_t$ than lenticulars of similar stellar masses. 

The uncertainties shown as error-bars in Fig.~\ref{fig:jt_M*_epns} are derived from the errors on the projected $j_p$ (see Sect.~\ref{sec:errors_jp}) and the width in the distributions of the correction factor $j_t(\lec 15R_e)/j_p (\lec a_\mathrm{max})$ from the simulated ETGs. The widths of these asymmetric distributions around the medians are estimated using their quartiles.

We fit the power-law in Eq.~\eqref{eq:j-M_powerlaw} to $j_{t,\mathrm{light}}-M_*$, $j_{t,\mathrm{mass}}-M_*$, and $j_{t,\mathrm{mass+IMF}} - M_\mathrm{*,var\;IMF}$, separating between the S0s and ellipticals; see Sect.~\ref{sec:fitting}. The results are reported in Table~\ref{tab:fit_jt}. 

For the ellipticals the slope in the mass-weighted case is $0.76\pm0.23$, while for the S0 this is $0.55\pm0.17$. Weighting by light or including the IMF gradients does not strongly impact the value of the slope. Only for the S0s, the slope decreases slightly from 0.54 to 0.45 when including IMF gradients.

The normalisation of the power-law at $10^{11}\MSUN{}$ is $2.45\pm0.10$ dex for the ellipticals in the mass-weighted case. This is a factor of two lower than for the S0s, and a factor of 9 lower than spiral galaxies (see also Sect.~\ref{jt-M*_spirals}). The normalisation is systematically higher in the light-weighted case by a factor of 1.2, while it decreases by a factor $\sim1.4$ when accounting for IMF gradients. 

\begin{table}[]
    \centering
    \begin{tabular}{lccc}
\hline\hline\noalign{\smallskip}
  \multicolumn{1}{l}{group of data} &
  \multicolumn{1}{c}{A} &
  \multicolumn{1}{c}{$\log_{10}j_0$} &
  \multicolumn{1}{c}{$\sigma_\bot$}\\
  \multicolumn{1}{l}{} &
  \multicolumn{1}{c}{} &
  \multicolumn{1}{c}{} &
  \multicolumn{1}{c}{[dex]}\\
\noalign{\smallskip}\hline\noalign{\smallskip}
    \textbf{Ellipticals} &&&\\
    $j_{t,\mathrm{light}} - M_*$ &  0.74$\pm$0.22  &  2.52$\pm$0.09 & 0.22\\
    $j_{t,\mathrm{mass}} - M_*$ & 0.76$\pm$0.23  &  2.45$\pm$0.10 & 0.24 \\
    $j_{t,\mathrm{mass+IMF}} - M_\mathrm{*,var\;IMF}$& 0.73$\pm$0.27  &  2.28$\pm$0.13 & 0.30 \\
\noalign{\smallskip}\hline\noalign{\smallskip}
    \textbf{Lenticulars} &&&\\
    $j_{t,\mathrm{light}} - M_*$ &  0.54$\pm$0.16  &  2.85$\pm$0.08 & 0.19\\
    $j_{t,\mathrm{mass}} - M_*$ & 0.55$\pm$0.17  &  2.78$\pm$0.08 & 0.20\\
    $j_{t,\mathrm{mass+IMF}} - M_\mathrm{*,var\;IMF}$ & 0.45$\pm$0.16 &  2.68$\pm$0.08 & 0.21 \\
\noalign{\smallskip}\hline\noalign{\smallskip}
    \textbf{Slow Rotators} &&&\\
    $j_{t,\mathrm{light}} - M_*$ &  0.49$\pm$0.30  &  2.67$\pm$0.14 & 0.17\\
    $j_{t,\mathrm{mass}} - M_*$ & 0.50$\pm$0.33 &  2.60$\pm$0.16 & 0.18\\
    $j_{t,\mathrm{mass+IMF}} - M_\mathrm{*,var\;IMF}$ & 0.47$\pm$0.34 &  2.48$\pm$0.18 & 0.19 \\
\noalign{\smallskip}\hline\noalign{\smallskip}
    \textbf{Fast Rotators} &&&\\
    $j_{t,\mathrm{light}} - M_*$ &  0.57$\pm$0.19  &  2.69$\pm$0.12 & 0.29\\
    $j_{t,\mathrm{mass}} - M_*$ & 0.59$\pm$0.21  &  2.62$\pm$0.10 & 0.30\\
    $j_{t,\mathrm{mass+IMF}} - M_\mathrm{*,var\;IMF}$ & 0.47$\pm$0.22 &  2.5$\pm$0.11 & 0.38 \\
    \end{tabular}
    \caption{Results of the fit of $j_t$ versus $M_*$ with the power-law in Eq.~\eqref{eq:j-M_powerlaw}. The rightmost column lists the unweighted orthogonal scatter $\sigma_\bot$ of the data points around the fitted relations. The errors on the parameters are the sum in quadrature of the error on the fit to the unweighted data-points, and the root-mean-square sigma of the distribution of parameters from fitting Monte Carlo simulations of the $j_t$ values sampled from their errors. See Sect.~\ref{sec:fitting}. }
    \label{tab:fit_jt}
\end{table}

For the ellipticals the scatter is relatively larger and the power-law  therefore less certain (Table~\ref{tab:fit_jt}). The vertical scatter $\sigma(\log j_t)$ is of order 0.3 (0.37 including IMF variations) and comparable to or slightly larger than the combined scatter of 0.31dex expected from the distribution of dark matter halo spin \citep[0.23dex,][]{2008MNRAS.391.1940M}, the stellar-mass-halo-mass relation \citep[0.15dex,][]{Moster2013_SMHM}, and the median error of $40\%j_t$ in the sAM measurements corresponding to 0.15dex. The small difference between the observed and expected values likely reflects the different formation and evolution paths characterising these objects. Table~\ref{tab:fit_jt} reports the orthogonal scatter $\sigma_\bot = \sigma(\log j_t)/\sqrt{1+A^2}$ with respect to the power-law fits to facilitate comparison with previous work. 

\subsubsection{Fitting and variations} \label{sec:fitting}

The results quoted in Table~\ref{tab:fit_jt} derive from a least square fit to the data-points without weighting them by their uncertainties, as also in previous work on ETGs. This is motivated by the fact that galaxies with lower angular momentum or higher stellar masses are also those with larger formal errors, both from the PN velocity fields and the from width of the distribution of the projection factors. Weighting by the errors would lead the fit to be completely driven by a handful of E-FRs with the smallest errors, biassing $\log j_0$. We tried to overcome this by imposing a minimum value for the uncertainties equal to the scatter of the data-points around the power-law. In this case, while the fitted parameters for the S0s are similar to the un-weighted case, for the Es the slope decreases by 10\% and the normalisation increases by 5\% because of still higher weight of lower-mass, faster-rotating ellipticals. 

Monte Carlo simulations with similar samples of measurements drawn from a power-law relation, with errors depending on $M_*$ and $j_t$ as in the observed sample, and typical intrinsic scatter were made to test different fit methods. The standard deviation in $A$ and $\log j_0$ for the elliptical galaxy sample were found to be typically $\sim0.2$ and $\sim0.1$dex, respectively, consistent with the errors given in Table~\ref{tab:fit_jt}. Biases in the mean were small for $A$ but can be substantial (up to 0.2dex) for $\log j_0$. The least biased results in the mean were obtained from unweighted fits. In the discussion above we have therefore quoted the results of the unweighted fit as less biased towards high-sAM galaxies. 

The fitted parameters listed in Table~\ref{tab:fit_jt} are for $j_t$ obtained with correction factors integrated out to 15$R_e$ (see Eq.~\ref{eq:jt_deproject_jp_ePNS} and Sect.~\ref{sec:fract_j_beyond_6Re}). Extending the outer boundary to integrate over all the bound particles of the simulated TNG ETGs leads to larger correction factors for the more massive ellipticals, up to a median 0.06 dex for the most massive SRs. This modest increase in the $j_t$ values at the high mass end determines a steeper slope of the $j_t-M_*$, from $\sim 0.75$ to $\sim 0.85$ for the elliptical galaxies, but leaves the normalisation unchanged. Hence integrating out to the virial radius, and therefore including the contribution of the ICL in the correction factors, does not strongly impact our conclusions.

\subsection{The $ j_{t} - M_{*} $ diagram for FRs and SRs}\label{jt-M*_rotatorclass}

Based on IFS studies revealing an ubiquity of rotating components in ETGs \citep{2011MNRAS.414.2923K, 2017ApJ...835..104V, 2018MNRAS.477.4711G}, the classification scheme for these galaxies has shifted from a morphological to a kinematic paradigm, distinguishing between FRs and SRs \citep{2007MNRAS.379..401E}. Figure~\ref{fig:jt_M*_epns} (left panel) shows that FRs and SRs are segregated in stellar mass but not in $j_t$. The $j_t$ values of the SRs are more uncertain than the FRs, because of the larger uncertainties on $j_p$ as well as the larger contribution of $j_t$ distributed beyond the radial coverage of our data (see Sect.~\ref{sec:fract_j_beyond_6Re}). 
The ePN.S FRs, on the other hand, show a larger scatter among galaxies at fixed stellar mass, with $j_{t}$ values differing by more than an order of magnitude at fixed $M_*$ (see also the values of $\sigma_\bot$ for the two families in Table~\ref{tab:fit_jt}). This wide range of $j_{t}$ is unlikely explainable by projection effects, which should be already accounted for by the dependence of the correction factor $C_i$ on the projected $\Lambda$ and $\varepsilon$. The differences in the measured $j_{t}$ is likely intrinsic and driven by differences in the bulge fractions among FRs.

A fit of the power-law in Eq.~\eqref{eq:j-M_powerlaw} to the FRs and SRs separately yields very similar relations, with nearly identical normalisation and slightly steeper slope for the FRs, despite the difference in the stellar mass range probed by the two samples (see Table~\ref{tab:fit_jt}).
Our results suggest that there is no fundamental difference in the sAM content of FRs and SRs, but only in the way this is distributed with radius.

\subsection{Comparison with previous work}\label{jt-M*_previouswork}
The right panel of Fig.~\ref{fig:jt_M*_epns} compares our results with the previous determination of the $j_t-M_*$ relation for ETGs in \cite{RomanowskyFall2012} and in \cite{Fall2013}, and for galaxies with different bulge fractions as derived in \cite{Fall2018}. These relations are obtained from $j_t$ values weighted with blue photometric bands, as the $j_{t,\mathrm{light}}$ in this work. The stellar masses in \citet[][]{RomanowskyFall2012} are derived from K-band photometry assuming a constant mass-to-light ratio $M_*/L_K=1$. \cite{Fall2013,Fall2018} revise $j_t-M_*$ relation using a color-dependent $M_*/L_K$, which returns $M_*$ comparable to ours (see Fig.~\ref{fig:ComparisonStellarMasses}), and which moves the position of the ellipticals slightly upwards by 0.05 dex compared to \cite{RomanowskyFall2012}. 
These determinations are closely comparable with the $j_{t,\mathrm{light}}-M_*$ relations derived in this work.

Our $j_{t,\mathrm{light}}-M_*$ relation for elliptical galaxies has a systematically lower normalisation compared to \citet[][]{RomanowskyFall2012} and \cite{Fall2013} by a factor $1.62$ and $1.67$ respectively. It is closer but still below the $j_t-M_*$ determination for pure bulges in \citet[][for the limit $\beta=1$, see Fig.~\ref{fig:jt_M*_epns}]{Fall2018}. 
The $j_{t,\mathrm{light}}-M_*$ relation for the S0s determined in the present work also has lower normalisation and shallower slope than that of \cite{RomanowskyFall2012}. These differences become more marked with our mass-weighted $j_t$ measurements.

A galaxy-by-galaxy comparison between the the $j_{t,\mathrm{light}}$ values derived here and in \cite{RomanowskyFall2012} is shown in Fig.~\ref{fig:comparisonRF12} and discussed in App.~\ref{sec:appc}: the $j_t$ values are systematically higher in \cite{RomanowskyFall2012} for the subset of galaxies in common. This is at least partially explained by their assumption of cylindrical velocity fields which would systematically overestimates $j_t$. An exception is NGC5128, for which the strong minor-axis rotation gives a non-negligible contribution to $j_t$ that is not accounted for in the major-axis based measure of \cite{RomanowskyFall2012}.
By taking into account the full two-dimensional kinematic information out to large radii and the effects of mass-weighting versus light-weighting, we find that elliptical galaxies have significantly lower $j_t$ than previously estimated.

\subsection{Comparison with spiral galaxies}\label{jt-M*_spirals}

The $j_t-M_*$ relation for spiral galaxies is well established, with different studies returning consistent values for both slope ($\sim 0.6$) and normalisation ($\sim 3.4$ dex at $10^{11}\MSUN{}$, e.g. \citealt{Fall1983, Fall2013, 2014ApJ...784...26O, 2018A&A...612L...6P,2022MNRAS.509.3751H}). This is because of the rapid convergence of the $j_p(\lec a)$ profiles within a few $R_e$ and the better constrained inclination angles compared to ETGs.
The pink line in Fig.~\ref{fig:jt_M*_epns} shows the $j_t-M_*$ relation for spiral galaxies from \citet{2023MNRAS.518.6340D}, which is based on IR light-profiles and therefore directly comparable with our $j_{t,\mathrm{mass}}-M_*$ relations. 

The comparison with the measurements for the ePN.S sample shows that, at fixed stellar mass, earlier morphological types have on average lower $j_t$, confirming the result of \cite{RomanowskyFall2012}. Even in the case of SRs, which have their largest sAM in the outskirts, the estimated $j_t$ do not converge to the values for spiral galaxies with similar stellar masses, even when including the contribution from ICL (see previous Section). 
We find that elliptical galaxies have 9 times (and up to 13 times when including IMF effects) less $j_t$ than spiral galaxies of the same stellar mass.

Another difference between elliptical and spiral galaxies is the larger scatter of the ellipticals in the $j_t-M_*$ diagram. The orthogonal scatter $\sigma_\bot$ in log-log space around their best fitting power-law is $\sim0.24$ (and increases up to 0.3 when including IMF variations). This is larger than the scatter typically measured for spirals, which ranges from 0.17-0.22, depending on the sample \citep[][]{2018A&A...612L...6P, 2022MNRAS.509.3751H, 2021MNRAS.507.5820D, 2023MNRAS.518.6340D}.

\begin{figure}
    \centering
    \includegraphics[width=\linewidth]{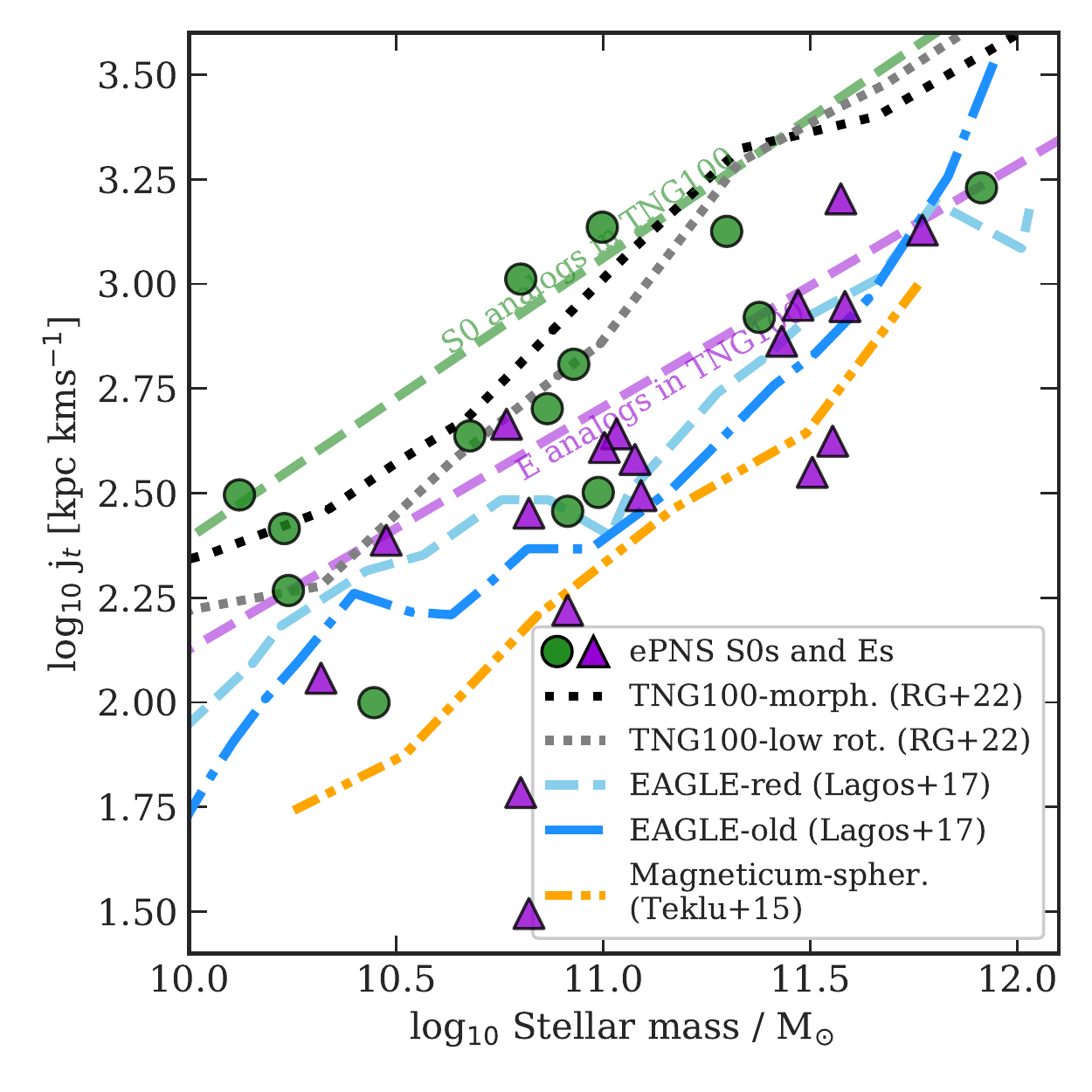}
    \caption{Comparison with simulations. The distribution of ePN.S galaxies on the $j_t-M_*$ plane (mass-weighted measurements with Chabrier IMF) is compared to the fit on their simulated analogs from TNG100 and to results from other works investigating ETGs in cosmological-hydrodynamical simulations (see text).}
    \label{fig:jt_ M*_epnsVSsimulations}
\end{figure}

\subsection{Comparison with simulations}

In this section, we compare the distribution of the ePN.S galaxies in the $j_t-M_*$ plane with that of simulated ETGs from TNG100 and from previous works using cosmological hydrodynamical simulations. For consistency, we consider only the mass-weighted measurements with constant Chabrier IMF. 
Figure~\ref{fig:jt_ M*_epnsVSsimulations} shows that overall the ETG samples from current cosmological hydrodynamical simulations give a good representation of the observed distribution of the ePN.S ETGs in the $j_t-M_*$ plane. 

We first discuss the comparison of the ePN.S galaxies with their analogs among the TNG100 ETGs. The TNG100 analogs were selected in Sect.~\ref{sec:jp_TNG_VS_ePNS} to have similar stellar masses, projected ellipticity, and rotational support as the ePN.S galaxies. For each ePN.S galaxy, we derive a median "simulated" $j_t$ from its analogs, which is used to obtain a corresponding $j_t-M_*$ relation for simulated ellipticals and lenticulars.

The TNG analog ETGs have overall similar properties as the ePN.S ETGs in the $j_t-M_*$ plane, with the analog S0s having systematically larger $j_t$ than the analog ellipticals and $j_t$ increasing with $M_*$ similarly as observed. Both families of TNG100 analogs have on average larger $j_t$ compared to the ePN.S galaxies, by a factor 1.8 for the ellipticals and 1.9 for the lenticulars, however within $\sim1$ standard deviation of the distribution as already quantified in Sect.~\ref{sec:jp_TNG_VS_ePNS} and Fig.~\ref{fig:jp_ePNSvsTNG}. 
This is likely related to the different radial distributions of $j_p$ between TNG and observed galaxies, see Fig.~\ref{fig:TNG1RE_A3D}. For this reason we
used ratios but not absolute $j_p$ values for the TNG galaxies in Eq.~\ref{eq:jt_deproject_jp_ePNS} for estimating the total $j_t$ of the ePN.S galaxies from their measured $j_p$.
It is unclear whether this offset is entirely due to differences in the physical properties between real and simulated galaxies, or whether it also reflects some bias in the sample selection not accounted for by our definition of analogs. The enlarged but still limited number statistics offered by the ePN.S sample does not allow us to investigate this conclusively.

With a different selection of TNG100 ETGs based either on the kinetic energy associated with ordered rotation, or on a neural network trained morphological classification, \cite{Rodriguez-Gomez2022} found median sAM values tracing roughly those of the ePN.S lenticulars (dotted gray gray lines in Fig.~\ref{fig:jt_ M*_epnsVSsimulations}), likely indicating a larger average contribution of disk-like components in their sample ETGs compared to the ePN.S ETGs (see Sects.~\ref{sec:ePNS_survey} and \ref{sec:TNG100_sample_selection_and_derivation_of_physical_quantities}). 

Blue lines in Fig.~\ref{fig:jt_ M*_epnsVSsimulations}) show the median stellar sAM measured within $5R_e$ in galaxies from EAGLE by \cite{Lagos2017}, selected to have the reddest colors, $(u-r)>2.2$, or alternatively the oldest ages in their simulated samples (mass-weighted stellar ages older than 9.5 Gyr). Their measurements follow well the location of the ePN.S ETGs. The smaller aperture used to integrate their sAM should not contribute to important systematic differences at least up to $M_*<10^{11.2}M_\odot$ where the large majority of ePN.S FRs has essentially converged sAM profiles already at 5$R_e$ (see Fig.~\ref{fig:jp_profiles}). The systematic offset with respect to the mean TNG100 values could at least partially be attributed to differences in the selection criteria, as the EAGLE galaxies are on average redder and older, and therefore likely to have lower sAM \citep{Lagos2017}.

Finally, \cite{Teklu2015ApJ...812...29T} analysed spheroidal galaxies from the Magneticum simulation selected based on their circularity parameter. Their sAM values are in good agreement with the ePN.S ellipticals (mean values shown by the orange line in Fig.~\ref{fig:jt_ M*_epnsVSsimulations})).

This comparison shows that the current generation of cosmological simulations reproduce well the properties of observed ETGs on the $j_t-M_*$ plane. The vertical offset among lines in Fig.~\ref{fig:jt_ M*_epnsVSsimulations} is most likely due to different disk-to-spheroid ratios in the different samples, given the sensitivity of $j_t$ to galaxy type, from sample selections and possibly also different galaxy formation models.

\section{The retained fraction of angular momentum}
\label{sec:retained_retention_fraction_fj}

\begin{figure}
    \centering
    \includegraphics[width=\linewidth]{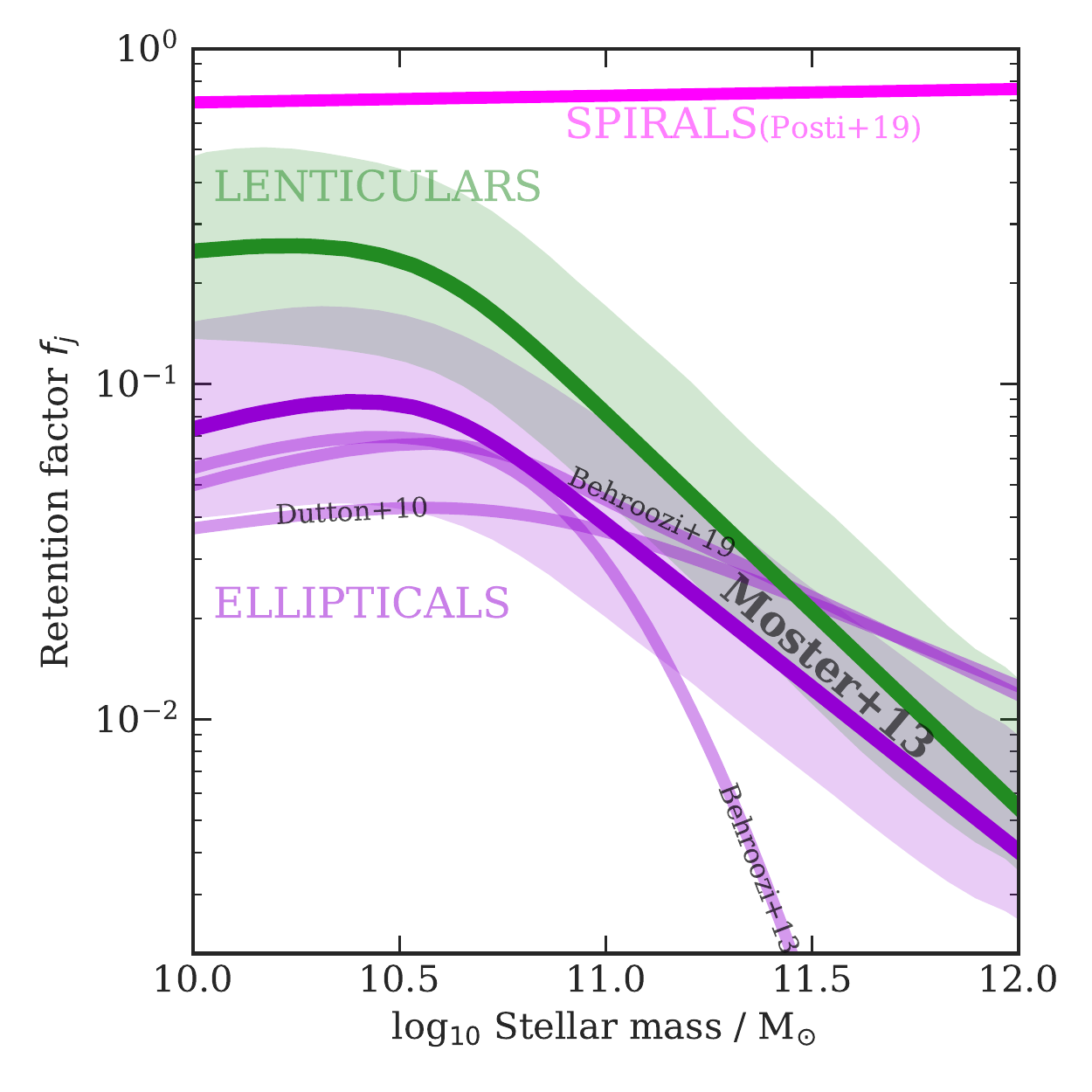}
    \caption{The ratio of the stellar to halo sAM as a function of stellar mass $f_j(M_*)$ for ETGs assuming the SMHM relation of \cite{Moster2013_SMHM}. Ellipticals and S0s are shown with thick purple and green lines, respectively. Colored bands show the scatter around the median given by the Monte-Carlo simulations described in the text. The thinner purple lines show the median $f_j(M_*)$ profiles using $f_M$ from \cite{Dutton2010_SMHM} for ETGs, \cite{Behroozi2013_SMHM}, and \cite{Behroozi2019_SMHM} for quiescent galaxies as labelled.
    For comparison, we also show the results of \cite{2019A&A...629A..59P} for spiral galaxies with a magenta line. 
    }
    \label{fig:retentionfactor_fj}
\end{figure}

In the previous section, we have derived the total sAM $j_t$ for the ePN.S galaxies which we now consider as our estimate of their total $j_*$.
The empirical $j_* - M_*$ relation of the stellar component in galaxies is  often interpreted in conjunction with the properties of their dark matter halos, by defining:
\begin{equation}
    f_j \equiv j_* \;/ \;j_h \qquad \mathrm{and} \qquad f_M \equiv M_*\;/\;M_h,
\label{eq:retentionfactor_fj_SMHM_f*}
\end{equation}
where $j_h$ and $M_h$ are sAM and mass of the dark halo.
$f_M$ is the stellar-mass-halo-mass (SMHM) relation, also referred to as the star formation efficiency, and $f_j$ is the retained fraction of angular momentum. 

The connection between the two components is based on the theoretical framework in which the stellar component "inherits" a fraction $f_j$ of the primordial $j_h$ exerted by tidal torques to the collapsing dark matter halos. The tidal torque theory predicts that $j_h \propto M_h^{2/3}$\citep{1969ApJ...155..393P, 1979MNRAS.186..133E} which, given Eq.~\eqref{eq:retentionfactor_fj_SMHM_f*}, translates into a relation for $j_*$ as a function of $M_*$:
\begin{equation}
j_* \propto f_j\;f_M^{-2/3}\;M_*^{2/3}.
\label{eq:j*proptoM*} 
\end{equation}
Since disk galaxies are observed to follow closely $j_* \propto M_*^{2/3}$, Eq.~\eqref{eq:j*proptoM*} leads to the result that the product $f_j \; f_M^{-2/3} \sim \mathrm{const}$, meaning that galaxies that are "efficient at forming stars are also efficient at retaining angular momentum" \citep{Posti2018A}. Previous results for disks report a retention factor between $0.8$ and $1$ \citep{Fall2013, 2018A&A...612L...6P,2021MNRAS.507.5820D, 2023MNRAS.518.6340D, 2023MNRAS.518.1002R}. For elliptical galaxies, this is more uncertain. 
In this section, we revisit the estimate of the retention fraction $f_j$ for ETGs given the $j_*-M_*$ derived in this work.

We follow the derivation from \cite{Posti2018A}, which assumes a Hubble constant $H_0 = 67.7\;\mathrm{km\;s^{-1}\;Mpc^{-1}}$ \citep{2016A&A...594A..13P} and adopts Navarro, Frenk, and White (\citeyear{1996ApJ...462..563N}) dark matter halos. This gives that the stellar $j_*$ can be written as 
\begin{equation}
j_* = \frac{77.4}{\sqrt{F_E(c)}} \; \left ( \frac{\lambda}{0.035} \right ) \; f_j\;f_{M}^{-2/3} \; \left( \frac{M_*}{10^{10} M_\odot } \right )^{2/3}\; \mathrm{kpc\;km\;s^{-1}}
\label{eq:j*vsM*_Posti}   
\end{equation}
where $\lambda$ is the halo spin parameter \citep{1969ApJ...155..393P} and $F_E(c)$ is a dimensionless factor that depends on the concentration parameter $c$ \citep{1998MNRAS.295..319M}. 

There are several models of the SMHM relation available in the literature. The current consensus is that $f_M$ has a characteristic "bell" shape, so that it peaks around $M_*\sim10^{10.5}M_\odot$ and decreases at lower and higher masses by $\sim2$ order of magnitudes. The decrease at high masses is usually interpreted as signature of AGN feedback reducing the star formation efficiency \citep{1998A&A...331L...1S, 2006MNRAS.365...11C, 2018ARA&A..56..435W}. Here, we consider the model of $f_M$ derived by \cite{Moster2013_SMHM} from abundance matching, which is a good approximation for ETGs \citep[][]{PostiFall2021}, although other models are qualitatively similar \citep[see, e.g.][and references therein]{Behroozi2019}. 

Figure~\ref{fig:retentionfactor_fj} shows the median retention factor $f_j$ and its scatter as a function of the $M_*$ for ellipticals and S0s separately. 
This is derived as in \citet{Posti2018A} by Monte-Carlo simulating $10^4$ galaxies with a uniform distribution in $M_*$ between $10^{10}$ and $10^{12} \MSUN{}$. Based on $M_*$, we assign to each galaxy a halo mass $M_h$ by sampling the SMHM relation of \citet{Moster2013_SMHM} with its scatter (0.15 dex), a halo concentration $c$ by sampling its log-normal distribution centered on the power-law dependence on $M_h$ found by \citet{2014MNRAS.441.3359D}, with scatter of 0.11 dex, and a spin parameter $\lambda$ sampled from a log-normal distribution with mean value $\lambda=0.035$ and scatter of 0.23 dex \citep{2008MNRAS.391.1940M} independent of $M_h$. Finally, we assign a $j_*$ value assuming a Gaussian distribution of $\log j_*$ values with mean value given by the mass-weighted power-laws fitted on the ellipticals and the S0s (Table~\ref{tab:fit_jt}) at $M_*$, and sigma equal to the median uncertainty on the $\log j_*$ measurements. The simulated $M_*$, $M_h$, $c$, $\lambda$, and $j_*$ are inserted in Eq.~\ref{eq:retentionfactor_fj_SMHM_f*} solved for $f_j$, to obtain the distribution of $f_j$ as a function of $M_*$. 

At the peak of the SMHM, that is for $M_*\sim10^{10}-10^{10.5}M_\odot$,
$f_j\sim0.25$ for the S0s and $\sim0.08$ for the ellipticals. 
At higher masses, as the star formation efficiency decreases, $f_j$ is reduced by 1.5 order of magnitudes at $10^{12}M_\odot$. 
It is worth noting that at high stellar masses the SMHM relations are more poorly constrained and the extrapolations to $M_*=10^{12}\MSUN{}$ of different models give results that can differ by more than an order of magnitude in $f_M$ because of the uncertainties in the stellar mass function at the high-mass end \citep[see, i.e., Fig.~34 in][and their discussion]{Behroozi2019}. To illustrate this uncertainty, Fig.~\ref{fig:retentionfactor_fj} also shows the $f_j(M_*)$ relations for the light-weighted $j_\mathrm{t,light}$ of the elliptical galaxies using the $f_M$ models from \cite{Dutton2010_SMHM}, \cite{Behroozi2013_SMHM}, and \cite{Behroozi2019_SMHM} for quiescent galaxies. 
This means that the retention factor $f_j$ of ETGs sharply decreases with $M_*$ in tandem with the star formation efficiency; the magnitude of this decrease, however, is highly uncertain. The differences between $f_j(M_*)$ at $10^{10}<M_*<10^{10.5} \MSUN{}$ instead reflect the differences in the peak of the star formation efficiency between models. 

\citet{RomanowskyFall2012} also find values $f_j\sim0.1$ for the ellipticals, assuming the SMHM from \cite{Dutton2010_SMHM}. In this case, the higher sAM in their data is compensated by the lower star formation efficiency assumed. Assuming \cite{Dutton2010_SMHM}, we find $f_j\sim0.04$ for low mass ellipticals and $f_j\sim0.01$ at the high mass end.

It is interesting to compare the $f_j(M_*)$ relation of the ETGs with that of spiral galaxies. The recent work of \cite{2019A&A...629A..59P}, \cite{PostiFall2021}, and \cite{2023MNRAS.518.6340D} has shown that, contrary to ETGs, the star formation efficiency of massive spirals increases monotonically with $M_*$ until reaching $f_M\sim f_b$ above $M_*=10^{11}\MSUN{}$. This implies that the angular momentum retention factor of spirals depends weakly on $M_*$ and it is close to $f_j\sim0.8-1$. Hence, the stellar component of these galaxies is consistent with the approximate conservation of the primordial sAM, in stark contrast with the behavior of ETGs which show declining $f_j$ with $M_*$ and $f_j<<1$.

The difference between ETGs and spirals can be explained by the different formation pathways that characterise these systems. On one side, massive spirals live in low mass dark matter halos \citep[e.g.,][]{PostiFall2021}. They follow a gradual evolution, with star formation activity sustained by smooth gas accretion and few minor mergers, and regulated by stellar winds, which promote the formation of high $j_*$ galaxies  \citep[e.g.,][]{Genel2015, Zavala2016, Rodriguez-Gomez2022}. 
On the other, massive ellipticals inhabit more massive structures. They form most of their stars from low angular momentum gas at early times while their strong AGN feedback inhibits the accretion of high $j$ gas at later times, determining the formation of low $j_*$ galaxies \citep{Genel2015,Zavala2016,Lagos2017, Rodriguez-Gomez2022}. In addition, ETGs undergo a further loss of $j_*$ through their merger activity \citep{Lagos2018_MergersAM}, which redistribute a fraction of their $j_*$ to the dark matter halo by dynamical friction \citep[e.g.,][]{1988ApJ...331..699B}. Since more massive ETGs host more massive supermassive black holes \citep{KormendyHo2013_BHandMorph} and undergo a larger number of massive, gas poor, mergers than low mass ETGs \citep{Rodriguez-Gomez2016,Rodriguez-Gomez2017, Pulsoni2021}, they are likely to "retain" a lower fraction of their primordial $j_*$ compared to lower mass ETGs, consistent with the declining trend of $f_j$ with $M_*$.

In addition, the fact that massive ETGs often live in massive environments further complicates the interpretation of our results and the connection between the properties of the stellar and the dark matter component. While the SMHM relation associates halo masses of $M_h\sim10^{15}\MSUN{}$ (approximately the mass of the Coma Cluster) to galaxies of $M_*\sim10^{11.5}\MSUN{}$, the $j_*$ measured in this work does certainly not integrate over the whole stellar component associated to the (cluster) dark matter halo. For example, our $j_*$ measurement does not include the contribution from satellite galaxies in the halo, which might contain a large fraction of the total $j_*$. Indeed, the fraction of the cluster stellar mass locked in the satellite galaxies can range from 20 to 90\% \citep{2022NatAs...6..308M}. Therefore, even though the measured $j_*$ converge within a number of effective radii to the total $j_*$ within the galaxies, it may not be representative of the total sAM of the whole stellar component contained in the halo.


\section{Conclusions}\label{sec:conclusions}

In this paper, we measure local $\lambda(a)$ and projected specific angular momentum (sAM) profiles in apertures, $j_p(\lec a)$, for a sample of 32 nearby early type galaxies (ETGs) from the ePN.S survey. These galaxies have stellar masses $M_*$ in the range $10^{10.1}- 10^{11.9} \MSUN{}$ and two-dimensional kinematic data out to a mean radius of $6$ effective radii ($6 R_e$). We use planetary nebulae as kinematic tracers of the stellar halos and IFS data for the central regions. This increases the number of ETGs for which the sAM can be estimated out to large radii by a factor of four over previous work. In addition, the full two-dimensional kinematic information allows us to do so without  assumptions on the geometry of the velocity fields and to include the contribution from minor axis rotation. 

In order to reconstruct the total sAM $j_*$ from the measured projected $j_p$, we use simulated ETGs from the IllustrisTNG simulation TNG100.
The TNG100 ETGs have velocity fields and projected ellipticity profiles qualitatively similar to the observed ETGs \citep{Pulsoni2020}. Simulated "ePN.S analogs" which are chosen to have similar ellipticity, rotational support, and stellar mass as the ePN.S ETGs, also have a similar distribution of $j_p(\lesssim 6R_e)$ as the ePN.S ETGs (see Fig.~\ref{fig:jp_ePNSvsTNG}). Our main results are:

(i) The kinematic diversity of the ETG stellar halos already explored in \cite{Pulsoni2018} is also visible, and quantified, through their large variety of $\lambda(a)$ and $j_p (\lec a)$ profiles (Figs.~\ref{fig:lambda_profiles} and \ref{fig:jp_profiles}). FRs can have rapid rotation at large radii or more or less steeply decreasing $\lambda(a)$. Elliptical FRs have reduced halo rotational support compared to S0s, which often contain extended disks or rapidly rotating stellar halos. 
SRs generally have increasing $\lambda$ with radius but the rotational support at large radii is still modest.
The contribution to $j_p$ from off-major-axis rotation increases with radius and stellar mass 
(Fig.~\ref{fig:Contribution_jxjy}), with the largest values ($\gtrsim60\%$) in the halos of the SRs.

(ii) The large radial extent of the PN data show that ETGs do not contain large mass fractions of high $j_*$ in their stellar halos. For the FRs, aperture $j_p(\lec a)$ profiles are approximately converged within the radial range of ePN.S data (typically $6R_e$), but this is not the case for SRs (Fig.~\ref{fig:jp_profiles}). We therefore use the simulated ETGs for estimating the total $j_*$ for the ePN.S galaxy sample from their measured $j_p(\lesssim 6R_e)$. In the TNG100 FRs $j_p(\lesssim 6R_e)$ has likewise essentially converged, but for the SRs $j_p$ increases by up to 40\% beyond 6 Re (Fig.~\ref{fig:jp_outside6Re}). For the projection factor $C_i = j_t/j_p$, relating the (galaxy integrated) projected $j_p$ to the three dimensional sAM $j_t$, we also use the TNG100 ETGs. We find that $C_i$ can be inferred from the observed projected ellipticity and rotational support with little scatter except for nearly face-on FRs (Fig.~\ref{fig:Cj_simulations}). The reconstructed $j_t$ values are our estimates of the total stellar sAM $j_*$ of the ePN.S ETGs.

(iii) We find that $j_p$ and $j_*$ increase with $M_*$ and confirm the well-known dependence of $j_*$ on morphology (Figs.~\ref{fig:jpM_diagram} and \ref{fig:jt_M*_epns}): at fixed $M_*$, ellipticals have lower sAM than S0s, which have lower sAM than spirals. However, we do not detect a significant difference between FRs and SRs: SRs lie at the high-$M_*$, high-$j_*$ end of the increasing trend with mass.

(iv) The total mass-weighted $j_p$ (and $j_*$) for ETGs are 0.07 dex lower than light-weighted sAM, due to colour gradients. Correcting for IMF gradients using results from ETG stellar population studies in the literature further reduces values by 0.2 dex relative to the light-weighted case, due to the additional stellar mass in the central regions. This mostly reduces the normalisation of the $j_*-M_*$ relation.

(v) A power-law fit to the distribution of ellipticals and S0s in the mass-weighted $j_*-M_*$ diagram gives a slope $0.55 \pm 0.17$ for the S0s and  $0.76\pm0.23$ for the ellipticals. Remarkably, like the dark matter halos, the most massive ETGs have the largest sAM. However, the normalisation is systematically lower: the ePN.S ellipticals have a factor two lower sAM than S0s and nine times lower than spirals. The gap between ellipticals and spirals increases up to a factor of 13 in the IMF-corrected case. The scatter increases from spirals and S0s to ellipticals, where it reaches 0.3dex in the mass-weighted and 0.37dex in the IMF-corrected case. This likely reflects the varied formation histories of ETGs as well as the scatter in the halo spin and in the stellar-mass-halo-mass relation.

(vi) In the theoretical framework in which stars inherit a fraction $f_j$ of the primordial sAM of their host halo, $j_h$, we estimate that at stellar masses in the range $10^{10} - 10^{10.5} M_\odot$ S0s retain $f_j\sim25\%$ of $j_h$, while ellipticals only $f_j<10\%$. For higher masses, this fraction strongly decreases with the stellar mass, by $\sim1.5$ orders of magnitudes at $M_*\sim 10^{12}\MSUN{}$ (Fig.~\ref{fig:retentionfactor_fj}). This strongly contrasts with the behavior of spirals which have $f_j\gtrsim0.8$ weakly dependent on $M_*>10^{10}\MSUN{}$ \citep[][]{2019A&A...629A..59P, 2023MNRAS.518.6340D}. 

Our results clearly show that the stellar sAM of ETGs increase with stellar mass roughly as $M_*^{2/3}$ but do not converge to those of the spirals in their stellar halos, confirming previous findings. 
However, the ratio of $j_*$ between ellipticals and spirals has increased with our analysis of the ePN.S sample from a factor of $\sim4.5$ to a factor of nine.
It is likely that, during their evolution, ETGs have lost a significant fraction of their sAM to the dark matter component through dynamical friction. In addition they may never have accreted some of the high $j$ baryons which could be retained in the baryonic component that did not condense into stars, such as the hot circumgalactic medium. Also, in the most massive systems, the high-$j_*$ stellar component associated with the dark matter halo are the satellite galaxies that have not yet merged with the central ETG. 
Therefore, while the decreasing trend of $f_j$ with $M_*$ is consistent with the increased gas-poor merger rate and AGN activity at higher masses, the actual magnitude of $f_j$ at $M_* > 10^{11}\MSUN{}$ is uncertain, due to the uncertainties in the star formation efficiency and in how representative the galaxy $j_*$ is of the sAM of the all the stars in the host halo.

\begin{acknowledgements}
  \\
  We thank the anonymous referee for their comments which improved the clarity of our manuscript.
  We thank C. Spiniello and T. Parikh for helpful discussions on IMF variations in ETGs. We are also grateful to R. Ragusa and M. Spavone for kindly providing us their photometric data in tabular form. C.P. is grateful to F. Hofmann for his support.
\end{acknowledgements}

\bibliographystyle{aa}
\bibliography{auto}

\include{table1}
\include{table2}

\begin{appendix}

\section{Kinematic profiles for the new MUSE velocity fields} \label{sec:MUSE_kin_profiles}

\begin{figure}
    \centering
    \includegraphics[width=4.2cm]{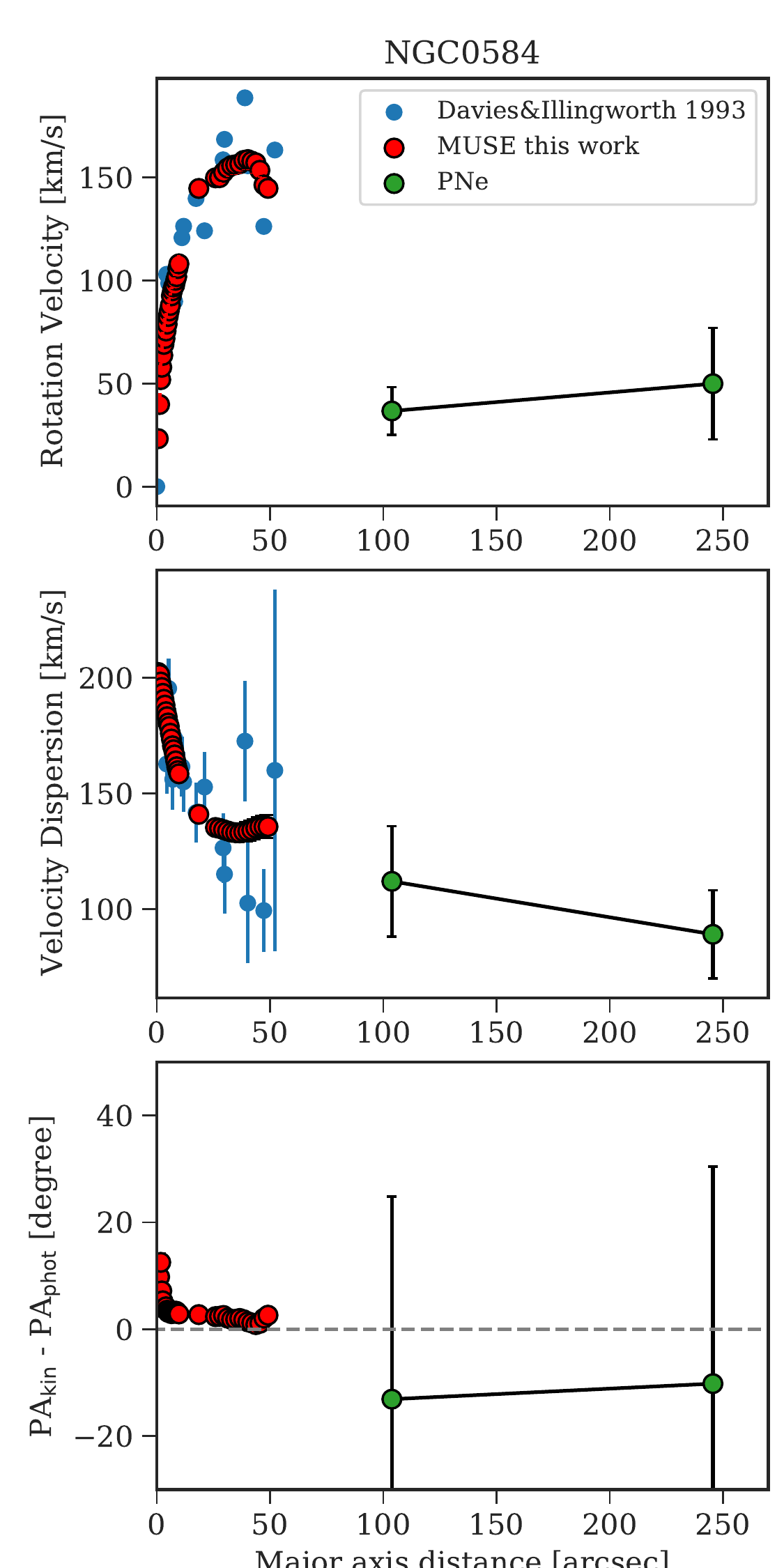}
    \includegraphics[width=4.2cm]{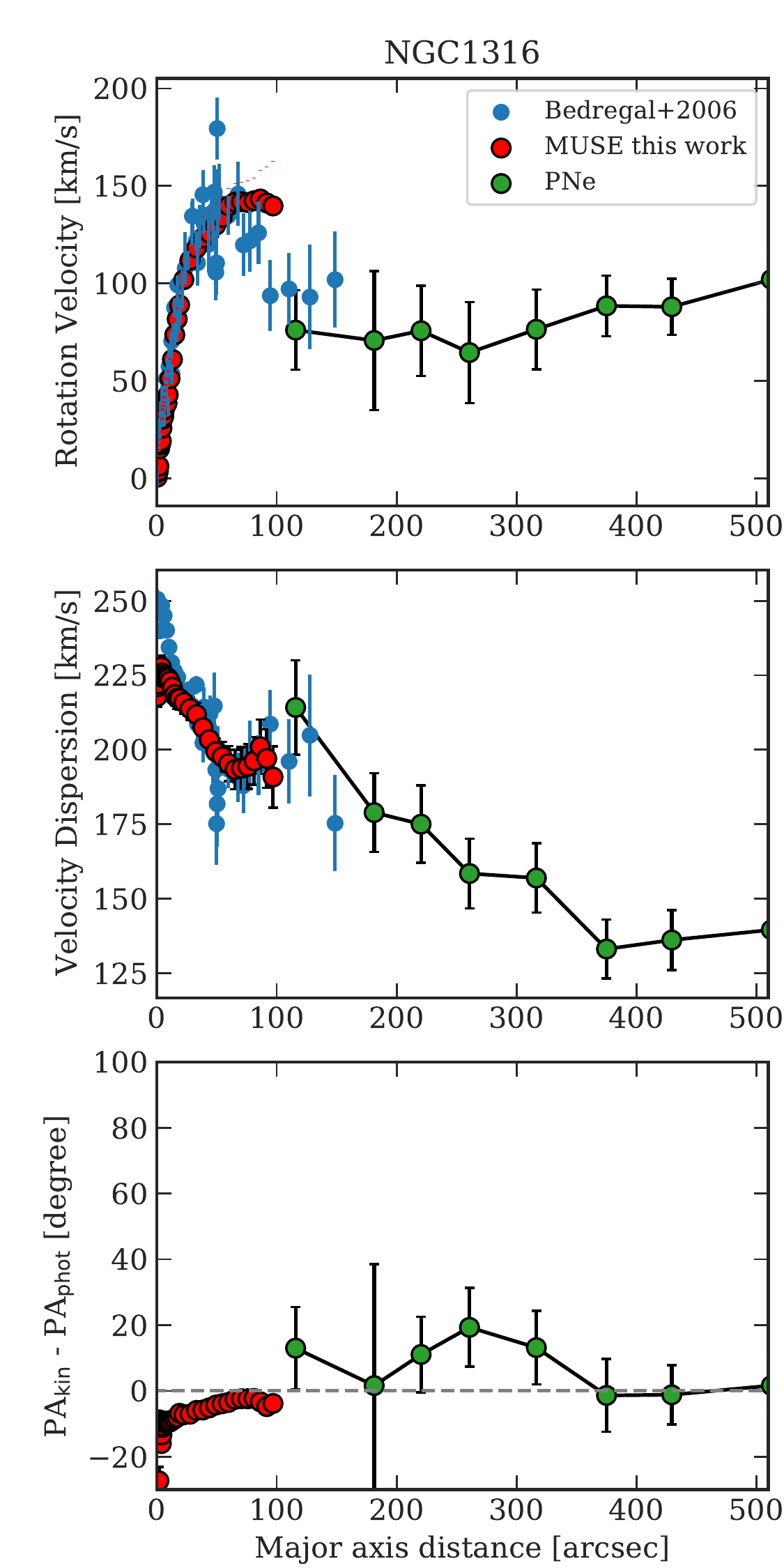}
    \caption{Kinematics profiles for the new MUSE velocity fields. The profiles from MUSE are shown with red symbols. Blue symbols show major axis slit data available in the literature \citep{1982ApJ...256..460K, 2000AJ....119..153S}; green circles show the PN kinematic profiles. }
    \label{fig:MUSE_kinematics1}
\end{figure}

In this section, we show the kinematic profiles extracted from the new MUSE velocity fields. The full 2D kinematic data will be made available in the future (Ennis et al. in preparation). For each galaxy, we show the major axis velocity profiles 
\begin{equation}
    V(a,\phi = PA_\mathrm{kin}) = V_\mathrm{rot}(a) + V_\mathrm{c3}(a),
\end{equation}
that is Eq.~\eqref{eq:harmonic_expansion} for $\phi = \PAkin$, the azimuthally averaged velocity dispersion profiles $\sigma(a)$ in elliptical annuli, and the misalignment between the kinematic position angle profile $\PAkin(a)$ and the mean photometric position angle $\langle PA_\mathrm{phot}\rangle$. We compare these kinematic profiles with slit data available in the literature and complement them with PN data at large radii. The different data sets agree well within the uncertainties. 

NGC4594 contains a thin, dusty disk seen close to edge-on. As already discussed in \cite{Pulsoni2018}, the disk dominates the major axis kinematics but not that of the PNe, which mainly follow the kinematics of the luminous bulge. This is demonstrated by the fact that the PN kinematics agrees well with the kinematic extracted from a slit aligned with the photometric major axis but offset by 30 arcsec, and therefore probing the bulge kinematics. The velocity dispersion profiles are azimuthally averaged, therefore MUSE data, PN, and offset slit are consistent with each other and systematically higher than the major axis velocity dispersion probing the disk.  

\begin{figure}
    \centering    \includegraphics[width=4.2cm]{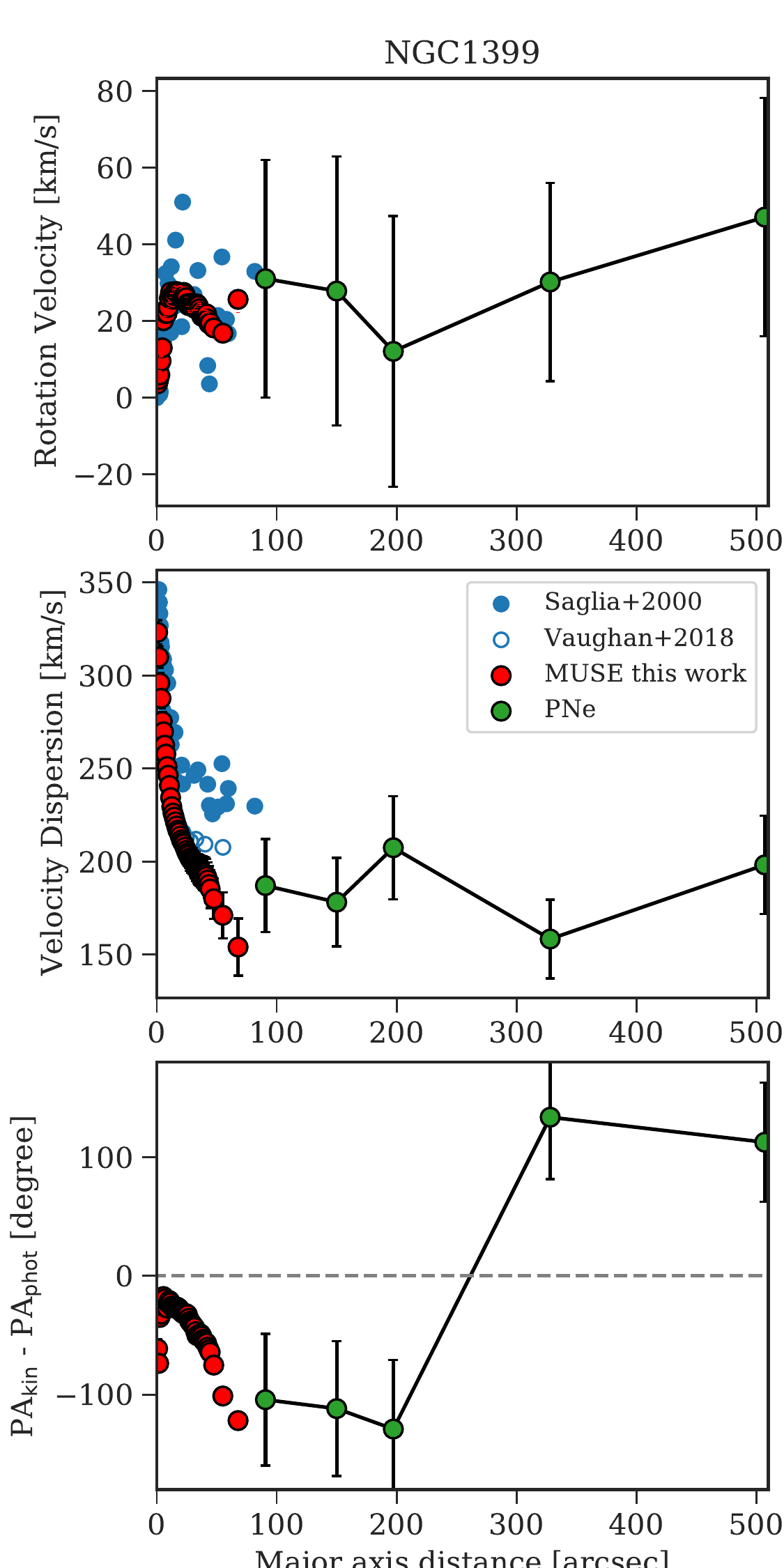}
    \includegraphics[width=4.2cm]{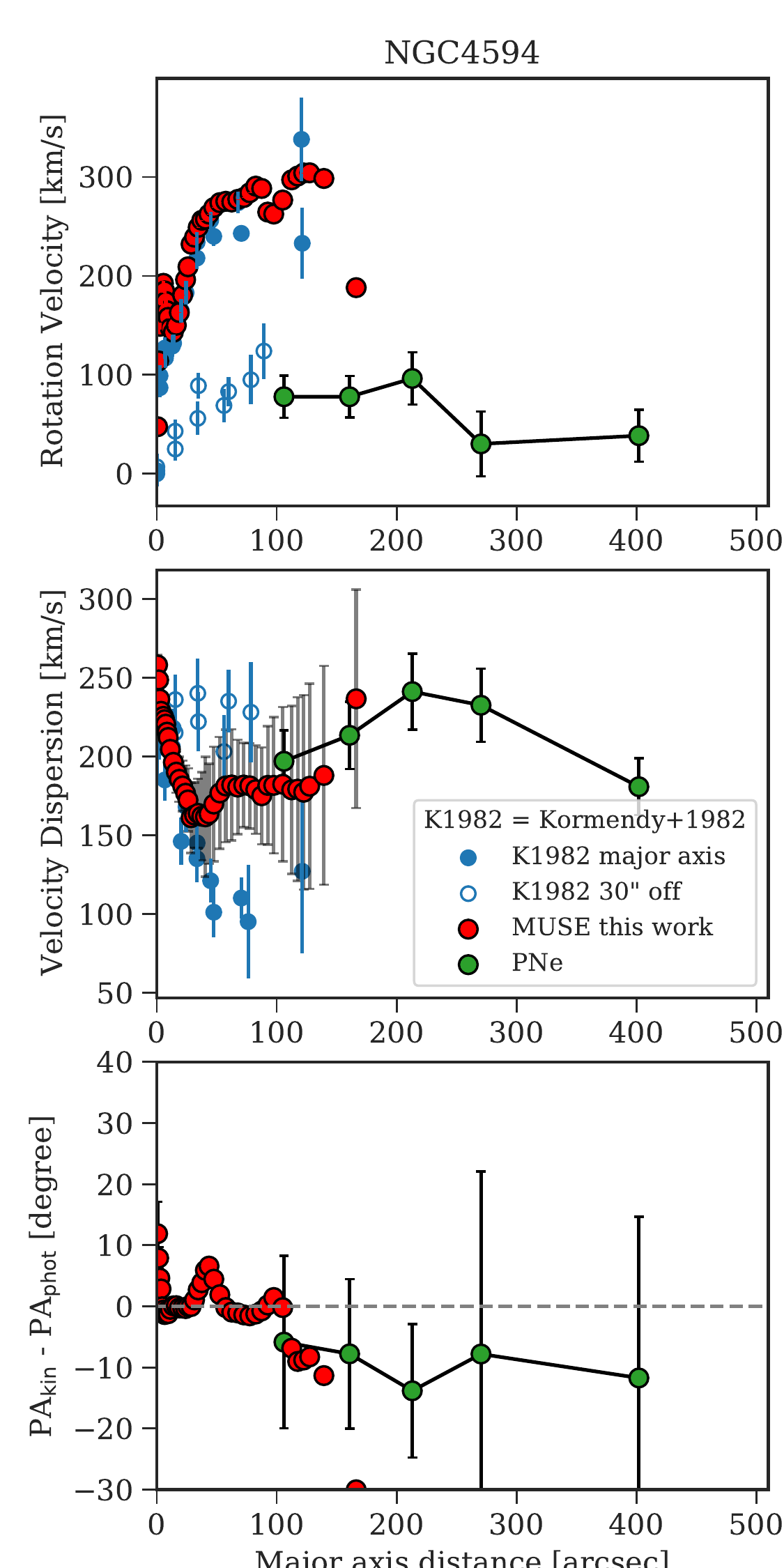}
    \caption{As in Fig.~\ref{fig:MUSE_kinematics1}. For NGC1399, we also show the velocity dispersion profile from \cite{2018MNRAS.479.2443V} based on different MUSE observations with open symbols.}
    \label{fig:MUSE_kinematics2}
\end{figure}

\section{The distribution of sAM in simulated TNG100 galaxies}\label{sec:TNG_AMradial_distribution}

In this section, we compare the sAM radial distribution of simulated TNG100 galaxies with observations. 
Previous works using IllustrisTNG galaxies found that their $j_t-M_*$ relation as well as its dependence on morphology agrees well with the observations \citep{2023MNRAS.518.6340D, Rodriguez-Gomez2022}. 

However, \cite{Pulsoni2020} found hints for a different radial distribution of angular momentum in the TNG100 galaxies compared to observations. The simulated galaxies in fact display shallower $V_{rot}/\sigma$ profiles with radius, which tend to peak at larger radii (at a median $3R_e$ compared to the median $1.3R_e$ in ePN.S), but comparable mass-size relation. This difference in the $V_{rot}/\sigma$ profile shapes could not be attributed to differences the sample selection or to resolution effects, as the effect was present also in galaxies from the higher resolution run of IllustrisTNG, TNG50.

\begin{figure}
    \centering
    \includegraphics[width=\linewidth]{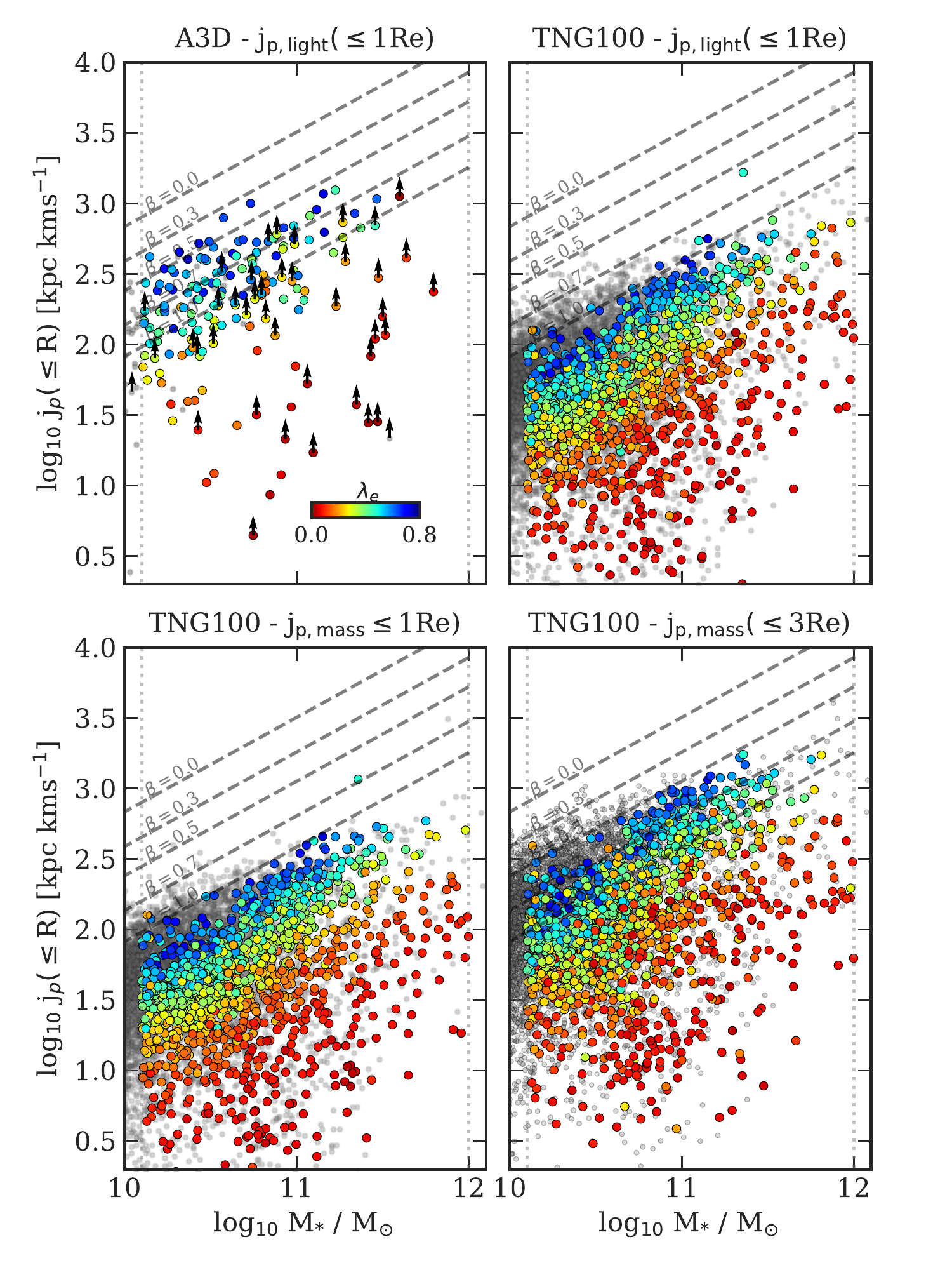}
    \caption{Projected sAM $j_p(\lec 1 R_e)$ in the TNG100 ETGs compared to Atlas3D galaxies. Colored symbols show the sample of ETGs, gray small symbols show all the other galaxies (namely the bluer and lower mass systems). The ETGs are color-coded according to the angular momentum parameter $\lambda_e$ integrated within $1R_e$. Arrows on the Atlas3D galaxies show values integrated out to $a<1R_e$, hence the show $j_p$ is likely a lower-limit estimate. Dashed black lines show the $j_t-M_*$ relations for different bulge fractions $\beta$ as derived by \cite{Fall2018}.
    The different $j_p-M_*$ relation between observed and simulated galaxies shows that the TNG100 ETGs have their angular momentum distributed at larger radii.}
    \label{fig:TNG1RE_A3D}
\end{figure}

Figure~\ref{fig:TNG1RE_A3D} compares the TNG100 ETGs with the Atlas3D ETGs, a survey that targets a volume and magnitude limited sample of 260 morphologically selected ETGs \citep{A3D_I_Cappellari2011}.
The Atlas3D ETGs have mostly red colors, nevertheless we apply the same color selection used for selecting the red sequence galaxies TNG100 to single out the few bluer objects ($g-r$ colors from the NASA-Sloan Atlas\footnote{http://www.nsatlas.org} or, when unavailable, $B-$ colors from the Hyperleda\footnote{http://www.leda.univ-lyon1.fr} catalog converted to $g-r$ using the relations in \citealt{Pulsoni2020}). The Atlas3D stellar masses are derived from the 2MASS K-band magnitudes as described in Sect.~\ref{sec:additional_data} for the ePN.S galaxies. Their projected angular momentum $j_p(\lec 1 R_e)$ is derived by applying Eq.~\eqref{eq:jp_def2} to the Atlas3D velocity fields and fluxes, integrated within elliptical apertures of semi-major axis $a = 1R_e$ (ellipticities and $R_e$ from \citealt{A3D_I_Cappellari2011} and \citealt{2011MNRAS.414..888E}).

The $j_p(\lec 1 R_e)-M_*$ diagram for the Atlas3D ETGs is shown in the top left panel of Fig.~\ref{fig:TNG1RE_A3D}. This is compared with the $j_p(\lec1 R_e)-M_*$ relation for the TNG100 ETGs observed along the z-axis of the simulation box: we show results for both $j_p(\lec1 R_e)$ weighted with fluxes (top right panel) and weighted with stellar masses (bottom left panel). We also include with small gray symbols all the other TNG100 galaxies at $z=0$ that are not part of the selected ETG sample.
The location of the TNG100 ETGs appears shifted towards lower $j_p(\lec 1 R_e)$ by more than 0.5 dex compared to Atlas3D in both light- and mass-weighted cases. 
Even the blue star-forming systems in TNG (in gray) barely reach the $j_p(\lec 1 R_e)$ values of the Atlas3D ETGs. 
This figure demonstrates that the difference between simulated and observed ETGs does not originate from the sample selection: galaxies with $j_p(\leq1R_e)$ as high as in Atlas3D are just not produced by the simulation.

The bottom right panel shows that we need to integrate $j_p$ out to $\sim 3R_e$ to move the TNG100 ETGs roughly into the same location as the Atlas3D galaxies. 
This result shows that even though the simulated TNG100 ETGs have total sAM converging to the observed values \citep[][]{2023MNRAS.518.6340D, Rodriguez-Gomez2022} and aperture values at $6R_e$ consistent with the ePN.S galaxies (Fig.~\ref{fig:jp_ePNSvsTNG}), their $j_p$ profiles increase more slowly with radius compared to observations in the central regions. In \citealt{Pulsoni2020}, we suggested that the shallower $j_p(\leq a)$ profiles in the TNG galaxies could be due to a too efficient conversion of the gas into stars that does not allow the gas to collapse to small enough radii before forming stars. The investigation of origin of this discrepancy is beyond the scope of this paper. 

\section{Comparison of stellar masses and specific angular momenta with previous work}
\label{sec:appc}

\begin{figure}[h]
    \centering
    \includegraphics[width=.8\linewidth]{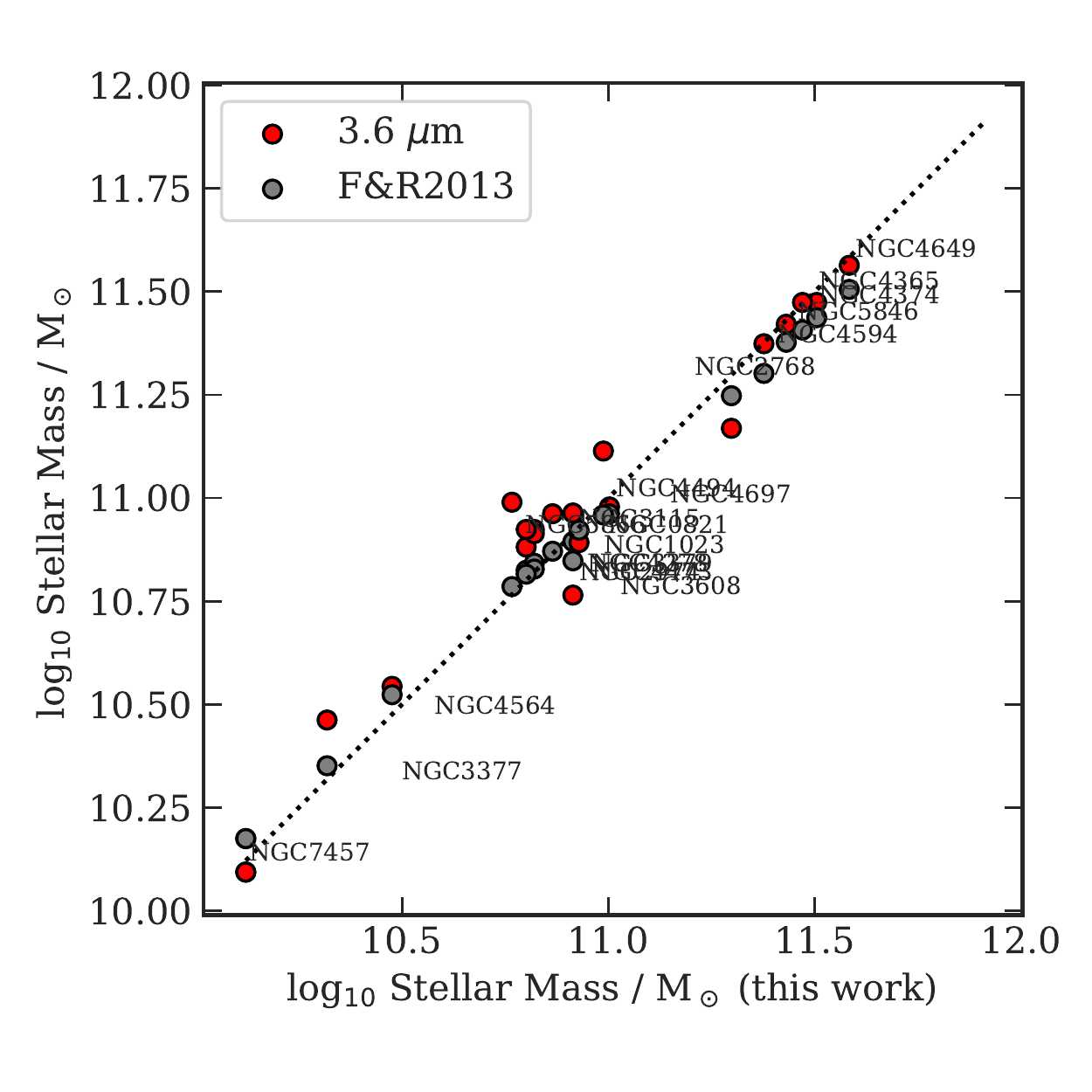}
    \caption{Comparison between the stellar masses adopted in this work, those determined by \cite{Fall2013}, and the stellar masses derived by \cite{Forbes2017_Spitzer} based on Spitzer 3.6 $\mu$m photometry. The measurements shown assume a constant Chabrier IMF.} 
    \label{fig:ComparisonStellarMasses}
\end{figure}

\begin{figure}[h]
    \centering
    \includegraphics[width=.8\linewidth]{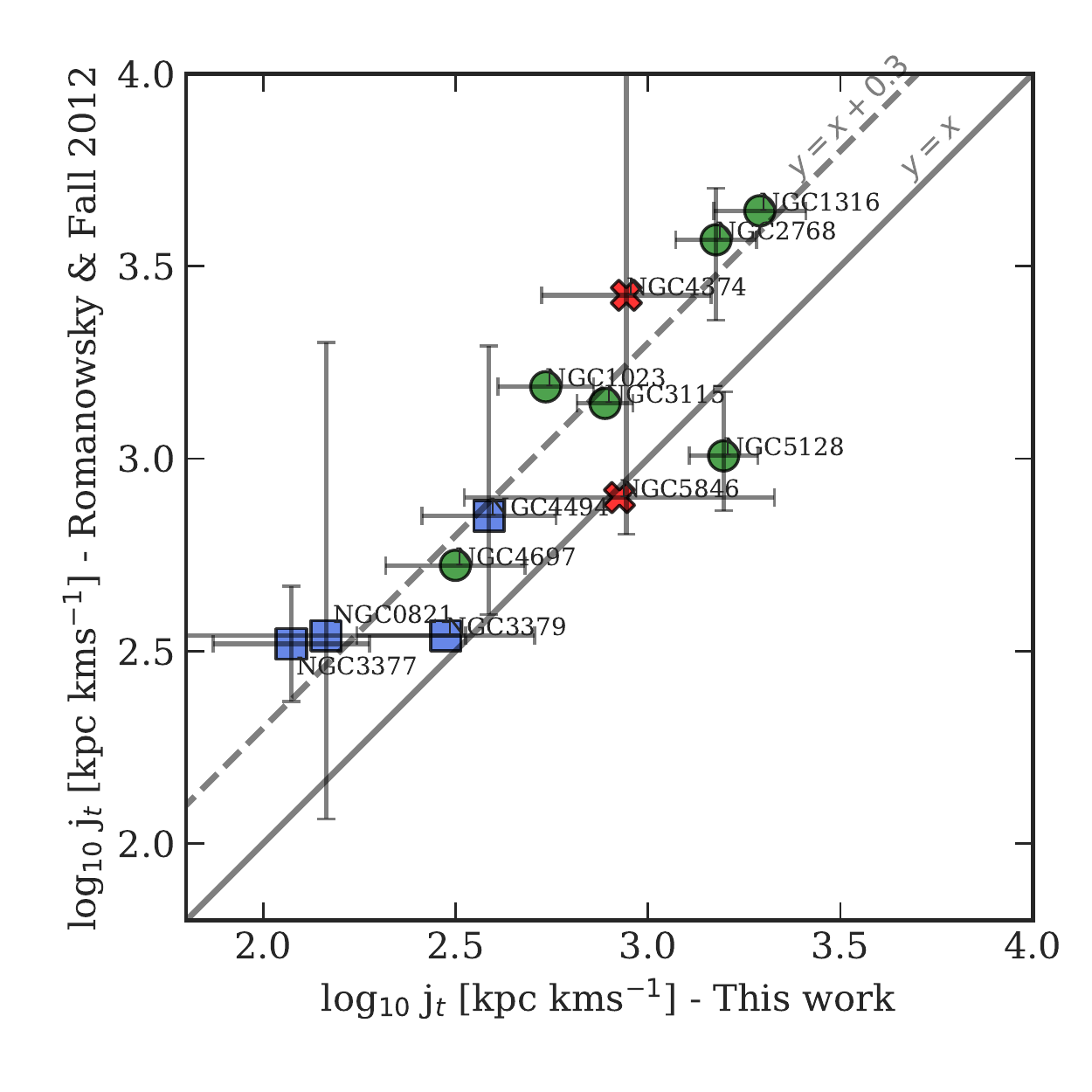}
    \caption{Galaxy-by-galaxy comparison of $j_t$ determinations of this work and \cite{RomanowskyFall2012} for the subset of ePN.S galaxies in common. Galaxies without error-bars on the y-axis are those for which \cite{RomanowskyFall2012} used the approximation $j_t \propto R_e V_{rot}(2R_e)$, see their Eq.~6.
    The discrepancies between measures can be partially explained by the assumption of cylindrical velocity fields in \cite{RomanowskyFall2012}, which lead to an overestimate of $j_t$. 
    Symbols as in Fig.~\ref{fig:jt_M*_epns}. }
    \label{fig:comparisonRF12}
\end{figure}

Figure~\ref{fig:ComparisonStellarMasses} compares different stellar mass determinations, assuming a universal Chabrier IMF. The stellar masses used in this work are based on magnitudes in the K-band and mass-to-light ratio $M_*/L_K$ from stellar population modelling. \cite{Fall2013} use a $M_*/L_K$ ratio based on color, also calibrated on stellar population models. The galaxy-by-galaxy agreement is excellent at intermediate masses, however at high and low masses the values of \cite{Fall2013} are slightly off as their $M_*/L_K$ - $M_K$ relation is slightly less steep. Figure~\ref{fig:ComparisonStellarMasses} also shows the comparison with the stellar masses from \cite{Forbes2017_Spitzer} using IR observations and $M_*/L_{3.6\mu m}$ based on age. In this case, the comparison with our determinations does not highlight any systematic effect. The mean variation between $M_*$ values from the three methods for the same galaxies is $0.05$ dex.

Figure~\ref{fig:comparisonRF12} compares the total sAM derived in this work (the light weighted $j_{t, \mathrm{light}}$) with the values from \cite{RomanowskyFall2012} for the same galaxies. 
The previous $j_t$ measurements are systematically larger by a median 0.3 dex (a factor 2).  This discrepancy can be at least partially explained by their assumption of cylindrical velocity field in galaxies with rotation concentrated along the major axis, overestimating $j_t$.

To elucidate this argument, we consider the two E-FRs NGC0821 and NGC3377, for which \cite{RomanowskyFall2012} used the same PN data set to trace the halo kinematics. We checked that the major-axis $V_{rot}$ profiles extracted here from the 2D stellar and PN kinematic data are in fact consistent with the major-axis profiles published in their Figs. 28 and 30. Yet the total sAM $j_t$ measured by \cite{RomanowskyFall2012} for these galaxies are more than a factor of two larger than the values obtained in the current analysis. 

The reconstruction of $j_t$ from the observable sAM involves 
additionally (i) an extrapolation from the radial coverage of the data to the whole extent of the galaxy, and (ii) a deprojection to the true three-dimensional $j_t$.
In the case of these two relatively low-mass E-FRs, (i) is not critical. The correction from the measured to the galaxy-integrated $j_p$ estimated from the TNG simulations is of order $10-20\%$ (see Fig.~\ref{fig:jp_outside6Re}), and the $j_p$ profiles are essentially converged within the radial coverage of the PNe (see Fig.~\ref{fig:jp_profiles}), as is also shown by the profiles in Figs. 28 and 30 of \cite{RomanowskyFall2012}. 
This means that the dominant difference in $j_t$ between the two studies must come from the different definitions of $j_p$ and the correspondingly different deprojection factors $C_i$.

The direct comparison of the $j_p$ values between the two studies is non-trivial because of their different definitions. In this work, $j_p$ is the projected sAM defined in Eq.~\ref{eq:jp_def1}. $C_i$ is defined to be the ratio between the 2D and the 3D sAM (Eq.~\ref{eq:Cj}) and is adjusted to each galaxy based on its flattening and rotational support (Sect.~\ref{sec:Ci_lambda_epsilon}). In \cite{RomanowskyFall2012}, $j^{RF}_p$ is a quantity derived from the major-axis kinematics which, if the galaxy has cylindrical rotation, is readily connectable to $j_t$ through its own "deprojection factor" $C_i^{RF}$. They use an inclination averaged value which accounting for inclination bias in the morphological classification of galaxies, and galaxies classified as ellipticals are assigned a uniform $C_i^{RF}$ corresponding to a mean inclination of 0.72 radians, i.e., about 40 degrees. 

Using only the major-axis rotation of NGC0821 and NGC3377, and deriving $j_p$ via Eq.~\ref{eq:jp_def1} assuming no dependence of $V_{rot}$ on the distance from the major axis $y_n$, as for an edge-on cylindrical rotator\footnote{This assumption is strictly true only for an edge-on cylindrical rotator. For NGC0821 and NGC3377, the Atlas3D data show velocity fields with rather high ellipticity, revealing that these galaxies are very inclined, but also that they are not cylindrical rotators.}, would increase the obtained $j_p$ by a factor 1.3 and 1.9, respectively. Adopting a deprojection factor $C_i$ for an average inclination of 40 degrees, gives a larger correction than the flattening of these galaxies would imply (a factor 3.5 rather than 2.4, see also Fig.~\ref{fig:Cj_inclinations}). This returns $j_t$ values of 281 kpc km/s for NGC0821 and of 378 kpc km/s for NGC3377, closer to the measurements of \cite{RomanowskyFall2012} which are $346^{+264}_{-165}$ kpc km/s for NGC0821 and $330\pm50$ kpc km/s for NGC3377.
This experiment shows that the cylindrical rotation assumption naturally leads to an overestimate of $j_t$ and can, at least in part, explain the differences seen in Fig.~\ref{fig:comparisonRF12}. Additionally the  inclination averaged $C_i^{RF}$ may amplify this for galaxies with inclined disk components.

Returning to Fig.~\ref{fig:comparisonRF12}, a notable exception is the merger NGC5128, for which we measure a higher $j_t$ compared to \cite{RomanowskyFall2012}. This galaxy has a large contribution to $j_t$ coming from minor axis rotation that is not accounted for in their major-axis-based measurements.

\section{Specific angular momentum measured from velocity fields versus particles}

\begin{figure}[h]
    \centering
    \includegraphics[width=0.8\linewidth]{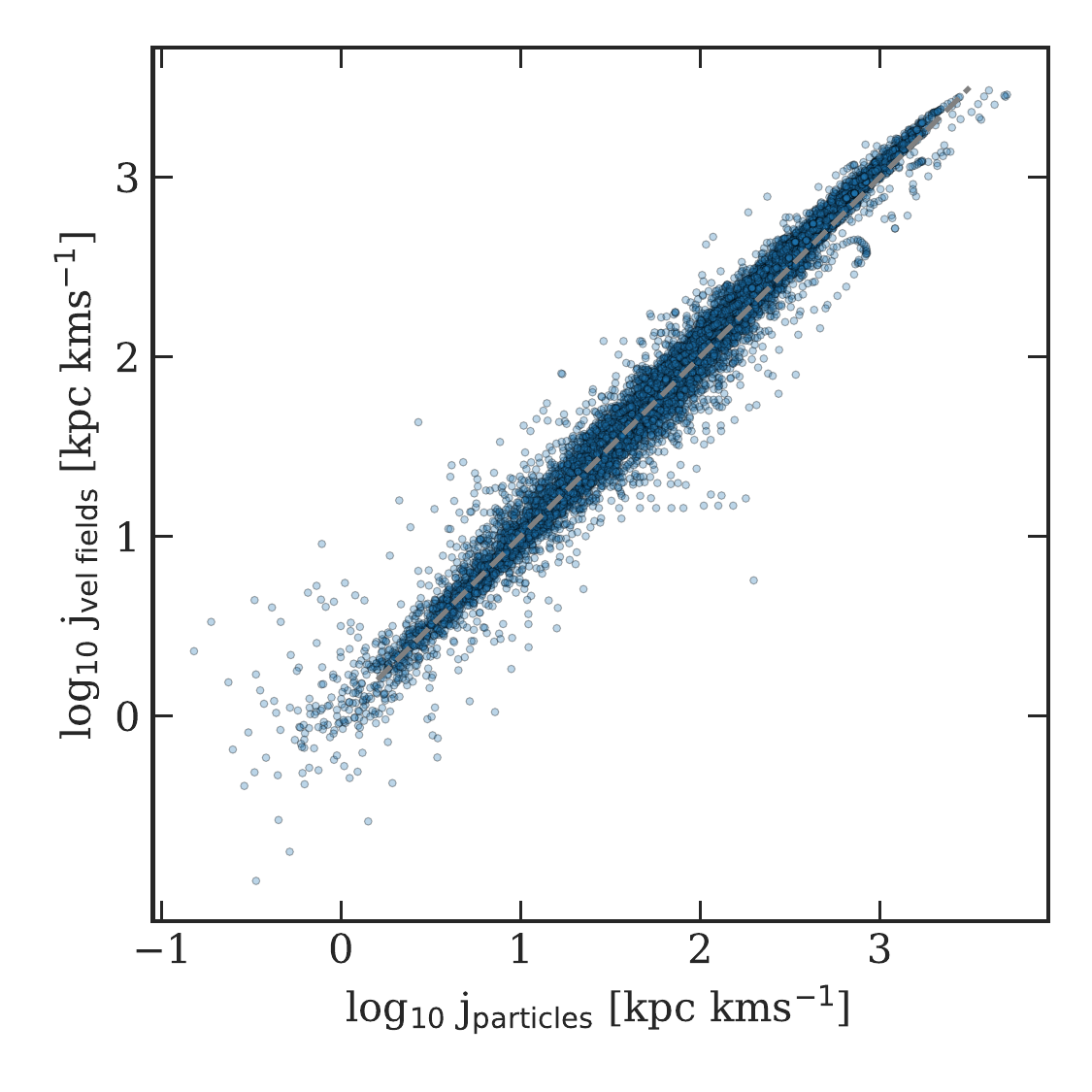}
    \caption{The sAM $j_p$ derived from smoothed velocity fields and compared with $j_p$ from discrete particles in simulated TNG100 ETGs at each radius.}
    \label{fig:TNG_jvfields_jparticles}
\end{figure}

\begin{figure}[h]
    \centering
    \includegraphics[width=.8\linewidth]{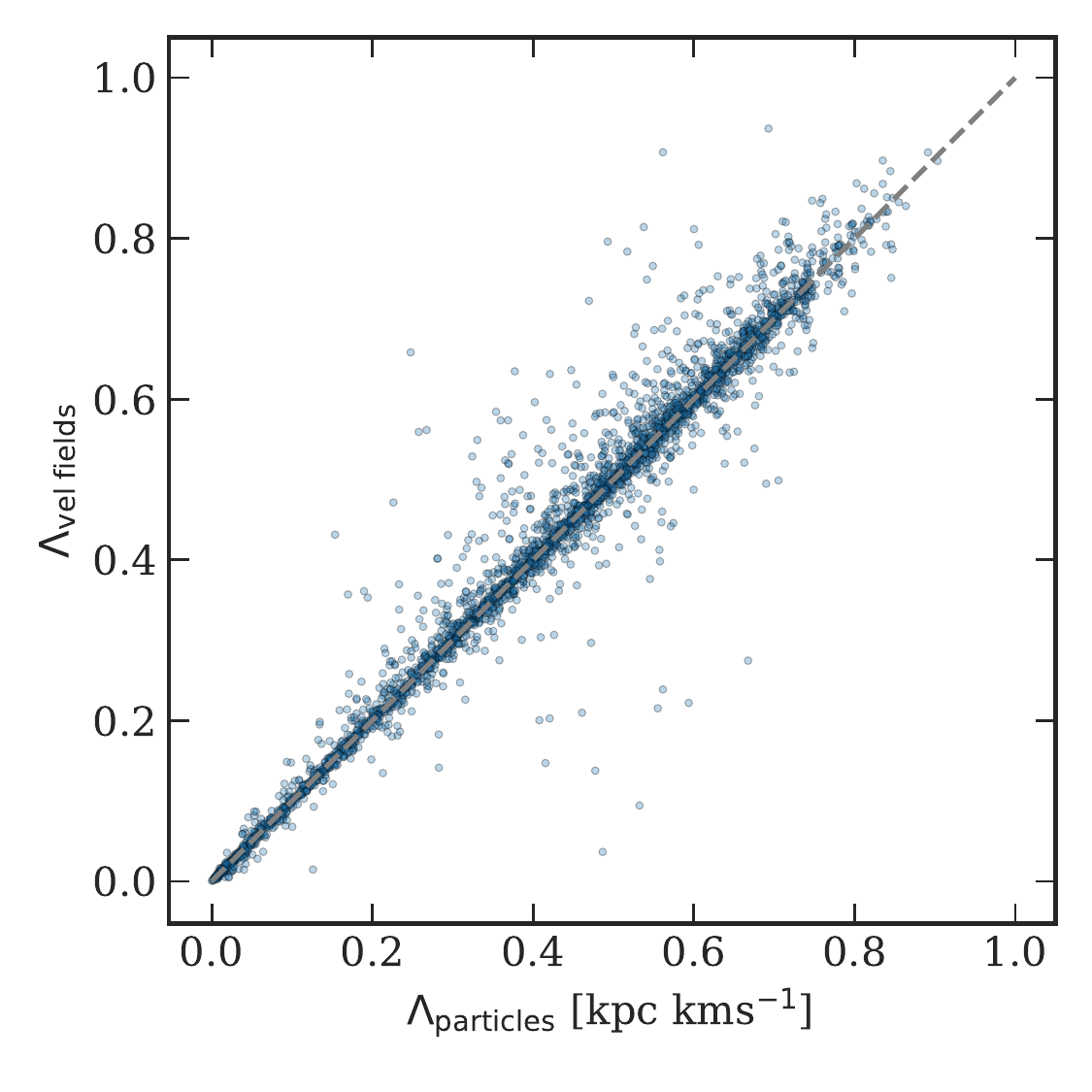}
    \caption{The parameter $\Lambda$ derived from smoothed velocity fields and compared with $\Lambda$ from discrete particles in simulated TNG100 ETGs at each radius.}
    \label{fig:TNG_Lambdavfields_Lambdaparticles}
\end{figure}

Figures~\ref{fig:TNG_jvfields_jparticles} and \ref{fig:TNG_Lambdavfields_Lambdaparticles} compare measurements of $j_p$ and $\Lambda$ in simulated TNG100 ETGs from projected mean velocity fields and from discrete particles. The results are shown for the 1327 selected ETGs projected along the z-axis of the simulation box. Each data-point in the figures is a local measurement within a galaxy. 

The good agreement between the two methods allows us to derive $j_p$ and $\Lambda$ directly from particles without deriving the mean velocity fields, and quickly extend the calculation to $1327\times100$ random line-of-sight projections. It also implies that we can consistently derive $\Lambda$ from the ePN.S mean velocity fields. 

\begin{figure}
    \centering
    \includegraphics[width=\linewidth]{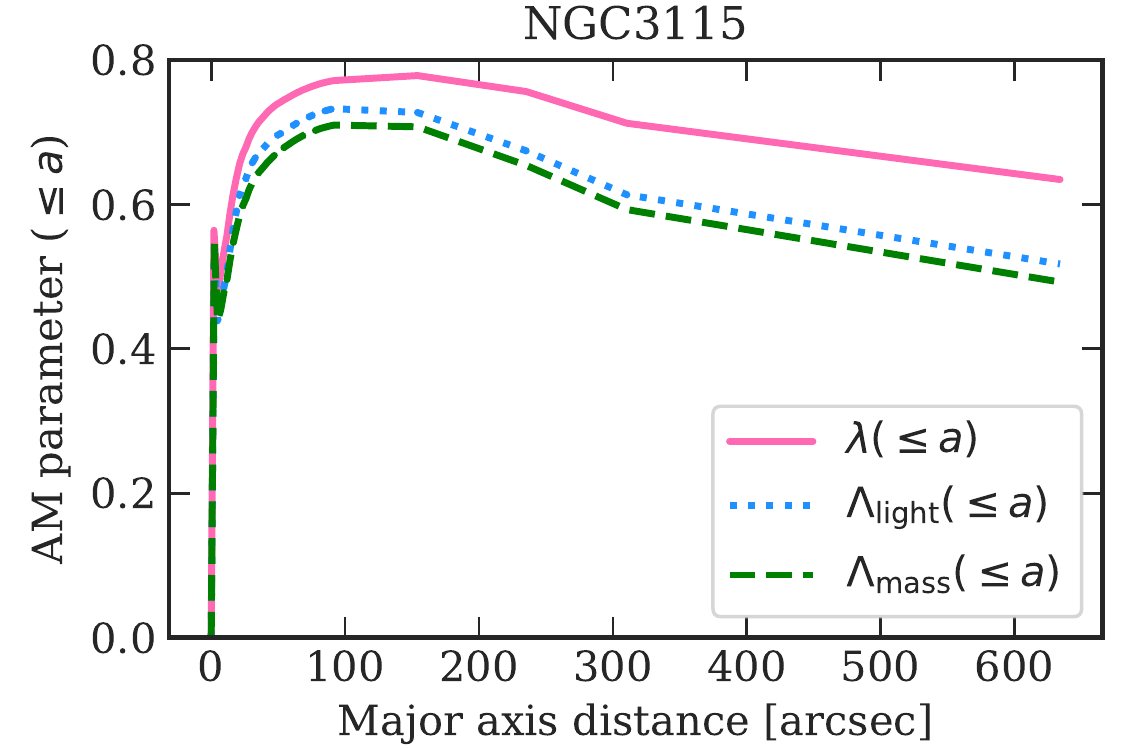}
    \caption{Angular momentum parameter profiles in NGC3115 quantifying its rotational support. Different lines show the aperture $\lambda(\lec a)$ profile, the light-weighted $\Lambda_\mathrm{light}(\lec a)$ profile, and the mass-weighted $\Lambda_\mathrm{mass}(\lec a)$ profile.}
    \label{fig:lambda_profiles_example_galaxy}
\end{figure}

Figure~\ref{fig:lambda_profiles_example_galaxy} shows the $\Lambda(\lec a)$ profile of an example galaxy, NGC3115. We show the comparison between light-weighted and mass-weighted $\Lambda(\lec a)$, derived using the blue and IR photometry respectively, and the comparison with the traditional $\lambda$ parameter defined by \citet{2007MNRAS.379..401E}.

  \end{appendix}

\end{document}

%% file: table1.tex
\begin{table*}
\caption[]{Summary and references of the kinematic and photometric data used}
\begin{center}
\begin{tabular}{llllll}
\hline\hline\noalign{\smallskip}
  \multicolumn{1}{l}{Galaxy} &
  \multicolumn{1}{l}{Type} &
  \multicolumn{1}{l}{Kinematics} &
  \multicolumn{1}{l}{Kinematics} &
  \multicolumn{1}{l}{Photometric} &
  \multicolumn{1}{l}{Spitzer}\\
  \multicolumn{1}{l}{NGC} &   
  \multicolumn{1}{l}{} & 
  \multicolumn{1}{l}{center} &  
  \multicolumn{1}{l}{in gap \tablefootmark{(a)}} &   
  \multicolumn{1}{l}{data  \tablefootmark{(d)}} &  
  \multicolumn{1}{l}{data  \tablefootmark{(e)}} \\
\noalign{\smallskip}\hline\noalign{\smallskip}

0584  & E-FR        &   MUSE (new) 			&    interp.    	&	(4)	 				&   0	\\
0821  & E-FR        &   Atlas3D    			&     SLUGGS    	&	(1);(2)				&   1	\\
1023  & S0.         &   Atlas3D      		&     SLUGGS    	&	 (6)   				&   1	\\
1316  & S0/merger   &   MUSE (new) 	 	 	&    interp.    	&	 (4);(13)  			&   0	\\
1399  & SR			&   MUSE (new) 			&     no gap    	&	  (4);(23) 			&	0	\\
2768  & S0			&   Atlas3D  			&     SLUGGS 		&	  (15);(19) 		&   1	\\
2974  & S0 &   Atlas3D   					&     SLUGGS 		&	  (4)  				&	1	\\
3115  & S0 &   MUSE (Gu{\'e}rou+2016)       &     SLUGGS 		&	  (4)   				&   1	\\
3377  & E-FR &   Atlas3D  				    &     SLUGGS    	&	 (1);(2);(3);(19)				&   1	\\
3379  & E-FR &   Atlas3D 				    &     no gap    	&	  (8);(9) 			    &   1	\\
3384  & S0 &   Atlas3D					    &     interp.   	&	  (9);(16);(20)  				&	0	\\
3489  & S0 &   Atlas3D					    &     interp.   	&	  (16)  				&   0	\\
3608  & SR &   Atlas3D                      &     SLUGGS    	&	 (2);(3) 				&   1	\\
3923  & SR &   PN only\tablefootmark{(b)}   &               	&	  (4);(18)  			&	0	\\
4278  & E-FR &   Atlas3D   					&     SLUGGS    	&	 (12);(19)  			&   1	\\
4339  & S0 &   Atlas3D						&     interp.   	&	 (13);(19)  			&	0	\\
4365  & SR &   Atlas3D						&     SLUGGS   		&	 (14)  				&	1	\\
4374  & SR &   Atlas3D						&     SLUGGS  		&	 (14)  				&   1	\\
4472  & SR &   Atlas3D						&     no interp.	&	 (14) 				&   0	\\
4473  & E-FR &   Atlas3D 					&     SLUGGS    	&	 (14);(17)  				&   1	\\
4494  & E-FR &   Atlas3D 					&     SLUGGS    	&	 (10)  				&   1	\\
4552  & SR &   Atlas3D   					&     no interp.	&	 (14) 				&   0	\\
4564  & E-FR &   Atlas3D 					&     SLUGGS    	&	 (14);(1)  			&   1	\\
4594  & S0 &   MUSE (new)					&     interp.   	&	  (4);(21)  		&   1	\\
4636  & SR &   Atlas3D  					&     no interp.	&	  (14) 				&	0	\\
4649  & E-FR &   Atlas3D					&     SLUGGS    	&	  (14) 				&   1	\\
4697  & S0 &   Atlas3D  					&     SLUGGS    	&	 (2);(11) 			&   1	\\
4742  & E-FR &   slit\tablefootmark{(c)}  	&               	&	  (4);(22) 			&	0	\\
5128  & S0/merger &   PN only\tablefootmark{(b)} &          	&	  (4);(24)  		&	0	\\
5846  & SR &   Atlas3D						&      SLUGGS   	&	 (5);(18) 			&   1	\\
5866  & S0 &   Atlas3D						&      SLUGGS   	&	  (16);(19) 		&	1	\\
7457  & S0 &   Atlas3D						&      SLUGGS   	&	  (16); (19) 		&	1	\\

\noalign{\smallskip}\hline
\end{tabular}
\tablefoot{\tablefoottext{a}{The procedure to estimate the kinematics at intermediate radii, between the radial coverage of the IFS and of the ePN.S data. In galaxies with “no gap” the velocity field is the juxtaposition of the IFS and PN velocity and velocity dispersion fields. “Interp” indicates galaxies where the velocities at intermediate radii are estimated from Eq.~\eqref{eq:harmonic_expansion}, with parameters given by a linear interpolation of the parameter fitted on the IFS field in the center and on the ePN.S field at large radii. In galaxies with SLUGGS dats available, we use $V_\mathrm{rot}(a)$ and $PA_\mathrm{kin}(a)$ profiles from the kinemetry analysis of \cite{2016MNRAS.457..147F}, while $V_\mathrm{s3}(a)$, $V_\mathrm{c3}(a)$ are the linear interpolation between the values fitted on the IFS and the PN data. In the three SRs with "no interp.", the comparison with long slit velocities or kinemetric $V_\mathrm{rot}$ from IFS data (\citealt{1997A&AS..126...15S}, \citealt{2017MNRAS.464..356V}, \citealt{2011RAA....11..909P}) assures the smooth continuity between the kinematic profiles from Atlas3D and ePN.S: the central regions of these galaxies are non-rotating as well as the most central PNe, and any contribution to the AM comes from the outskirts of these galaxies. In these cases we simply consider the combination of the Atlas3D and ePN.S velocity fields without adding any interpolation.}\\
 \tablefoottext{b}{For NGC3923 and NGC5128, we only use PN data, as discussed in Sect.~\ref{sec:kinematic_data_center}.}\\
 \tablefoottext{c}{For NGC4742, only major axis slit data from are available from \cite{1983ApJ...266...41D}. We build a 2D velocity field in the center using Eq.~\eqref{eq:harmonic_expansion}, where $V_\mathrm{rot}(a)$ is given by the measured major axis velocities from the slit, the three parameters $V_\mathrm{s3}(a)$, $V_\mathrm{c3}(a)$, and $PA_\mathrm{kin}(a)$ are given between a linear interpolation between 0 at $a=0$ and the measured values for the PN velocity field. The 2D velocity dispersion field is instead build using the major-axis slit data and assuming azimuthally constant $\sigma$.}\\
 \tablefoottext{d}{References for the blue-light photometric data: (1) \citealt{1994A&AS..104..179G}, (2) \citealt{2005AJ....129.2138L}, (3) \citealt{1987MNRAS.226..747J}, (4) \citealt{2011ApJS..197...22L}, (5) \citealt{2000A&AS..144...53K}, (6) \citealt{2008MNRAS.384..943N}, (7) \citealt{2007A&A...467.1011S}, (8) \citealt{1990AJ.....99.1813C}, (9) \citealt{2022FrASS...952810R}, (10) \citealt{2009MNRAS.393..329N}, (11) \citealt{2008MNRAS.385.1729D}, (12) \citealt{1990AJ....100.1091P}, (13) \citealt{1994A&AS..106..199C}, (14) \citealt{2009ApJS..182..216K}, (15) \citealt{2009ApJS..181..135H}, (16) \citealt{2013MNRAS.432.1768K}, (17) \citealt{1990A&AS...86..429C}, (18) \citealt{2017A&A...603A..38S}, (19) \citealt{1993A&AS...98...29M}, (20) \citealt{2007AN....328..562M}, (21) \citealt{2012MNRAS.423..877G}, (22) \citealt{1995AJ....110.2622L}, (23) \citealt{2016ApJ...820...42I}, (24) \citealt{2022A&A...657A..41R}.}\\
 \tablefoottext{e}{Galaxies with available 3.6$\mu$m Spitzer data are marked with 1, 0 otherwise.}}
 
\end{center}
\label{tab:kinematics_AND_photometry_data} 
\end{table*}

%% file: table2.tex
\begin{table*}
\caption[]{Stellar masses, projected sAM integrated out to the radial coverage of the ePN.S data, and correction factors based on the TNG100 ETGs to derive the total sAM from the ePN.S measurements.}
\begin{center}
\begin{tabular}{ccccccccccccc}
\hline\hline\noalign{\smallskip}
  \multicolumn{1}{l}{Galaxy} &
  \multicolumn{1}{l}{log$_{10} M_\mathrm{*}/M_\odot$ } &
  \multicolumn{1}{l}{log$_{10} M_\mathrm{*}/M_\odot$ } &
  \multicolumn{1}{l}{$j_\mathrm{p,light}$} &   
  \multicolumn{1}{l}{$j_\mathrm{p,mass}$}&   
  \multicolumn{1}{l}{$j_\mathrm{p,mass+IMF}$}&   
  \multicolumn{1}{l}{median(corr)}& 
  \multicolumn{1}{l}{median(corr)} & 
  \multicolumn{1}{l}{NOTES} & 
  \\ 
  \multicolumn{1}{l}{} &   
  \multicolumn{1}{l}{Chabrier} &   
  \multicolumn{1}{l}{var IMF} &   
  \multicolumn{1}{l}{kpc km/s} &   
  \multicolumn{1}{l}{kpc km/s} &   
  \multicolumn{1}{l}{kpc km/s} &   
  \multicolumn{1}{l}{light} &   
  \multicolumn{1}{l}{mass}&   
  \multicolumn{1}{l}{} \\ 
  
  \multicolumn{1}{l}{} &   
  \multicolumn{1}{l}{(1)} &   
  \multicolumn{1}{l}{(2)} &   
  \multicolumn{1}{l}{(3)} &   
  \multicolumn{1}{l}{(4)} &   
  \multicolumn{1}{l}{(5)} &   
  \multicolumn{1}{l}{(6)} &   
  \multicolumn{1}{l}{(7)}&   
  \multicolumn{1}{l}{} \\ 
\noalign{\smallskip}\hline\noalign{\smallskip}
NGC0584 & 11.03 & 11.16 & 203 $\pm$ 16 & 176 $\pm$ 35 & 153 $\pm$ 27 & 2.5  $_{- 0.17 } ^{+ 0.35 }$ & 2.49 $_{- 0.17 } ^{+ 0.36 }$  &   \\ [0.20cm]
NGC0821 & 10.91 & 11.02 & 50  $\pm$ 18 & 58  $\pm$ 26 & 47  $\pm$ 18 & 2.91 $_{- 0.57 } ^{+ 1.11 }$ & 2.88 $_{- 0.56 } ^{+ 1.17 }$  &   \\ [0.20cm]
NGC1023 & 10.86 & 10.94 & 261 $\pm$ 32 & 242 $\pm$ 30 & 222 $\pm$ 32 & 2.08 $_{- 0.04 } ^{+ 0.05 }$ & 2.08 $_{- 0.04 } ^{+ 0.05 }$  &   \\ [0.20cm]
NGC1316 & 11.91 & 12.06 & 670 $\pm$ 80 & 581 $\pm$ 130& 488 $\pm$ 95 & 2.91 $_{- 0.54 } ^{+ 1.08 }$ & 2.92 $_{- 0.56 } ^{+ 1.08 }$  &   \\ [0.20cm]
NGC1399 & 11.55 & 11.70 & 73  $\pm$ 72 & 63  $\pm$ 73 & 49  $\pm$ 59 & 6.49 $_{- 2.57 } ^{+ 4.75 }$ & 6.65 $_{- 2.73 } ^{+ 4.84 }$  &   \\ [0.20cm]
NGC2768 & 11.3  & 11.45 & 688 $\pm$ 72 & 595 $\pm$ 63 & 470 $\pm$ 72 & 2.18 $_{- 0.14 } ^{+ 0.44 }$ & 2.24 $_{- 0.2  } ^{+ 0.52 }$  &   \\ [0.20cm]
NGC2974 & 10.8  & 10.87 & 445 $\pm$ 15 & 405 $\pm$ 11 & 382 $\pm$ 15 & 2.53 $_{- 0.12 } ^{+ 0.22 }$ & 2.53 $_{- 0.12 } ^{+ 0.21 }$  &   \\ [0.20cm]
NGC3115 & 10.93 & 11.01 & 347 $\pm$ 25 & 282 $\pm$ 22 & 259 $\pm$ 25 & 2.23 $_{- 0.13 } ^{+ 0.32 }$ & 2.28 $_{- 0.14 } ^{+ 0.44 }$  &   \\ [0.20cm]
NGC3377 & 10.32 & 10.32 & 57  $\pm$ 11 & 55  $\pm$ 8  & 55  $\pm$ 8 & 2.09 $_{- 0.07 } ^{+ 0.11 }$ & 2.09 $_{- 0.07 } ^{+ 0.1  }$  &  no IMF corr \\ [0.20cm]
NGC3379 & 10.82 & 10.95 & 54  $\pm$ 12 & 52  $\pm$ 11 & 41  $\pm$ 12 & 5.48 $_{- 2    } ^{+ 2.21 }$ & 5.49 $_{- 1.97 } ^{+ 2.21 }$  &   \\ [0.20cm]
NGC3384 & 10.68 & 10.76 & 225 $\pm$ 5  & 195 $\pm$ 35 & 157 $\pm$ 24 & 2.23 $_{- 0.07 } ^{+ 0.07 }$ & 2.22 $_{- 0.08 } ^{+ 0.08 }$  &   \\ [0.20cm]
NGC3489 & 10.45 & 10.45 & 52  $\pm$ 8  & 45  $\pm$ 12 & 45  $\pm$ 12& 2.21 $_{- 0.1  } ^{+ 0.15 }$ & 2.22 $_{- 0.1  } ^{+ 0.14 }$  &  no IMF corr \\ [0.20cm]
NGC3608 & 10.77 & 10.83 & 215 $\pm$ 30 & 180 $\pm$ 25 & 156 $\pm$ 30 & 2.53 $_{- 0.3  } ^{+ 0.51 }$ & 2.57 $_{- 0.39 } ^{+ 0.56 }$  &   \\ [0.20cm]
NGC3923 & 11.57 & 11.71 & 667 $\pm$ 129& 578 $\pm$ 165& 499 $\pm$ 137& 2.76 $_{- 0.39 } ^{+ 0.81 }$ & 2.76 $_{- 0.39 } ^{+ 0.77 }$  &   \\ [0.20cm]
NGC4278 & 10.82 & 10.92 & 15  $\pm$ 12 & 7   $\pm$ 9  & 2   $\pm$ 12 & 4.54 $_{- 1.28 } ^{+ 1.91 }$ & 4.19 $_{- 1.13 } ^{+ 1.67 }$  &   \\ [0.20cm]
NGC4339 & 10.24 & 10.24 & 38  $\pm$ 4  & 33  $\pm$ 7  & 33  $\pm$ 7& 5.52 $_{- 1.28 } ^{+ 1.38 }$ & 5.56 $_{- 1.32 } ^{+ 1.36 }$  &  no IMF corr \\ [0.20cm]
NGC4365 & 11.5  & 11.58 & 141 $\pm$ 32 & 111 $\pm$ 25 & 91  $\pm$ 32 & 3.19 $_{- 0.63 } ^{+ 1.16 }$ & 3.2  $_{- 0.63 } ^{+ 1.15 }$  &   \\ [0.20cm]
NGC4374 & 11.47 & 11.65 & 207 $\pm$ 45 & 209 $\pm$ 45 & 167 $\pm$ 45 & 4.25 $_{- 1.41 } ^{+ 2.28 }$ & 4.25 $_{- 1.4  } ^{+ 2.26 }$  &   \\ [0.20cm]
NGC4472 & 11.77 & 11.91 & 570 $\pm$ 149& 494 $\pm$ 173& 426 $\pm$ 163& 2.72 $_{- 0.42 } ^{+ 0.73 }$ & 2.72 $_{- 0.43 } ^{+ 0.72 }$  &   \\ [0.20cm]
NGC4473 & 10.8  & 10.94 & 26  $\pm$ 14 & 25  $\pm$ 14 & 24  $\pm$ 14 & 2.41 $_{- 0.35 } ^{+ 0.56 }$ & 2.39 $_{- 0.32 } ^{+ 0.65 }$  &   \\ [0.20cm]
NGC4494 & 11.0  & 11.11 & 103 $\pm$ 17 & 108 $\pm$ 19 & 92  $\pm$ 17 & 3.76 $_{- 1.01 } ^{+ 1.61 }$ & 3.77 $_{- 0.99 } ^{+ 1.62 }$  &   \\ [0.20cm]
NGC4552 & 11.08 & 11.20 & 160 $\pm$ 35 & 139 $\pm$ 43 & 101 $\pm$ 30 & 2.71 $_{- 0.45 } ^{+ 0.8  }$ & 2.74 $_{- 0.47 } ^{+ 0.82 }$  &   \\ [0.20cm]
NGC4564 & 10.48 & 10.63 & 165 $\pm$ 6  & 119 $\pm$ 3  & 104 $\pm$ 6  & 2.04 $_{- 0.05 } ^{+ 0.06 }$ & 2.05 $_{- 0.06 } ^{+ 0.06 }$  &   \\ [0.20cm]
NGC4594 & 11.38 & 11.55 & 336 $\pm$ 55 & 296 $\pm$ 38 & 264 $\pm$ 55 & 2.82 $_{- 0.42 } ^{+ 0.55 }$ & 2.81 $_{- 0.41 } ^{+ 0.55 }$  &   \\ [0.20cm]
NGC4636 & 11.09 & 11.15 & 73  $\pm$ 34 & 63  $\pm$ 36 & 58  $\pm$ 33 & 4.91 $_{- 1.22 } ^{+ 3    }$ & 4.93 $_{- 1.22 } ^{+ 3.17 }$  &   \\ [0.20cm]
NGC4649 & 11.58 & 11.71 & 260 $\pm$ 38 & 233 $\pm$ 31 & 188 $\pm$ 38 & 3.86 $_{- 1.31 } ^{+ 4.02 }$ & 3.79 $_{- 1.22 } ^{+ 4    }$  &   \\ [0.20cm]
NGC4697 & 10.99 & 11.12 & 111 $\pm$ 20 & 108 $\pm$ 19 & 94  $\pm$ 20 & 2.83 $_{- 0.51 } ^{+ 1.33 }$ & 2.95 $_{- 0.55 } ^{+ 1.36 }$  &   \\ [0.20cm]
NGC4742 & 10.23 & 10.23 & 98  $\pm$ 5  & 85  $\pm$ 16 & 85  $\pm$ 16& 3.07 $_{- 0.35 } ^{+ 0.35 }$ & 3.05 $_{- 0.35 } ^{+ 0.39 }$  &  no IMF corr \\ [0.20cm]
NGC5128 & 11.0  & 11.03 & 568 $\pm$ 50 & 492 $\pm$ 101& 478 $\pm$ 88 & 2.77 $_{- 0.49 } ^{+ 0.95 }$ & 2.77 $_{- 0.5  } ^{+ 0.99 }$  &   \\ [0.20cm]
NGC5846 & 11.43 & 11.50 & 157 $\pm$ 62 & 134 $\pm$ 53 & 115 $\pm$ 62 & 5.38 $_{- 2.01 } ^{+ 3.3  }$ & 5.43 $_{- 2.   } ^{+ 3.34 }$  &   \\ [0.20cm]
NGC5866 & 10.91 & 10.98 & 260 $\pm$ 34 & 141 $\pm$ 12 & 128 $\pm$ 34 & 2.01 $_{- 0.04 } ^{+ 0.09 }$ & 2.03 $_{- 0.05 } ^{+ 0.08 }$  &   \\ [0.20cm]
NGC7457 & 10.12 & 10.12 & 164 $\pm$ 5  & 124 $\pm$ 4  & 124 $\pm$ 4  & 2.47 $_{- 0.14 } ^{+0.37  }$ & 2.53 $_{- 0.17 } ^{+ 0.27 }$  & no IMF corr \\ [0.20cm]
					  	
\noalign{\smallskip}\hline
\end{tabular}
\tablefoot{(1) Stellar masses derived as described in Sect.~\ref{sec:additional_data} assuming a Chabrier IMF; (2) stellar masses corrected for IMF gradients as described in Sect.~\ref{sec:ePN.S_jp_profiles_mass_variableIMF}; (3) light-weighted projected $j_p$, see Sect.~\ref{sec:ePN.S_jp_profiles_light}; (4) mass-weighted projected $j_p$ assuming constant IMF with radius, see Sect.~\ref{sec:ePN.S_jp_profiles_mass_constantIMF}, and (5) with variable IMF, see Sect.~\ref{sec:ePN.S_jp_profiles_mass_variableIMF}. (6 and 7) Median and quartiles of the distribution of correction factors $j_t(\leq 15R_e) / j_p(\leq a_{max})$ estimated on the TNG100 ETGs, selected based on $\Lambda_\mathrm{light}$ or $\Lambda_\mathrm{mass}$, see Sect.~\ref{sec:ePNS_jt_M*_relation}.
 }
 
\end{center}
\label{tab:Mass_AM} 
\end{table*}